\documentclass[12pt]{article}

\usepackage[makeroom]{cancel}
\usepackage[normalem]{ulem}
\usepackage{amsmath,amssymb}
\usepackage{slashed}
\usepackage{xcolor} 
\usepackage{graphicx}

\usepackage{cite}
\usepackage{pdflscape}
\usepackage{multirow}
\usepackage[thinlines]{easytable}
\usepackage{hyperref}
\usepackage{url}
\usepackage[capitalise]{cleveref}

\newcommand{\fref}[1]{Fig.~\ref{fig:#1}} 
\newcommand{\eref}[1]{Eq.~\eqref{eq:#1}}

\newcommand{\aref}[1]{Appendix~\ref{app:#1}}
\newcommand{\sref}[1]{Section~\ref{sec:#1}}
\newcommand{\tref}[1]{Table~\ref{tab:#1}}

\newcommand{\nn}{\nonumber \\}  
\newcommand{\nnl}{\nonumber \\}

\newcommand{\beq}{\begin{equation}} 
\newcommand{\eeq}{\end{equation}} 
\newcommand{\ba}{\begin{array}}  
\newcommand{\ea}{\end{array}} 
\newcommand{\bea}{\begin{eqnarray}}  
\newcommand{\eea}{\end{eqnarray} }  
\newcommand{\be}{\begin{eqnarray}}  
\newcommand{\ee}{\end{eqnarray} }  
\newcommand{\bal}{\begin{align}}
\newcommand{\eal}{\end{align}}   
\newcommand{\bi}{\begin{itemize}}  
\newcommand{\ei}{\end{itemize}}  
\newcommand{\ben}{\begin{enumerate}}  
\newcommand{\een}{\end{enumerate}}  
\newcommand{\bc}{\begin{center}}
\newcommand{\ec}{\end{center}} 
\newcommand{\bt}{\begin{table}}
\newcommand{\et}{\end{table}}  
\newcommand{\btb}{\begin{tabular}}
\newcommand{\etb}{\end{tabular}}  
\newcommand{\bvec}{\left ( \ba{c}}
\newcommand{\evec}{\ea \right )}
\def\beqa{\begin{eqnarray}}
\def\eeqa{\end{eqnarray}}

\newcommand{\cO}{{\mathcal O}}

\newcommand{\gev}{\mathrm{GeV}}

\newcommand{\eps}{\epsilon}

\newcommand{\eP}{\epsilon_P}
\newcommand{\eL}{\epsilon_L}
\newcommand{\eR}{\epsilon_R}

\begin{document}

\begin{titlepage}

\vspace*{-2cm}
\begin{flushright}
IFIC/21-52\\
FTUV-21-1201.4107\\
LA-UR-21-31791 \\
\vspace*{2mm}
\end{flushright}

\begin{center}
\vspace*{15mm}

\vspace{1cm}
{\LARGE \bf
 
Semileptonic tau decays \\ 
beyond the Standard Model
} 
\vspace{1.4cm}

\renewcommand{\thefootnote}{\fnsymbol{footnote}}
{Vincenzo~Cirigliano$^a$, David D\'iaz-Calder\'on$^b$, Adam~Falkowski$^c$, Mart\'{i}n~Gonz\'{a}lez-Alonso$^{b}$, Antonio Rodr\'iguez-S\'anchez$^{c}$}
\renewcommand{\thefootnote}{\arabic{footnote}}
\setcounter{footnote}{0}

\vspace*{.5cm}
\centerline{$^a${\it Theoretical Division, Los Alamos National Laboratory, Los Alamos, NM 87545, USA}}
\centerline{${}^{b}$ \it Departament de F\'isica Te\`orica, IFIC, Universitat de Val\`encia - CSIC,}
\centerline{\it Apt.  Correus 22085, E-46071 Val\`encia, Spain}
\centerline{$^c$ {\it Universit\'{e} Paris-Saclay, CNRS/IN2P3, IJCLab, 91405 Orsay, France}}

\vspace*{.2cm}

\end{center}

\vspace*{10mm}
\begin{abstract}\noindent\normalsize

Hadronic $\tau$ decays are studied as probe of new physics.
We determine the dependence of several inclusive and exclusive $\tau$ observables on the Wilson coefficients of the low-energy effective theory describing charged-current interactions between light quarks and leptons.  
The analysis includes both strange and non-strange decay channels. 
The main result is the likelihood function for the Wilson coefficients in the tau sector, based on the up-to-date experimental measurements and state-of-the-art theoretical techniques. 
The likelihood can be readily combined with inputs from other low-energy precision observables. 
We discuss a combination with nuclear beta, baryon, pion, and kaon decay data. 
In particular, we provide a comprehensive and  model-independent description of the new physics hints in the combined dataset, which are known under the name of the Cabibbo anomaly.

\end{abstract}

\end{titlepage}
\newpage 

\renewcommand{\theequation}{\arabic{section}.\arabic{equation}} 

\tableofcontents
\newpage

\section{Introduction}

Hadronic tau decays provide a unique laboratory to study fundamental physics~\cite{Pich:2013lsa,Schael:2005am}. 
In the past they have been mainly used to extract fundamental Standard Model (SM) parameters or to learn about low-energy hadronic physics.
In particular, {\em inclusive} tau decay observables play a role in the determination of the strong coupling constant~\cite{Braaten:1991qm,Boito:2014sta,Pich:2016bdg},  the strange quark mass, or the $V_{us}$ entry of the Cabibbo-Kobayashi-Maskawa (CKM) 
matrix~\cite{Gamiz:2002nu,Gamiz:2004ar}. 
They also provide  a valuable QCD laboratory, where chiral low-energy constants or properties of the QCD vacuum can be extracted with high precision  through dispersion relations~\cite{Boito:2015fra,Gonzalez-Alonso:2016ndl}. 
In what concerns {\em exclusive} tau decay channels, the two-body decays $\tau\to\pi \nu_\tau, K \nu_\tau$ are under firm theoretical control. 
Their key non-perturbative parameters, the pion and kaon decay constants,  are now precisely calculated in lattice QCD~\cite{Aoki:2021kgd}.  
On the other hand, exclusive modes with two or more hadrons in the final state  are much harder to predict within QCD with high accuracy.  

Whenever hadronic uncertainties can be brought under sufficient control, tau decays can also serve as useful probe of new particles and interactions beyond the Standard Model (BSM).
There are several immediate motivations for such studies.
One is the so-called {\em CKM unitarity problem}, or more generally the {\em Cabibbo anomaly}.
Different observables in kaon, pion, tau, and nuclear beta  decays point to mutually inconsistent values of the Cabibbo angle (if interpreted in the SM context)~\cite{Seng:2018yzq,Grossman:2019bzp,Coutinho:2019aiy}.  
Hadronic tau decays provide a valuable input about BSM models that can successfully address the tensions in the existing data. 
Another motivation is provided by the recent anomalies in $B \to D^{(*)} \tau \nu_\tau$ decays~\cite{HFLAV:2019otj}, which hint at new physics coupled to tau leptons. 
More generally, it is theoretically plausible that violation of lepton-flavor universality observed in $b \to s \mu \mu (e e)$ transitions~\cite{LHCb:2021trn,LHCb:2017avl} has a counterpart in the tau sector.
Many BSM models addressing these B-meson anomalies predict couplings of the new particles to the light quarks (up, down, strange), especially if they involve flavor symmetry as an organizing principle. 
Hadronic tau decays may provide important information about such models.

With the exception of the $\tau\to\pi(K) \nu_\tau$ channels, 
the BSM perspective has been rarely explored in the tau literature so far (but see~\cite{Bernard:2007cf,Garces:2017jpz,Cirigliano:2017tqn}). 
In a recent letter~\cite{Cirigliano:2018dyk}, we have embarked onto an unprecedented comprehensive analysis of the BSM reach of hadronic tau decays.
That analysis was based on an effective field theory (EFT) approach to new physics in  charged-current interactions~\cite{Cirigliano:2009wk,Cirigliano:2012ab}, 
rooted in the broader framework of the Standard Model EFT (SMEFT)~\cite{Buchmuller:1985jz}.  
In Ref.~\cite{Cirigliano:2018dyk}  we performed a quantitative analysis  of non-strange inclusive and exclusive  $\tau$ decays  and showed that the resulting constraints  on the EFT parameters (Wilson coefficients)  encoding new physics are very competitive and quite complementary to the ones obtained from  electroweak precision observables  and the LHC. 

With the present  manuscript  we continue our study of hadronic $\tau$ decays as probe of new physics.
First, we provide the fine-grained details that led to the results in Ref.~\cite{Cirigliano:2018dyk}, which we update with improved calculations and with current values for the experimental and theoretical inputs. 
We also extend the framework to the strange sector, which leads to novel results, in particular for the inclusive $\tau \to s$ decays. 
Finally, we combine our results for strange and non-strange hadronic tau decays with the results obtained with $d\to u\ell\nu_\ell$ and $s\to u\ell\nu_\ell$ transitions  
in Ref.~\cite{Gonzalez-Alonso:2016etj}  (which we update to include recent developments).   
This exercise leads us to the most comprehensive analysis to date of new physics effects in the charged-current transitions involving the light quarks. 
The combination is particularly relevant to frame   
the possible BSM explanations to the Cabibbo anomaly together with all the related low-energy observables. 
We provide the final combined likelihood for the low-scale EFT Wilson coefficients and perform a first exploration of some of the preferred directions in the space of BSM couplings.

The paper is organized as follows. 
In Section \ref{sec:theory} we briefly introduce the theoretical framework that we use in the rest of this work. 
The phenomenological study starts in Section~\ref{sec:TauToPnu}, where we apply the formalism to translate updated results in $\tau\to \pi\nu_\tau, K\nu_\tau$ decays into new physics bounds. 
The sensitivity of two-hadron decays to potential new physics effects is studied in Section~\ref{sec:TauToPPnu}, with special emphasis on those channels for which the limitations on the predictive power can be overcome, $\tau\rightarrow \pi\pi\nu_{\tau}$, $\tau\rightarrow \eta\pi\nu_{\tau}$ and, up to a certain extent, $\tau\rightarrow K\pi\nu_{\tau}$. 
In Section \ref{sec:inclusive} we discuss inclusive tau decays. 
We extend the traditional SM framework, based on dispersion relations, to describe also potential new physics effects. We study the associated phenomenology, including improvements of the results of Ref.~\cite{Cirigliano:2018dyk} and extension to the inclusive strange sector. 
We recapitulate the obtained $\tau$ bounds and perform the combination with the $d\to u\ell\nu_\ell$ and $s\to u\ell\nu_\ell$ transitions in Section~\ref{sec:RecapAndFlavor}, where we also show some important applications of the combined likelihood. 
Our final conclusions and remarks are given in Section~\ref{sec:conclusions}.
Additional technical details and results are shown in Appendices.

\section{Theoretical framework}
\label{sec:theory}
\setcounter{equation}{0}

We work in the framework of a low-energy EFT where the degrees of freedom are the light quarks ($u$, $d$, $s$), charged leptons ($e$, $\mu$, $\tau$), neutrinos ($\nu_e$, $\nu_\mu$, $\nu_\tau$), gluon, and photon. 
The remaining particles of the SM have been integrated out, in particular the surviving gauge symmetry is  $U(1)_{\rm em}\times SU(3)_C$.  
We assume the absence on any exotic degrees of freedom with masses below $\sim 2$~GeV; in particular we do not consider right-handed neutrinos here. 
This framework is referred to as the WEFT (or WET, or LEFT) in the literature. 
For the sake of this paper we focus on the subset of the  Lagrangian describing the leading order effective charged-current weak interactions between quarks and leptons. 
We parametrize these interactions as~\cite{Cirigliano:2009wk}:\footnote{We have not included wrong-flavor neutrino interactions~\cite{Cirigliano:2012ab}. 
These do not interfere with the SM amplitude and thus contribute to the observables only at $\mathcal{O}(\epsilon_X^2)$, 
except in neutrino oscillation observables~\cite{Falkowski:2019xoe}.}
\begin{eqnarray}
\label{eq:leff1} 
{\cal L}_{\rm eff} 
&=& - \frac{G_\mu V_{uD}}{\sqrt{2}}  \Bigg[
\Big(1 + \epsilon_L^{ D\ell}  \Big) \bar{\ell}  \gamma_\mu  (1 - \gamma_5)   \nu_{\ell} \cdot \bar{u}   \gamma^\mu (1 - \gamma_5 ) D
+  \eR^{D\ell}  \   \bar{\ell} \gamma_\mu (1 - \gamma_5)  \nu_\ell    \cdot \bar{u} \gamma^\mu (1 + \gamma_5) D
\nonumber\\
&&+~ \bar{\ell}  (1 - \gamma_5) \nu_{\ell} \cdot \bar{u}  \Big[  \epsilon_S^{D\ell}  -   \eP^{D\ell} \gamma_5 \Big]  D
+{1 \over 4} \hat \epsilon_T^{D\ell} \,   \bar{\ell}   \sigma_{\mu \nu} (1 - \gamma_5) \nu_{\ell}    \cdot  \bar{u}   \sigma^{\mu \nu} (1 - \gamma_5) D
\Bigg]+{\rm h.c.}, 
\end{eqnarray}
where $D=d,s$ is the down-type quark flavor, $\ell=e,\mu,\tau$ is the lepton flavor, and $\sigma^{\mu \nu} = i\,[\gamma^\mu, \gamma^\nu]/2$.
The normalization is provided by the  Fermi constant $G_\mu=1.16638 \times 10^{-5}~\gev^{-2}$ measured in muon decay.
$V_{ud}$ and $V_{us}$ are elements of the unitary CKM matrix, and they are positive and real by convention.
Consequently, the two are not independent, but instead  are tied by the  unitarity relation $V_{ud}^2+V_{us}^2=1$.\footnote{More precisely  $V_{ud}^2+V_{us}^2 +|V_{ub}|^2 =1$ but, given $|V_{ub}| = 3.82(24) \times 10^{-3}$~\cite{ParticleDataGroup:2020ssz}, $V_{ub}$ has a negligible effect on this relation.}
The effects of physics beyond the SM are parametrized by the Wilson coefficients $\epsilon_X^{q \ell}$.
The main goal of this paper is to derive novel constraints on new physics in the tau sector, and construct a likelihood function for  $\epsilon_X^{D\tau}$.  

The Wilson coefficients $\epsilon^{D\tau}_X$ are renormalization scale and scheme dependent~\cite{Gonzalez-Alonso:2017iyc}. Numerical values shown in this work are obtained at $\mu= 2 \, \mathrm{GeV}$ in the $\overline{MS}$ scheme. 
This choice is convenient mainly because it is the standard one used by the lattice community to give their results, which we use as inputs in our approach.
Note that $\epsilon_X^{D\ell}$ are in general complex parameters, but the sensitivity of the observables considered in this work to their imaginary parts  is very small (with some exceptions that will be mentioned explicitly). Thus, the results hereafter implicitly refer to the real parts of $\epsilon_X^{D\ell}$, unless otherwise stated. We added a hat on the tensor Wilson coefficient to stress the fact that it differs by a factor of four with the notation of our previous work on tau decays~\cite{Cirigliano:2018dyk}. The normalization used in this work is such that BSM models producing tensor interactions give typically similar contribution to $\hat \epsilon_T^{D\ell}$, $\epsilon_S^{D\ell}$ and $\epsilon_P^{D\ell}$~\cite{Dorsner:2016wpm,Angelescu:2021lln}.

In the presence of general new physics, observables never probe the CKM elements directly. 
Instead, they always probe certain combinations of $V_{uD}$ and $\epsilon_X^{D \ell}$.
For this reason it is convenient to define ``polluted" CKM elements that relate in a more straightforward way to observables, and which can be assigned numerical values based on available experimental data~\cite{Descotes-Genon:2018foz}.
We define
\begin{equation}
\hat V_{uD}=\left(1 + \epsilon_L^{De} + \epsilon_R^{De} \right)\,V_{uD}~ . 
\label{eq:VuDNPs}
\end{equation}
The point of this definition is that the vector currents coupling electrons to light quarks  depend only on $\hat V_{uD}$ and not on $\epsilon_X^{D \ell}$.  
Consequently, $\hat V_{ud}$ and $\hat V_{us}$ can be readily extracted,  respectively, from nuclear decays and $K\to \pi e \nu_e$~\cite{Gonzalez-Alonso:2016etj}. 
In our analysis of hadronic tau decays we will use the numerical values
\begin{equation}
\label{eq:hatvalues}
\hat V_{ud}=0.97386(40)\; ,\quad\quad\quad
\hat V_{us} = 0.22306(56)\; .
\end{equation}
These values are extracted from $d\to u e\bar{\nu}_e$ transitions 
and $K_{e 3}$ decays (taking into account possible nonstandard effects), as explained in detail in~\sref{LEFFE}.

\section{$\tau\to \pi\nu_\tau, K\nu_\tau$}
\label{sec:TauToPnu}
\setcounter{equation}{0}

The single-hadron channels, $\tau\to \pi\nu_\tau, K\nu_\tau$, are the only hadronic decays of the tau lepton that are widely perceived as sensitive new physics probes (see e.g. Refs.~\cite{Pich:2013lsa,Filipuzzi:2012mg}), especially through ``theoretically clean" ratios such as $\Gamma(\tau\to\pi \nu_\tau)/\Gamma(\pi\to \mu \nu_\mu)$ where the main QCD contributions cancel. The separate branching ratios are also powerful probes because the QCD effects are captured by a single quantity, the pion and kaon decay constants $f_{\pi,K}$, which can be calculated accurately in lattice QCD~\cite{Aoki:2021kgd}. 

The width of these channels in the presence of non-standard interactions is given by~\cite{Rodriguez-Sanchez:2018dzw}
\bea
\Gamma(\tau \to P \nu_\tau) 
&=& \frac{m_\tau^3 f_P^2 G_\mu^2 |\hat{V}_{uD}|^2}{16\pi} \left( 1- \frac{m_P^2}{m_\tau^2} \right)^2 (1 + \delta_{\rm RC}^{(P)} ) \left( 1+ 2\,\delta_{\rm BSM}^{(P)} \right)~\\
&=& \hat{\Gamma}(\tau \to P \nu_\tau)_{\rm SM} \left( 1+ 2\,\delta_{\rm BSM}^{(P)} \right)~, 
\eea
where
\bea
\label{eq:deltaBSMP}
\delta_{\rm BSM}^{(P)} ~= ~\eL^{D\tau} - \eL^{De} - \eR^{D\tau} - \eR^{De} - \frac{B_0^D}{m_\tau} \eP^{D\tau}~.
\eea
Here $D=d,s$ for $P=\pi,K$ respectively, $B_0^D$ is a short notation for the ratio $m_P^2/(m_u+m_D)$, 
$f_P$ is the pseudoscalar decay constant, and $\delta^{(P)}_{\rm RC}$ are the radiative corrections (RC). The hat in $\hat{\Gamma}(\tau \to \pi \nu_\tau)_{\rm SM}$ reminds that the ``polluted" CKM element $\hat{V}_{uD}$ was used.
Let us note that the huge chiral enhancement of the pseudoscalar piece in $\pi (K)\to \ell\nu_\ell$ is not present here due to the large tau mass ($B_0^D/m_\tau \sim 1$).  

Combining the PDG values of the branching ratios (BR) with the tau lifetime~\cite{ParticleDataGroup:2020ssz} we find the following experimental values
\bea
\Gamma(\tau \to \pi \nu_\tau)_{\rm exp} &=& 2.453(12)\times10^{-13}\,\rm{GeV}~,\\
\Gamma(\tau \to K \nu_\tau)_{\rm exp} &=& 1.578(23)\times10^{-14}\,\rm{GeV}~,
\eea
with a $4.2\%$ correlation that we will take into account. 
The 0.5\% uncertainty in $\tau \to \pi \nu_\tau$ is dominated by the BR error, but with a small contribution from the lifetime error, whereas the $1.4 \%$ error in $\tau \to K \nu_\tau$ is entirely dominated by the BR error.

For the calculation of the SM prediction, we use $f_{K^\pm}/f_{\pi^\pm}=1.1932(21)$~\cite{Aoki:2021kgd,Dowdall:2013rya,Carrasco:2014poa,Bazavov:2017lyh,Miller:2020xhy,Follana:2007uv,Bazavov:2010hj,Durr:2010hr,Blum:2014tka,Durr:2016ulb,Bornyakov:2016dzn,Blossier:2009bx}
and $f_{\pi^\pm}=130.2(8)$ MeV~\cite{Aoki:2021kgd,Follana:2007uv,Bazavov:2010hj,Blum:2014tka}
For the radiative corrections we use $\delta_{\rm RC}^{(\pi)}=+1.94(61)\%$ and $\delta_{RC}^{(K)}=+2.04(62)\%$ which we obtain by combining the recent calculation of the RC to the ratio $\tau\to P\nu_\tau / P\to \mu\nu_\mu$~\cite{Arroyo-Urena:2021nil} 
and those to $P\to \mu\nu_\mu$ from chiral perturbation theory~\cite{Cirigliano:2007xi,Rosner:2015wva}.
We see that the RC themselves cannot be neglected, but their uncertainties are subleading compared with the $f_P$ and experimental ones.  
Altogether we find
\bea
\label{eq:GammaSM}
\hat{\Gamma}(\tau \to \pi \nu_\tau)_{\rm SM} &=& 2.458(34)\times10^{-13}\,\rm{GeV}~,\\
\hat{\Gamma}(\tau \to K \nu_\tau)_{\rm SM} &=& 1.584(24)\times10^{-14}\,\rm{GeV}~,
\eea
with a correlation of $\rho=0.73$. 

Putting the SM and experimental results together gives the following 68\%~CL results
\bea
\label{eq:epsilonpi}
\delta^{(\pi)}_{\rm BSM} &=& \eL^{d\tau} - \eL^{de} - \eR^{d\tau} - \eR^{de} - \frac{B_0^d}{m_\tau} \eP^{d\tau} = -(0.9\pm 7.3)\times 10^{-3}~,\\
\label{eq:epsilonK}
\delta^{(K)}_{\rm BSM} &=& \eL^{s\tau} - \eL^{se} - \eR^{s\tau} - \eR^{se} - \frac{B_0^s}{m_\tau} \eP^{s\tau} = -(2\pm 10)\times 10^{-3}~,
\eea
with a 51\% correlation. 
The small difference in the $\tau\to\pi\nu_\tau$ constraint with Ref.~\cite{Cirigliano:2018dyk} is due to the new input used for the radiative corrections~\cite{Arroyo-Urena:2021nil}. The slightly smaller error in Ref.~\cite{Roig:2019rwf} for the $\tau\to K \nu_\tau$ constraint is obtained using the $f_{K^\pm}$ FLAG average. The latter includes calculations where the QCD scale is set using the experimental $f_{\pi^\pm}$ value, which is polluted by BSM effects in the general EFT setup. For this reason we have used instead the lattice calculations of $f_{\pi^\pm}$ and $f_{K^\pm}/f_{\pi^\pm}$ as inputs in our analysis.

Equivalently, the new physics bounds obtained above are simply the result of comparing the value of $V_{ud} \,(V_{us})$ obtained from $\tau \to \pi \nu_\tau (K \nu_\tau)$ with $\hat{V}_{ud}$ and $\hat{V}_{us}$, which are obtained from $d\to u e\bar{\nu}_e$ transitions 
and $K_{e 3}$ decays. 
More explicitly:
\bea
V_{us}^{\tau\to K\nu} &=& 
\hat{V}_{us}
\left(1 + \delta_{\rm BSM}^{(K)} \right) ~,\\
\left[\frac{V_{us}}{V_{ud}}\right]^{\frac{\tau\to K\nu}{\tau\to \pi \nu}} &=& 
\frac{\hat{V}_{us}}{\hat{V}_{ud}}
\left(1 + \delta_{\rm BSM}^{(K)} - \delta_{\rm BSM}^{(\pi)} \right)~.
\eea
Thus, our results make it possible to understand which specific BSM effects we are probing when we compare these different $V_{us}$ extractions. 

We discuss now briefly the uncertainty sources. The error decomposition for the $\tau \to \pi\nu_\tau$ bound is
\bea
7.3\times 10^{-3} = (2.5_{\rm exp}\pm 6.1_{f_{\pi^\pm}} \pm 3.0_{\rm RC} \pm 0.4_{{\hat{V}_{ud}}})\times 10^{-3}~,
\eea
{\it i.e.}, the error is dominated by the $f_{\pi^\pm}$ uncertainty. 
Improved future determinations of this quantity are therefore crucial to search for new physics in this process. 

The error decomposition for the $\tau \to K\nu_\tau$ bound is
\bea
10\times 10^{-3} = (7.2_{\rm exp}\pm 6.1_{f_{\pi^\pm}}\pm 3.0_{\rm RC}\pm 2.5_{\hat{V}_{us}}\pm 1.8_{f_{K^\pm}/f_{\pi^\pm}})\times 10^{-3}~,
\eea
{\it i.e.}, in this channel the experimental error dominates, but closely followed by the $f_{K^\pm}$ (via $f_{\pi^\pm}$) uncertainty. 
Thus, a combined experimental and lattice effort is needed to make significant progress in the BSM bound from $\tau\to K\nu_\tau$ given above. 
Finally we note that the RC and $\hat{V}_{us}$ errors are also not negligible.

Let us stress that the analysis above includes the ratio $\Gamma(\tau\to K\nu_\tau) / \Gamma(\tau\to \pi\nu_\tau)$, fully taking into account that its SM prediction is better known thanks to the precise lattice calculation of the $f_{K^\pm}/f_{\pi^\pm}$ ratio. This is indeed the origin of the significant correlation between the bounds in~\eref{epsilonpi} and~\eref{epsilonK}. We note that a further reduction in the $f_{K^\pm}/f_{\pi^\pm}$ uncertainty will have a minor impact in the BSM bounds above. This is in contrast with meson decays, where experimental measurements are more precise and the $f_{K^\pm}/f_{\pi^\pm}$ uncertainty plays a major role.

Likewise, once we combine the above tau-decay bounds with those obtained from pion and kaon decays, which we will do in~\sref{RecapAndFlavor}, our final likelihood will take into account that stringent BSM constraints can be obtained from ``theoretically clean'' ratios of observables where the $f_{\pi,K}$ dependence cancels out, such as $\Gamma(\tau\to\pi \nu_\tau)/\Gamma(\pi\to \mu \nu_\mu)$. This is once again reflected in significant correlations between tau and meson decay bounds due to common $f_{\pi,K}$ uncertainties.

Let us briefly discuss the expected impact of future lattice calculations and new data from facilities such as Belle-II. 
Major improvements are not expected in $f_{\pi,K}$~\cite{Cerri:2018ypt}, in part because decreasing further the scale setting error is challenging, and in part because of a lack of motivation. Our results show that the latter is actually not a good reason and we encourage efforts to improve these quantities, which would also improve BSM bounds (or $V_{us}$ determinations) extracted from $K\to\mu\nu_\mu$. Nonetheless we expect some modest improvement. 
Improvements in the experimental determination of the $\tau\to\pi\nu_\tau (K\nu_\tau)$ branching ratio seem also possible with the arrival of Belle-II (or even with the existing BaBar and Belle data, see e.g. Ref.~\cite{Lueck-talk-ICHEP2018}). Indeed the current PDG result is dominated by a BaBar measurement~\cite{Aubert:2009qj}.

\section{$\tau \to PP'\nu_\tau$}
\label{sec:TauToPPnu}
\setcounter{equation}{0}

The decay of $\tau$  into two pseudoscalar mesons ($\tau^-\to\nu_\tau P^-P^{'0}$)   is mediated in the SM by the vector current. 
In presence of new physics,  scalar and tensor operators can contribute as well.  
The relevant hadronic matrix elements can be parametrized in terms of appropriate from factors as follows~\cite{Pich:2013lsa} 
(as usual, $D$ stands for a down-type quark, $d$ or $s$)
\bea
\langle P^- P^{'0}|\,\bar D \gamma^\mu u \, |0\rangle\;   &= &
\; C_{PP'}\,\left\{  \left( p_--p_0-\frac{\Delta_{PP'}}{s}\, q\right)^\mu\, F^{PP'}_V(s)\, +\,  \frac{\Delta_{PP'}}{s}\, q^\mu\; F^{PP'}_S(s)\right\} 
\\
\langle P^- P^{'0}|\,\bar D    u \, |0\rangle\;   &= &  - C_{PP'} \,   \frac{ \Delta_{PP'}}{m_{D} - m_u} \, F_S^{PP'} (s) 
\\
\langle P^- P^{'0}|\,\bar D  \sigma^{\mu \nu}   u \, |0\rangle\;   &= &  - i  \ 
\Big( p_-^\mu p_0^\nu - p_-^\nu p_0^\mu \Big)    \   F_T^{PP'} (s) ~, 
\eea
where $p_-^\mu$ and $p_0^\mu$ are the momenta of the charged and neutral pseudoscalars, 
$q^\mu = (p_-+p_0)^\mu$ and $s=q^2$.
In the matrix element of the vector current, 
the two Lorentz structures correspond to $J^P=1^-$ and $0^+$ transitions. The
scalar contribution is suppressed by the mass-squared difference $\Delta_{PP'} = m_{P^-}^2-m^2_{P^{'0}}$ because the vector current is conserved in the limit of equal quark masses. 
The  normalization coefficients $C_{PP'}$  (chosen so that the vector form factor satisfies  $F^{PP'}_V(0) = 1$, except for the $\pi\eta$ one, which vanishes in the isospin limit) are given by:
\be
C_{\pi\pi} = C_{\pi\eta} = \sqrt{2}\, ,\qquad
C_{K\bar K} = -1\, ,\qquad
C_{K\pi} = \frac{1}{\sqrt{2}}\, ,\qquad
C_{\pi \bar K} = -1\, ,\qquad
C_{K\eta_8} = \sqrt{\frac{3}{2}}\, .
\ee

New physics effects modify the $\tau^-\to\nu_\tau P^-P^{'0}$ decay rate  in several ways: 
(i) $\epsilon_L^{D\tau} + \epsilon_R^{D\tau}$ (the shift in the vector current) modifies the overall normalization; 
(ii) the effect of the tensor coupling $\hat{\epsilon}_T^{D\tau}$ cannot be absorbed in any SM piece and contributes with  a different kinematic dependence; 
(iii) finally,  the effect of the scalar coupling $\epsilon_S^{D\tau}$ can be absorbed in the redefinition 
$F_S(s) \to  F_S(s)  ( 1 +   \epsilon_S^{D\tau}   \ s / (m_\tau ( m_{D} - m_u))$.
Explicitly, the hadronic invariant-mass distribution including new physics effects to first order 
is given by~\cite{Garces:2017jpz,Miranda:2018cpf,Rodriguez-Sanchez:2018dzw}
\bea
\label{PP_spectrum}
\frac{d\Gamma^{}}{d s}   &=& \left[ \frac{d \hat \Gamma^{}}{d s} \right]_{\rm SM}  \bigg( 1 + 2 (\epsilon_L^{D\tau} + \epsilon _R^{D\tau} -  \epsilon_L^{De} - \epsilon _R^{De}) 
 +  a_S (s) \, \epsilon_S^{D\tau}  + 
a_T (s) \, \hat{\epsilon}_T^{D\tau}   +  {\cal O}(\epsilon^2)  \bigg)
\qquad  
\\
\left[ \frac{d \hat \Gamma^{}}{d s} \right]_{\rm SM}   &=&    \frac{G_\mu^2 |\hat{V}_{uD}|^2 m_\tau^3}{768\pi^3}\;
S_{EW}^{\mathrm{had}}\; C_{PP'}^2\,
\biggl(1-\frac{s}{m_\tau^2}\biggr)^{2}\;
\nonumber \\
& \times &    \Biggl\{  \biggl(1+2\,\frac{s}{m_\tau^2} \biggr)\, \lambda_{PP'}^{3/2}\: |F_V^{PP'}(s)|^2
+ 3\,\frac{\Delta_{PP'}^2}{s^2}\,\lambda_{PP'}^{1/2}\:
 |F_S^{PP'}(s)|^2 \Biggr\}\, ,
\\
a_S(s)  &=&  \frac{ 6 |F_S(s)|^2  \frac{s \Delta_{PP'}^2}{(m_{D} - m_u) m_\tau}}{
3 |F_S(s)|^2   \Delta_{PP'}^2   + |F_V(s)|^2  (1 + 2 \frac{s}{m_\tau^2})   \lambda(s,m_P^2,m_{P'}^2)}
\\
a_T(s)  &=& \frac{3}{c_{PP'}}  \   \frac{ \frac{s}{m_\tau}  {\rm Re} \left(F_V(s)   F_T^* (s) \right)  \lambda (s,m_P^2, m_{P'}^2) 
}{
3 |F_S(s)|^2   \Delta_{PP'}^2   + |F_V(s)|^2  (1 + 2 \frac{s}{m_\tau^2})   \lambda(s,m_P^2,m_{P'}^2)} , 
\eea
where $\lambda_{PP'}\equiv\lambda(s,m_{P^-}^2,m^2_{P'^{0}} )/s^2$ 
and  the hat in $[d\hat{\Gamma}/ds]_{\rm SM}$ indicates, once again, 
that the $\hat{V}_{ud}$ value was used.\footnote{The K\"all\'en function is defined as usual: $\lambda(x,y,z)\equiv x^2+y^2+z^2-2xy-2xz-2yz$.}  
$S_{EW}^{\mathrm{had}}=1.0157(3)$ accounts for the short-distance
electroweak corrections~\cite{Marciano:1988vm,Braaten:1990ef,Erler:2002mv}. 
Long-distance electromagnetic corrections and isospin-breaking contributions are channel dependent and have been  studied  for the $\pi\pi$~\cite{Cirigliano:2002pv,Davier:2009zi} 
and $K\pi$~\cite{Antonelli:2013usa,Flores-Baez:2013eba} final states. 
Additional angular and kinematic distributions  (which have not been measured yet) have been presented in Refs.~\cite{Miranda:2018cpf,Garces:2017jpz} including BSM effects. 
We next discuss the new physics constraints that can be obtained in various channels.

\subsection{$\tau \to \pi\pi\nu_\tau$}

This channel  has sensitivity only to the vector ($\epsilon^{d\tau}_L + \epsilon^{d\tau}_R$) and the tensor ($\hat \epsilon_T^{d\tau}$) contributions, 
due to the  fact that  $a_S (s) \sim  \Delta_{\pi \pi}/s \ll 1$ 
across the whole physical region  $4 m_\pi^2  < s < m_\tau^2$.  
Therefore, the expressions  in Eq.~(\ref{PP_spectrum}) reduce to:
\bea
\label{pipi_spectrum}
\left[ \frac{d \hat \Gamma_{\pi \pi}}{d s \ \ \  } \right]_{\rm SM}   &=&    \frac{G_{\mu}^2 |\hat{V}_{ud}|^2 m_\tau^3}{384\pi^3}\;
S_{EW}^{\mathrm{had}}\;  
\biggl(1-\frac{s}{m_\tau^2}\biggr)^{2}\;
     \biggl(1+2\,\frac{s}{m_\tau^2} \biggr)\, \lambda_{\pi\pi}^{3/2}\: |F_V^{\pi \pi}(s)|^2  \ G_{EM} (s)
\\
a_T(s)  &=& \frac{3}{\sqrt{2}}  \  \frac{ {\rm Re} \left(F^{\pi \pi}_V(s)   F_T^{\pi \pi*} (s) \right)}{|F_V^{\pi \pi}(s)|^2 }
\,  
\frac{\frac{s}{m_\tau}}{1 + 2 \frac{s}{m_\tau^2}}
\eea
where $G_{EM} (s)$ represents the long-distance radiative corrections~\cite{Cirigliano:2002pv}.    
In order to  constrain the  BSM couplings,  one needs to know the vector form factor
$F_V^{\pi \pi}(s)$    (controlling the SM amplitude)  and tensor form factor
$F_T^{\pi \pi}(s)$   (controlling the ``BSM leverage arm"  $a_T(s)$).   
The uncertainty in  $F_V^{\pi \pi}(s)$ ultimately limits the strength of the bounds on BSM couplings, 
while  the  requirement on the uncertainty on  $F_T^{\pi \pi}(s)$  is less stringent. 
Since  $F_{V,T}^{\pi \pi}(s)$  involve non-perturbative QCD dynamics, they are hard to predict in a model-independent way, 
and we discuss below our strategy to obtain reliable form factors.

Extracting   $F_V^{\pi \pi}(s)$  from the $\pi \pi$ invariant mass distribution in  $\tau\to\pi\pi\nu_\tau$ is not feasible at the moment, 
as this distribution is  potentially contaminated by new physics contributions. 
We note, however,  that   $F_V^{\pi \pi}(s)$  can be extracted  from the  process 
$e^+e^-\to\pi^+\pi^-$, after the proper inclusion of isospin-symmetry-breaking corrections 
(see Refs.~\cite{Davier:2013sfa,Davier:2017zfy,Keshavarzi:2018mgv,Jegerlehner:2011ti} and references therein). 
The crucial point here is that  new physics effects (associated with the scale $\Lambda \gg 2~{\rm GeV}$) can be entirely neglected  in $e^+e^-\to\pi^+\pi^-$  at  energy $\sqrt{s} \ll \Lambda$ due to  the  electromagnetic nature of this process. 
In this context one can benefit from past studies that exploited this isospin relation to extract from both
$\tau\to\pi\pi\nu_\tau$   and  $e^+e^-\to\pi^+\pi^-$     data the $\pi\pi$ component of the lowest-order hadronic   
vacuum polarization contribution to the muon $g-2$, usually denoted by $a_\mu^{\rm{had,LO}}[\pi\pi]$. (This approach 
implicitly assumes  the absence of BSM effects, which however may contaminate the $\tau$ data.)
While these studies entail an extraction of $F_V^{\pi \pi} (s)$  by averaging various $e^+e^-$ datasets, 
here we chose not to use the full spectral information but rather perform a simpler analysis based 
on the particular weighted integrals of 
$d \Gamma_{\pi \pi} /d s$,     
corresponding to  $a_\mu^{\rm{had,LO}}[\pi\pi]$.

We begin by defining 
\be
a_\mu^{\tau}  \equiv   \int_{4 m_\pi^2}^{m_\tau^2}  \, ds \   W_{a_\mu} (s)  \ \frac{d \Gamma_{\pi \pi}}{d s \ \ \  } ~, 
\label{eq:WI1}
\ee
where  the weight factor $W_{a_\mu} (s)$ is~\cite{Eidelman:1995ny,Brodsky:1967sr,Bouchiat:1957zz}
\begin{eqnarray}
    W_{a_\mu}(s)
    &\equiv& 
    \frac{32\pi\alpha_0^2}{G_\mu^2|\hat{V}_{ud}|^2m_\tau^5}
    \left(\frac{s}{m_\tau^2}\left(1-\frac{s}{m_{\tau}^2}\right)^2\left(1+\frac{2s}{m_{\tau}^2}\right)\right)^{-1}\,
    \,\int_0^1 dx \frac{x^2(1-x)}{x^2+\frac{s}{m_{\mu}^2}(1-x)}
    ~\xi(s)~, \quad  
    \end{eqnarray}
where $\alpha_0$ is the fine structure constant and
$\xi(s)$ encodes the radiative corrections and isospin breaking  effects~\cite{Cirigliano:2001er,Cirigliano:2002pv,Zhang:2009ag,Davier:2013sfa,Aoyama:2020ynm}. 

In absence of new physics, 
the spectral integral defined by $a_\mu^{\tau}$ gives 
the $\pi\pi$ component of the lowest-order hadronic vacuum polarization contribution to the muon $g-2$, namely $a_\mu^{\rm{had,LO}}[\pi\pi]$. 
Moreover, still assuming no BSM contributions, $a_\mu^\tau$  
should coincide within errors with the corresponding quantity $a_\mu^{ee}$   obtained from $e^+ e^- \to \pi^+ \pi^-$ data, 
assuming isospin-breaking effects and their uncertainty are properly taken into account.

On the other hand,  in presence of new physics  one has 
\bea
a_\mu^{\tau} &  = &  
\int_{4 m_\pi^2}^{m_\tau^2}  \, ds \   W_{a_\mu} (s)  \ \left(\frac{d \hat{\Gamma}_{\pi \pi}}{d s \ \ \  }\right)_{\mathrm{SM}}
\times 
\bigg( 1 + 2 (\epsilon_L^{d\tau} + \epsilon _R^{d\tau} -  \epsilon_L^{de} - \epsilon _R^{de}) 
+ a_T (s) \, \epsilon_T^{d\tau}    \bigg)
\nonumber \\
&=&    a_\mu^{ee}  \Big( 1 + 2 (\epsilon_L^{d\tau} + \epsilon _R^{d\tau} -  \epsilon_L^{de} - \epsilon _R^{de})  \Big) + 
\hat{\epsilon}_T^{d\tau} \,   \int_{4 m_\pi^2}^{m_\tau^2}  \, ds \   W_{a_\mu} (s)  \ \frac{d \hat{\Gamma}_{\pi \pi}}{d s \ \ \  } \,  a_T(s)~, 
\eea
which leads to 
\bea
\label{eq:pipi0}
\frac{a_\mu^{\tau}\!-\!a_\mu^{ee}}{2\,a_\mu^{ee}} \! &=&\! \eL^{d\tau}\!-\!\eL^{de}\!+\!\eR^{d\tau}\!-\!\eR^{de} + c_T \,\hat{\epsilon}_T^{d\tau}
\\
\label{eq:cT}
c_T &=& 
\frac{1}{2} \   \frac{  \int_{4 m_\pi^2}^{m_\tau^2}  \, ds \   W_{a_\mu} (s)  \ \frac{d \hat{\Gamma}_{\pi \pi}}{d s \ \ \  } \,  a_T(s) 
}{
\int_{4 m_\pi^2}^{m_\tau^2}  \, ds \   W_{a_\mu} (s)  \ \frac{d \hat{\Gamma}_{\pi \pi}}{d s \ \ \  }  
}~.
\eea

To estimate the coefficient  $c_T$ multiplying $\epsilon_T$  in~\eref{pipi0}, we   use  
a relatively simple  form of the vector form factor based on analyticity, unitarity,  chiral symmetry, and the high-momentum asymptotic 
behavior of QCD~\cite{Guerrero:1997ku}, as well as a dispersive parameterization based on data (see Ref.~\cite{Celis:2013xja} and references therein).\footnote{At the precision needed we can ignore isospin-breaking and new physics contaminations.}  
We treat the tensor form factor as follows: 
\begin{itemize}
\item First, we  assume  that the proportionality of the tensor and vector form factors, which is exact in the  elastic region~\cite{Cirigliano:2017tqn,Miranda:2018cpf},   
holds over the whole $s$ region allowed by kinematics, namely 
\bea\label{eq:tensorshapepipi}
F_T^{\pi \pi} (s) =    F^{\pi \pi}_T(0)   F_V^{\pi \pi} (s)~. 
\eea
Note that this proportionality also holds in the resonance chiral theory framework~\cite{Ecker:1988te}, assuming dominance of the lowest lying state.
Since in the elastic region $s \leq 1$~GeV${}^2$  the form factors  $F_{V,T}^{\pi \pi}(s)$  are dominated by the $\rho$ resonance and 
fall off rapidly for $s>1$~GeV${}^2$,   this approximation is quite reasonable   
(see Ref.~\cite{Pich:2013lsa} and references therein).  Moreover, since the weight $W_{a_\mu} (s)$ falls off rapidly  with $s$,  
the $s > 1$~GeV${}^2$ region, likely to involve inelastic effects,  contributes only about 2\% to the integrals in Eq. (\ref{eq:cT}).   
Variations due to different parameterizations of the vector form factors are also at the few per-cent level.
Based on this, we conservatively assign a 10\% uncertainty to $c_T$, due to inelastic effects. 
\item Second, we   use the lattice QCD result of Ref.~\cite{Baum:2011rm} for $F^{K \pi}_T(0)$
and the $SU(3)$ relation~ $F^{\pi \pi}_T(0) =  2  F^{K \pi}_T (0)$  to determine $F^{\pi \pi}_T(0) =  1.87 (7)$~GeV$^{-1}$,\footnote{Note that Ref.~~\cite{Baum:2011rm} uses a different normalization for the tensor form factor. }
consistently with Ref.~\cite{Hoferichter:2018zwu}. 
The  relative sign  $F^{\pi \pi}_T(0)/F^{\pi \pi}_V(0) >0$  can be fixed by studying the ratio of form factors 
in the resonance chiral theory and imposing the appropriate QCD asymptotic constraints~\cite{Mateu:2007tr,Cata:2008zc}.
Overall, the form factor normalization brings in another  uncertainty of about $3.5\%$ for $c_T$. 
Combining linearly the two uncertainties in $c_T$, we arrive at  $c_T=0.43(8)$.
\end{itemize}

In order to use \eref{pipi0} to bound the new physics couplings,  we need precise input on $a_\mu^\tau$ and $a_\mu^{ee}$. 
For  $a_\mu^{ee}$, in the spirit of Ref.~\cite{Aoyama:2020ynm} we merge the two model-independent evaluations of Refs.~\cite{Davier:2019can,Keshavarzi:2019abf} quoting conservative uncertainties according to the prescription of Ref.~\cite{Aoyama:2020ynm},\footnote{Explicitly, the prescription is: (i) use as central value the arithmetic mean of the two results; 
(ii) assign as `experimental error' the largest of the two quoted experimental errors; 
(iii) assign as `systematic error' the uncertainty related to the tension between the BABAR and KLOE data~\cite{Davier:2019can,Aoyama:2020ynm}.} 
finding  $a_\mu^{ee} = (506.1 \pm 1.9_{\rm exp} \pm 2.8_{\rm syst}) \times10^{-10}$.
For $a_\mu^\tau$  we use as baseline value the data-based evaluation $a_\mu^{\tau} = (516.2 \pm 3.6)\times10^{-10}$  from Ref.~\cite{Davier:2013sfa}.
With the above input we find 
\bea
\label{eq:pipi}
\! \eL^{d\tau}\!-\!\eL^{de}\!+\!\eR^{d\tau}\!-\!\eR^{de} +\!0.43 (8)\,\hat{\epsilon}_T^{d\tau} \!=\! (10.0 \! \pm \! 4.9) \! \times \!10^{-3} \! , 
\eea
which implies a sub-percent level sensitivity to new physics effects.~\footnote{A similar but more conservative treatment of isospin breaking corrections (for which the associated uncertainties are estimated to be more than  $50\%$ of their total size) is performed in Ref.~\cite{Miranda:2020wdg}, leading to  $a_{\mu}^{\tau}=514.6\, {}^{+5.7}_{-4.8}$.
This value leads to 
$\eL^{d\tau}\!-\!\eL^{de}\!+\!\eR^{d\tau}\!-\!\eR^{de} +\!0.43 (8)\,\hat{\epsilon}_T^{d\tau} \!=\!  (8.4 \! \pm \! 6.4) \! \times \!10^{-3}$, not changing the qualitative result of sub-percent sensitivity to new physics couplings. }
The $\sim\!\!2\sigma$ tension with the SM reflects the long-standing disagreement between $e^+e^-$ and $\tau$ data sets~\cite{Davier:2017zfy}.
Ref.~\cite{Jegerlehner:2011ti} argued that this disagreement can be 
removed by considering the effect of $\rho^0$-$\gamma$ mixing,  
which is present in $e^+ e^- \to \pi \pi$ data but not in the charged-current $\tau$ data. 
This effect is however model-dependent and  may be impacted by  significant uncertainties, 
not yet assessed~\cite{Aoyama:2020ynm}. We  therefore  stick with the analysis of Ref.~\cite{Davier:2013sfa} 
and  expect that lattice QCD will soon provide new insights on the size and uncertainty of isospin-breaking corrections 
entering in $a_\mu^\tau$~\cite{Aoyama:2020ynm,Bruno:2018ono}. 

We note that another constraint on new physics couplings can be obtained  
by studying  the branching ratio $B_{\pi \pi} = \Gamma(\tau \to \pi \pi \nu_\tau)/\Gamma_\tau$~\cite{Zhang:2009ag,Jegerlehner:2011ti,Pich:2013lsa}.
The analysis parallels the one described above, with the replacements 
$a_\mu \to B_{\pi \pi}$ and $W_{a_\mu} (s) \to  1/\Gamma_\tau$ in Eqs.~(\ref{eq:WI1})-(\ref{eq:cT}), and 
uses the isospin-rotated spectral function extracted from $e^+ e^- \to \pi \pi$ data. 
The resulting constraint, however, is almost degenerate to  Eq.~(\ref{eq:pipi}) 
and suffers  from larger uncertainties because the flat weight corresponding to the BR samples a region of the spectral 
function with relatively larger uncertainties. We therefore do not include this constraint in our analysis. 

We conclude this subsection by 
noting that the constraint obtained above can be strengthened by directly looking at the $s$-dependence of the spectral functions (instead of the $a^\tau_\mu$ integral), which would also allow us to disentangle the vector and tensor interactions.
Moreover, note that the $a_\mu^{\tau,ee}$ uncertainties include a scaling factor due to internal inconsistencies of the various datasets~\cite{Davier:2017zfy}, which hopefully will decrease in the future. In fact, new analyses of the $\pi\pi$ channel are expected from CMD3, BABAR, and possibly Belle-2~\cite{Davier:2017zfy,Keshavarzi:2018mgv}. 
Finally,   lattice QCD calculations of the isospin rotation needed to  relate $e^+ e^- \to \pi^+ \pi^-$ to $\tau \to \pi \pi \nu_\tau$ data are being  performed~\cite{Bruno:2018ono}, 
and will contribute to reducing the uncertainty in this step of our analysis.  
All in all, we can expect a significant improvement in precision with respect to the result in~\eref{pipi} in the near future.

Another interesting possibility is to extract bounds on the tensor interaction from its effect on the $\pi \pi$ invariant mass distribution in  $\tau\to\pi\pi\nu_\tau$~\cite{Miranda:2018cpf}. We note that the experimental data should be analyzed including {\it simultaneously} the tensor coefficient and the free parameters of the vector form factor in the chosen parametrization. 
As an initial exercise, Ref.~\cite{Miranda:2018cpf} analyzed the $\pi\pi$ distribution fitting only the tensor coupling and using values for the QCD parameters that were obtained from the same $\pi\pi$ distribution neglecting non-standard terms. 
The obtained per-mil level bound 
illustrates the maximum sensitivity that can be obtained from a proper analysis.\footnote{Let us note a possible weakness of this approach. Since the current parametrizations of the form factors are not fully derived from first principles, it can become challenging to assess whether a potential deviation from data really comes from new physics or from an incomplete parametrization.\label{foot:param}}

\subsection{$\tau \to \pi\eta\nu_\tau$}
\label{sec:pieta}

As pointed out in Ref.~\cite{Garces:2017jpz} the  $\tau\to\eta\pi\nu_\tau$ channel can provide useful information since  
the non-standard scalar contribution is enhanced with respect to the (very suppressed) SM one. 
Because of this, one can obtain a nontrivial constraint on $\epsilon_S^{d \tau}$ even though both SM and BSM contributions are hard to predict with high accuracy. 

The  $\tau \to \pi\eta\nu_\tau$ decay mode proceeds only through isospin-violation  in the SM (see Ref.~\cite{Pich:2013lsa} and references therein), 
with the branching fraction expected at the $10^{-5}$ level.  This mode has not  yet  been observed experimentally and we use the experimental limit on the branching fraction to bound the BSM couplings. 
Following Ref.~\cite{Garces:2017jpz} we write the new physics dependence of the branching ratio in the form 
\be
 \frac{   {\rm BR}_{\rm exp}  (\tau\to\eta\pi\nu_\tau) }{\widehat {\rm BR}_{\rm SM}  (\tau\to\eta\pi\nu_\tau)}  
 =  1 + \alpha \,\epsilon^{d\tau}_S   + \gamma   \, (\epsilon^{d\tau}_S)^2 ~, 
\label{eq:etapi}
\ee
where the SM prediction $\widehat {\rm BR}_{\rm{SM}} (\tau\to\eta\pi\nu_\tau)$ is estimated to be in the interval $[0.3, 2.1] \times 10^{-5} $~\cite{Escribano:2016ntp}.
The coefficients $\alpha$ and $\gamma$ are estimated to be in the ranges 
$\alpha \in  [3,8]  \times  10^2$~\cite{Garces:2017jpz}  and $\gamma \in  [0.7,1.75]  \times  10^5$~\cite{Roig:private}. 
These large coefficients  can be understood  by recalling that for this decay mode  $a_S(s) \sim  s/( m_\tau (m_d - m_u))$. 
Exceptionally we retain the quadratic terms in $\epsilon^{d\tau}_S$, because it dominates in the parameter region where the bound is set. 
On the other hand, we ignore the dependence on $\epsilon_L^{d\tau}$, $\epsilon_R^{d\tau}$, and $\hat \epsilon_T^{d\tau}$, because their coefficients are not enhanced,  and thus their effects are irrelevant given the current experimental and theoretical precision. 
We use the experimental limit ${\rm BR}_{\rm exp}  (\tau\to\eta\pi\nu_\tau) < 9.9 \times 10^{-5}$ at 95\%~CL~\cite{delAmoSanchez:2010pc,ParticleDataGroup:2020ssz}.  
Using the most conservative values for the SM prediction, as well as for $\alpha$ and $\gamma$ (within their respective ranges),  we find the 68\% CL interval: 
\begin{equation}
\epsilon^{d \tau}_S  \in   (-0.021,0.010) . 
\label{eq:eta}
\end{equation}
The likelihood is highly non-gaussian due to the quadratic dependence. The bound above is much weaker than the one obtained in the original work of Ref.~\cite{Garces:2017jpz} because the latter did not take into account the large theory uncertainties affecting the SM prediction and the $\alpha$ and $\gamma$ parameters.

The bounds from $\tau \to \pi \eta \nu_\tau$ will significantly improve if theory or experimental uncertainties can be reduced. The latter will certainly happen with the arrival of Belle-II, which is actually expected to provide the first measurement of the SM contribution to this channel~\cite{Belle-II:2018jsg,Petar:1369,Moussallam:2021flg} 
(see also Ref.~\cite{Hayasaka:2009zz} for Belle results). 
Improvement on the theoretical side will be possible with lattice QCD calculations of the relevant form factors. 
Finally, note that  $\tau \to \pi \eta \nu_\tau$  is one of two  probes considered  in this work with a significant sensitivity (via $\cO(\eps_S^2)$ effects) to the imaginary part of $\epsilon^{D\tau}_X$ coefficients (the other probe being $\tau \to K \pi \nu_\tau$, sensitive to ${\rm Im} \hat{\epsilon}_T^{s\tau}$).
Allowing for a complex $\epsilon^{D\tau}_X$, the bound in \eref{eta} refers to the real part, and simultaneously we obtain $|{\rm Im}  \left( \eps^{d \tau}_S \right)| < 0.014$.

\subsection{$\tau \to K \pi\nu_\tau$}

For this $\Delta S=1$ mode the situation is more involved compared to the analogous   $\Delta S=0$ case ($\tau \to \pi \pi \nu_\tau$).  
The SM amplitude is controlled by the vector from factor $F_V^{K \pi} (s)$ and a small but now non-negligible 
contribution from the scalar form-factor $F_S^{K \pi} (s)$,   which contributes to the decay rate at the \% level. 
Once BSM couplings are turned on, the channel $\tau \to K \pi\nu_\tau$ is mostly sensitive to  
the vector combination  $\epsilon_{L+R}^{s\tau}-\epsilon_{L+R}^{se}$ and the tensor coupling $\hat{\epsilon}_T^{s\tau}$. 
So to obtain \%-level bounds one needs  \%-level  predictions of  $F_{V}^{K \pi} (s)$ 
and a  less precise  determination  of the scalar and tensor form factors 
$F_{S,T}^{K \pi} (s)$  as well.  

For the tensor form factor,  as shown in  Ref.~\cite{Cirigliano:2017tqn},  in the elastic region  
unitarity enforces the  proportionality  
$F_T^{K \pi } (s) =    F^{K \pi}_T(0)   F_V^{K  \pi } (s)$. This relation can be extended to the whole physical region 
to a good approximation, due to the  dominance of the elastic channel  through the $K^*(892)$ resonance. 
In fact, O(1) violation of the above relation are expected in the $K^*(1410)$ region, 
where however both kinematics and the fall-off of $F_V^{K \pi} (s)$ conspire to make the effect only a few \%~\cite{Cirigliano:2017tqn}. 
So the problem is  reduced to obtaining a reliable and BSM-free parameterization for the vector and scalar form factors.

One possible way to achieve precise determinations of $F_{V,S}^{K \pi} (s)$ 
is to 
use dispersion parameterizations available in the literature (see for example Ref.~\cite{Antonelli:2013usa} and 
references therein)  and fix the subtraction constants and other parameters by matching to lattice QCD, rather than 
fitting to $\tau \to K \pi \nu_\tau$  data.
This removes the possible BSM contamination at the price of probably having larger uncertainties.
Reaching \%-percent level bounds on the BSM couplings with this approach might not be possible anytime soon. 

Another possibility would be to  invoke the same   strategy  used for the
$\tau \to \pi \pi \nu_\tau$ channel  and use $e^+e^-$ data (through an $SU(3)$ rotation) 
to obtain the vector from factor, neglecting the \%-level contribution from the scalar form factor. 
The main error here would be the $SU(3)$-breaking corrections and a bound on the new physics 
coefficients at the level of ${\cal O}(0.1)$ could be possible.  

Using one of the approaches outlined above, one should be able to obtain 
constraints on ${\rm Re} (\epsilon_i^{s \tau})$ at the 5-10\% level. 
Such an analysis is however beyond the scope of this paper and we leave it   to future work.   
On the other hand,  we note that  the CP-violating component   ${\rm Im} (\hat{\epsilon}_T^{s \tau})$  
could produce a BSM contribution to the  CP asymmetry in $\tau^-  \to K_S \pi^- \nu_\tau$, 
whose measured value~\cite{BABAR:2011aa}  is in tension with the SM prediction
\cite{Grossman:2011zk,Bigi:2005ts}  at the 2.8-$\sigma$ level.
As shown in Ref.~\cite{Cirigliano:2017tqn},  explaining the tension would require 
$|{\rm Im} (\hat{\epsilon}_T^{s \tau})| \sim 0.20$,  
while the neutron EDM provides 
via loop effects a much stronger bound at the level of 
$|{\rm Im} (\hat{\epsilon}_T^{s \tau})|  \lesssim  4\times 10^{-5} $.

Finally, one can extract bounds on the scalar and tensor interactions from their effect on the $K \pi$ invariant mass distribution~\cite{Rendon:2019awg,Gonzalez-Solis:2020jlh}. As in the $\tau\to\nu_\tau \pi\pi$ channel, we note that the experimental data should be analyzed including {\it simultaneously} the nonstandard terms and the free parameters of the vector and scalar form factors in the chosen parametrization. The $\sim1\%$-level bounds obtained in Ref.~\cite{Rendon:2019awg,Gonzalez-Solis:2020jlh} in a BSM fit (without fitting the QCD parameters simultaneously) illustrate the maximum sensitivity that can be obtained from a proper analysis (see however footnote \ref{foot:param}).

It is particularly simple and interesting to discuss the $\tau\to \nu_\tau K\pi$ channels when $\epsilon_{S,T}^{s\tau}=0$. In that case, the SM extraction of the form factors from normalized kinematic distribution is correct. Thus, we can use the associated SM prediction of the BRs~\cite{Antonelli:2013usa} to constrain the vector combination of couplings, which simply produces an overall rescaling:
\bea
BR(\tau\to\nu_\tau K\pi)_{\rm exp} = \widehat{BR}(\tau\to\nu_\tau K\pi)_{\rm SM} \left( 1 + 2\epsilon_{L+R}^{s\tau}-2\epsilon_{L+R}^{se}\right)
\eea
where $\widehat{BR}_{\rm SM}$ is the SM prediction calculated using $ |\hat V_{us}|$ from $K_{e3}$~\cite{Antonelli:2013usa}. Using the experimental values from Ref.~\cite{HFLAV:2019otj} and combining the $\tau\to\nu_\tau K^-\pi^0$ and $\tau\to\nu_\tau \bar{K}^0\pi^-$ channels, we find:
\bea
\label{eq:KpiBound}
\epsilon_{L+R}^{s\tau}-\epsilon_{L+R}^{se} + f(\epsilon_S^{s\ell},\hat{\epsilon}_T^{s\ell}) &=& -0.008 \pm 0.019~,
\eea
where $f(\epsilon_S^{s\ell},\hat{\epsilon}_T^{s\ell})$ ($\ell=\mu,\tau$) is just a symbolic term to remind us that we do not know the form of the bound if those coefficients are present. The approach assumes implicitly that $\epsilon_{S,T}^{s\mu}=0$ because Ref.~\cite{Antonelli:2013usa} includes $K_{\mu 3}$-shape data in their analysis of the form factors.\footnote{There's no such problem with $K_{e 3}$-shape data, since the linear contribution from chirality-flipping operators is negligible in that case due to the smallness of the electron mass~\cite{Gonzalez-Alonso:2016etj}.} This assumption could be avoided redoing the analysis of Ref.~\cite{Antonelli:2013usa} without including $K_{\mu 3}$-shape data, which would lead to a larger SM uncertainty and hence a weaker BSM bound. 

We close this section by noting that $\tau\to \nu_\tau K^- K^0, \nu_\tau K^-\eta, \nu_\tau K^-\eta'$ can be also used to probe nonstandard interactions~\cite{Gonzalez-Solis:2019lze,Gonzalez-Solis:2020jlh}, although large theoretical uncertainties prevent the current extraction of stringent constraints.

\section{Inclusive decays}\label{sec:inclusive}
\setcounter{equation}{0}

In contrast with exclusive decays, the predictive power of analytic methods for inclusive decays does not rely on our knowledge of the different form factors. Even when having a limited theoretical knowledge about the hadronic dynamics, dependent on the internal degrees of freedom, very precise predictions can be made when integrating over them. A precise value for the strong coupling can be obtained from non-strange spectral functions~\cite{Braaten:1991qm,Davier:2013sfa,Boito:2014sta,Pich:2016bdg}, as well as valuable information from QCD in the non-perturbative regime (for example see Refs.~\cite{Boito:2015fra,Gonzalez-Alonso:2016ndl,Pich:2021yll}). Likewise, the same approach can be used to extract a precise value of $|V_{us}|$ from strange spectral functions~\cite{Gamiz:2007qs,Hudspith:2017vew}.

Following the change of perspective adopted in Ref.~\cite{Cirigliano:2018dyk} and in this work, we do not take for granted the validity of the SM and use those theoretical methods to determine SM parameters. Instead, we take them as external inputs that should come from determinations insensitive to BSM effects within our general EFT assumptions (as it will be the case for $\alpha_{s}$ or $f_{\pi^\pm}$), or from determinations where the BSM contamination is known in terms of non-standard couplings (this will be the case for $G_\mu$, $V_{ud}$ or $V_{us}$). The (dis)agreement between the SM predictions of hadronic-tau-decays observables and the experimental results can be then directly translated into bounds for the non-standard couplings.

\subsection{Non-strange decays}
\label{sec:nonstrangedecays}

The decay of a $\tau$ lepton into a neutrino and a hadronic state $n$ with total momentum $p_{n}$ can proceed through the various quark currents of the Lagrangian of Eq.~(\ref{eq:leff1}). The quantum numbers of both the final hadronic states and the quark currents set useful restrictions on the possible sources of those decays. In this section we work with final states without strangeness, which can only be mediated by the nonstrange part ($D=d$) of the Effective Lagrangian in Eq. (\ref{eq:leff1}).

The hadronic invariant mass distribution of a hadronic $\tau$ decay ($\tau \rightarrow n \, \nu_{\tau}$) can be written as the product of a trivial leptonic part and a hadronic exclusive spectral function that depends on the nonperturbative dynamics, namely~\cite{Rodriguez-Sanchez:2018dzw}
\bea
\frac{d\Gamma^{(n)}}{dq^2} &=& \sum_{J_1 J_2} L_{J_1 J_2}(q) \rho^{(n)}_{J_{1}J_{2}}(q)~,\\
\rho^{(n)}_{J_{1}J_{2}}(q)&=& (2\pi)^{3} \int d\phi_{n} \, \delta^{4}(p_{n}-q) \, \langle n | J_{1} | 0 \rangle \langle 0 | J_{2}^{ \dagger} | n \rangle \, ,
\eea
where $q^\mu$ is the momentum transfer and $d\phi_n\equiv \prod_{i} \frac{d^{3}p_{i}}{(2\pi)^3  2 E_{i}}$ is the differential hadronic phase space element and $J_{1}$ and $J_{2}$ are quark currents. It can be proven that the inclusive spectral functions obtained summing over all possible channels, $\rho_{J_{1} J_{2}}(q)\equiv\sum_{n} \rho^{(n)}_{J_{1} J_{2}}$, are equal to the imaginary part of two-point correlation functions of quark currents~\cite{Weinberg:1995mt,deRafael:1997ea,Gonzalez-Alonso:2010vnm}
\bea
\rho_{J_{1}J_{2}}(q)=\frac{1}{\pi}\mathrm{Im}\Pi_{J_{1}J_{2}}(q)~,
\eea
where
\begin{equation}
\Pi_{J_{1}J_{2}}(q)\equiv\int d^{4}x \, e^{-iqx} \, \langle 0 | T(J_{1}(x) \, J_{2}^{\dagger}(0)) |0 \rangle \, .
\end{equation}
As a result the inclusive differential decay width can be written in terms of a few correlators. Finally, the analytic properties of the latter make possible to calculate integrated moments of the former (dispersion relation). 

A priori this important result applies only to the fully inclusive non-strange channel. However, it is well-known that {\it within the SM} the same approach works as well for less inclusive quantities, namely for the vector and axial components. Let us briefly review the argument and extend it by including also non-standard currents. 
As shown in~\sref{TauToPnu}, vector, scalar and tensor currents do not contribute to the one-meson mode, $\tau\to P\nu_\tau$. On the other hand, axial and pseudoscalar currents do not contribute to the two-meson mode ($\tau\to PP'\nu_\tau$), whereas the scalar-current contribution is also absent in the isospin limit, {\it cf.}~\sref{TauToPPnu}. For the rest of channels, one can make use of $G$-parity, a combination of isospin and charge conjugation that forbids the production of the different hadronic channels either through vector and tensor or through axial, scalar and pseudoscalar currents, with the well-known exception of the  $\pi K\bar{K}$ modes, whose G-parity is not well-defined and has to be decomposed using theoretical input, with the associated uncertainty. Thus, the usual $V$ and $A$ separation made by experimental collaborations within the SM~\cite{Davier:2013sfa} can be reinterpreted as a $(V,T)$ and $(A,P,S)$ separation in our BSM setup. 

Neglecting contributions of order  $(\epsilon^{d\tau}_{i})^2$ leaves only the ($VV,VT$) and ($AA,AP,AS$) correlators in the so-called vector and axial channels, respectively. Additionally, the $AS$ correlator is zero due to parity considerations. Finally 
partial conservation of the axial current relates the $P$ matrix elements with the longitudinal parts of the $A$ one, connecting $\Pi_{AP}(q)$ with the longitudinal part of $\Pi_{AA}(q)$\cite{Bijnens:2003rc}. Taking this into account we can calculate the normalised invariant mass-squared distributions, $dN^{V/A}_\tau / ( N_\tau ds ) = \tau_\tau\, d\Gamma^{V/A}/ds$, where $\tau_\tau$ is the tau lifetime. 
We find~\cite{Gonzalez-Alonso:2010vnm,Rodriguez-Sanchez:2018dzw}
\begin{align}
\frac{dN^{V}_{\tau}}{N_{\tau}ds}
=
\kappa_d\,\left(1-\frac{s}{m_{\tau}^2}\right)^2
&\Bigg[\left(1+\frac{2s}{m_{\tau}^{2}}\right)(1+2\epsilon_{L+R}^{d\tau})\operatorname{Im}\Pi_{VV}^{(1+0)}(s)
+ 6\, \hat \epsilon^{d\tau}_T \, \frac{\operatorname{Im}\Pi_{VT}(s)}{m_{\tau}S_{EW}^{\mathrm{had}}}\Bigg] \, , \label{eq:distV}  
\\[0.7cm] \nonumber
\frac{dN^{A}_{\tau}}{N_{\tau}ds}
=
\kappa_d\,\left(1-\frac{s}{m_{\tau}^2}\right)^2
&\Bigg[\left(1+\frac{2s}{m_{\tau}^{2}}\right)(1+2\epsilon_{L+R}^{d\tau}-4\epsilon_{R}^{d\tau})\operatorname{Im}\Pi_{AA}^{(1+0)}(s) \nnl &-\frac{2s}{m_{\tau}^{2}}\left(1+2\epsilon_{L+R}^{d\tau}-4\epsilon_{R}^{d\tau}+\epsilon^{d\tau}_{P}\frac{m_{\tau}}{S_{EW}^{\mathrm{had}}(m_{u}+m_{d})}\right)\operatorname{Im}\Pi_{AA}^{(0)}(s)\Bigg] \, , \label{eq:distA}
\end{align}
where $\kappa_d\equiv 12 \pi |V_{ud}|^2 \hat{B}_{e} S_{EW} / m_{\tau}^2$ 
and $\epsilon_{L+R}^{d\ell}\equiv \epsilon_L^{d\ell}+ \epsilon_R^{d\ell}$. The VV, AA and VT correlators have been Lorentz-decomposed as follows:
\bea\label{eq:eqcurrcorr}
&&i \int d^{4}x\; e^{iqx} \;
\langle 0 |T[J^{\mu}(x)J^{\nu \dagger}(0)]|0\rangle
\; =\;
(-g^{\mu\nu}q^{2}+q^{\mu}q^{\nu})\; \Pi^{(1)}_{JJ}(q^{2})
+ q^{\mu}q^{\nu}\;\Pi^{(0)}_{JJ}(q^{2}) \, ,\\
&&i \int d^{4}x\; e^{iqx} \; 
\langle 0 |T[V^{\mu}(x)T^{\alpha\beta \dagger}(0)]|0\rangle
\; =\;
i(g^{\mu\alpha}q^{\beta}-g^{\mu\beta}q^{\alpha})\; \Pi_{VT}(q^{2})\, ,
\eea
where $J=\{V,A\}$, $V^{\mu}=\bar{d}\gamma^{\mu}u$, $A^{\mu}=\bar{d}\gamma^{\mu}\gamma^{5}u$ and $T^{\mu\nu}=\bar{d}\sigma^{\mu\nu}u$. In~\eref{distV} we also took into account that $\mathrm{Im}\Pi_{VV}^{(0)}$ vanishes in the isospin limit. 
The factor $S_{EW}\equiv S_{EW}^{\mathrm{had}} / S_{EW}^{\mathrm{lep}} = 1.0201(3)$ contains the renormalization-group-improved electroweak correction to the semileptonic decay, including a next-to-leading order resummation of large logarithms
\cite{Marciano:1988vm,Braaten:1990ef,Erler:2002mv}. Following the usual conventions, we include the radiative correction to the purely leptonic process $\tau\to e \nu_\tau\bar{\nu}_e$, denoted by $S_{EW}^{lep}$, in the $S_{EW}$ factor as well as in the $\hat{B}_e$ factor, defined by
\begin{equation}
\label{eq:Bprimee}
\hat{B}_{e}
\equiv  \Gamma\left(\tau\to \nu_\tau e\bar{\nu}_e (\gamma) \right)_{\rm SM} \tau_\tau
= \frac{G_\mu^2\tau_{\tau}}{192\pi^{3}}m_{\tau}^{5}S_{EW}^{lep}=0.17778(31)
\, ,
\end{equation}
which in the SM limit corresponds with the  branching ratio of the decay {$\tau \rightarrow e\nu_{\tau}\bar{\nu}_{e} (\gamma)$}. We remind that $S_{EW}^{lep} = \left(1+\frac{3}{5}\frac{m_{\tau}^{2}}{M_{W}^{2}}\right) \left( 1+\frac{\alpha(m_{\tau})}{2\pi} \left(\frac{25}{4}-\pi^{2}\right)\right)$ up to negligible terms of order $m_{\mu}^{2}/M_{W}^{2}$, $m_e^{2}/m_\tau^{2}$, and 2-loop corrections~\cite{Pich:2013lsa}.

The large scale dependence of $\hat{\epsilon}_{T}^{d\tau}$~\cite{Gonzalez-Alonso:2017iyc} is cancelled in the expression above by that of the $VT$ correlator, which we study later in this work. On the other hand we will safely approximate $S_{EW}^{\mathrm{had}}\approx 1$ in the tensor term.

Finally, taking into account that the spin-0 part of the axial correlator can be safely approximated by the pion pole, one obtains the following predictions for the experimentally extracted spectral functions
$\rho_{V/A}^{\rm exp}(s)$:
\begin{align}\nonumber
\rho_{V}^{\rm exp}(s)
&\equiv\frac{dN_{\tau,V}}{N_{\tau}ds}\frac{m_{\tau}^{2}}{12\pi^2|\hat{V}_{ud}|^{2}\hat{B}_{e}S_{EW}}
\left(1-\frac{s}{m_{\tau}^{2}}\right)^{-2}\left(1+\frac{2s}{m_{\tau}^{2}}\right)^{-1}
\\&=\left(1+2\epsilon_{L+R}^{d\tau}-2\epsilon_{L+R}^{de}\right)\frac{1}{\pi}\operatorname{Im}\Pi_{VV}^{(1+0)}(s)+6 \,\hat \epsilon^{d\tau}_{T}\left(1+\frac{2s}{m_{\tau}^{2}}\right)^{-1}\frac{\operatorname{Im}\Pi_{VT}}{\pi\,m_{\tau} }(s) \; ,
\label{eq:rhoVexp}
\\[8pt] \nonumber
\rho_{A}^{\rm exp}(s>s_{th})
&\equiv\frac{dN_{\tau,A}}{N_{\tau}ds}\frac{m_{\tau}^{2}}{12\pi^2|\hat{V}_{ud}|^{2}\hat{B}_{e}S_{EW}}
\left(1-\frac{s}{m_{\tau}^{2}}\right)^{-2}\left(1+\frac{2s}{m_{\tau}^{2}}\right)^{-1}
\\&=\left(1+2\epsilon_{L+R}^{d\tau}-2\epsilon_{L+R}^{de}-4\epsilon_{R}^{d\tau}\right)\frac{1}{\pi}\operatorname{Im}\Pi_{AA}^{(1+0)}(s)  \, ,
\label{eq:rhoAexp}
\end{align}
where $s_{th}=4m_\pi^2$ is the continuum threshold. Note that the electron-flavor Wilson Coefficient $\epsilon_{L+R}^{de}$ has appeared due to the use of the phenomenological value $\hat{V}_{ud}$, {\it cf.} \eref{hatvalues}.

In the absence of BSM effects, the experimental spectral functions coincide with the QCD spectral functions,   $\frac{1}{\pi}\operatorname{Im}\Pi_{V/A}^{(1+0)}$, as shown in 
Eqs.~(\ref{eq:rhoVexp}-\ref{eq:rhoAexp}). 
That is, of course, the rationale for the standard experimental definition in terms of the differential distributions. However, that relation is spoiled by new physics effects. 

Eqs.~(\ref{eq:rhoVexp}-\ref{eq:rhoAexp}) connect the accurately-known experimental distributions,  $\rho_{V/A}^{\rm exp}(s)$, the BSM Wilson Coefficients, and QCD correlation functions. As a consequence, precise theoretical knowledge of the latter, which in principle only depend on $\alpha_{s}$ and the quark masses, would immediately translate into stringent BSM bounds. However, our theoretical knowledge of the imaginary parts of the correlators 
is limited, since perturbative QCD is known not to be valid below $s\sim1 \, \mathrm{GeV}^2$, especially in the Minkowskian axis, where the experimental data lie. 
Fortunately, the situation is different for integrals of the imaginary parts of the correlators, which can be calculated with accuracy using the Operator Product Expansion (OPE) of the corresponding correlators $\Pi_{\mathcal{J}\mathcal{J'}}$. This allows one to predict theoretically the value of weighted integrals of the experimental spectral functions~\cite{Shifman:1978bx}. In order to derive such dispersion relations, we 
integrate Eqs.~(\ref{eq:rhoVexp}-\ref{eq:rhoAexp}) multiplied by a monomial weight function $w(s/s_0)=(s/s_0)^n$, which gives
\bea
\label{eq:masterformula}
\boxed{
    I_{V\pm A}^{\rm exp}
- I_{V\pm A}^{\rm SM} 
=
2\left(\epsilon_{L+R}^{d\tau}-\epsilon_{L+R}^{de}\right)~I_{V\pm A}^{\rm SM}
\mp\, \,4\epsilon_R^{d\tau}~I_A^{\rm SM}
+6\, \hat \epsilon_{T}^{d\tau}\,I_{VT}
\; ,}
\eea
where we have omitted the dependence on $s_0$ (upper integration limit), $n$ (the weight function) and $\mu$ (renormalization scale) of the various $I_i$ integrals to lighten the notation. These objects are defined by
\bea
I_{J}^{\rm exp} (s_0;n) &\equiv& \int^{s_{0}}_{s_{th}}\frac{ds}{s_{0}}\left(\frac{s}{s_{0}}\right)^{n} \rho_{J}^{\rm exp}(s)~,\\
I_{J}^{\rm SM} (s_0;n) &\equiv& 
\int^{s_{0}}_{s_{th}}\frac{ds}{s_{0}}\left(\frac{s}{s_{0}}\right)^{n}\frac{1}{\pi}\operatorname{Im}\Pi_{J}^{(1+0)}~,\\
\label{eq:IVTdef}
I_{VT}(s_0;n;\mu)  &\equiv& \int_{s_{th}}^{s_{0}}\frac{ds}{s_{0}}\left(\frac{s}{s_{0}}\right)^{n}
\left(1+\frac{2s}{m_{\tau}^{2}}\right)^{-1}\frac{\operatorname{Im}\Pi_{VT}}{\pi\,m_{\tau}}(s)~,
\eea
where $J=V,A,V\pm A$ and once again $s_{th}=4 m_{\pi}^{2}$. 
The master formula in~\eref{masterformula} and the $I_i$ definitions can be trivially generalized to any analytic weight function.

The $I^{\rm exp}_{V\pm A}$ integrals are calculated using the latest ALEPH spectral functions~\cite{Davier:2013sfa} and represent the experimental input in our analysis. We take into account in this work the correlations between bins and between channels.

For the calculation of the SM prediction, $I^{\rm SM}_{V\pm A}$, we follow the standard approach~\cite{Braaten:1991qm}: the integral of the imaginary part of the correlator along the real axis is related to the contour integral of the OPE of the correlator, which is a function of the strong coupling constant $\alpha_s$, the quark masses and the so-called QCD vacuum condensates ${\cal O}_{2n}$. As a result of that calculation one obtains
\bea
\label{eq:VplusAprediction}
I_{V+A}^{\rm SM} (s_0;n) &=& -\frac{f_{\pi^\pm}^{2}}{s_{0}}\left(\frac{m_{\pi}^{2}}{s_{0}}\right)^{n}
+2 A_{P}^{(n)}(s_{0})
-\frac{\mathcal{O}^{V+A}_{2(n+1)}}{(-s_{0})^{n+1}}
+\delta^{\rm DV}_{V+A}(s_{0};n)~,\\
\label{eq:VminusAprediction}
I_{V-A}^{\rm SM} (s_0;n) &=& +\frac{f_{\pi^\pm}^{2}}{s_{0}}\left(\frac{m_{\pi}^{2}}{s_{0}}\right)^{n}
-\frac{\mathcal{O}^{V-A}_{2(n+1)}}{(-s_{0})^{n+1}}
+\delta^{\rm DV}_{V-A}(s_{0};n)~.    
\eea
The details of this derivation as well as those associated to the calculation of each term in Eqs. (\ref{eq:VplusAprediction})-(\ref{eq:VminusAprediction}) are presented in~\aref{inclSMprediction}. Here we simply discuss the main elements of these expressions in a qualitative way:
\bi
\item $A_{P}^{(n)}(s_{0})$ is the purely perturbative contribution, which is only present in the $V+A$ channel. We calculate it using $\alpha_{s}(M_{Z}^{2})=0.1184(8)$ from the lattice~\cite{Aoki:2021kgd,Maltman:2008bx,PACS-CS:2009zxm,McNeile:2010ji,Chakraborty:2014aca,Bruno:2017gxd,Bazavov:2019qoo,Cali:2020hrj,Ayala:2020odx}. 
\item The $\mathcal{O}_{2(n+1)}$ condensates parametrize the small non-perturbative contributions from the OPE power corrections, and their numerical values will be discussed below. 
$\delta^{\rm DV}_{V\pm A}(s_{0};n)$ denotes the so-called quark-hadron Duality Violations, which parameterize the error introduced by approximating the correlator by its OPE. These contributions are small for large $s_0$ values and will be estimated from the $s_0$-dependence of the dispersive relation. 
\item Finally, it is worth mention the origin of the $f_{\pi^\pm}$ terms in the SM predictions, which might be surprising since the observables, $I_{V\pm A}^{\rm exp}$, do not include the one pion channel (the integral starts at $s_{th}=4m_\pi^2$). Its contribution appears nonetheless in the SM prediction, $I_{V\pm A}^{\rm SM}$, due to the analytic properties of the correlators, which relate different regions in the complex plane. 
Equivalently, we have to substract the one-pion channel (the $f_{\pi^\pm}$ term) because the dispersive method predicts the {\it total} non-strange integral. 
We use the $N_f=2+1$ lattice average $f_{\pi^\pm}=130.2(8) \, \mathrm{MeV}$~\cite{Aoki:2021kgd}, from Refs.~\cite{Blum:2014tka,Bazavov:2010hj,Follana:2007uv}, as in~\sref{TauToPnu}. 
\ei

Now we discuss the calculation of the nonstandard terms, {\it i.e.}, the ones in the RHS in Eq.~(\ref{eq:masterformula}). First we note that, up to quadratic BSM contributions, we can approximate $I_J^{\rm SM}\approx I_J^{\rm \rm exp}$, which we calculate using the ALEPH data~\cite{Davier:2013sfa}.\footnote{In Ref.~\cite{Cirigliano:2018dyk} the $I_J^{\rm SM}$ integrals were instead calculated theoretically using a dispersion relation, like in the SM terms. Our current approach, $I_J^{\rm SM}\approx I_J^{\rm exp}$, gives a simpler and more precise estimate. The numerical impact of this change on our final results will be negligible, since SM and experimental values are both precisely known and in agreement.} The experimental error is typically around 1$\%$ and thus its impact on the nonstandard terms can be neglected. Finally, the coefficient of the tensor contribution,
$I_{VT}$, is calculated using a dispersion relation, in analogy to the SM contribution (see~\aref{VTintegral}). The $I_{VT}$ error is more significant and will be kept in the analysis.

It is convenient to work with the $V+A$ and $V-A$ channels (instead of $V$ and $A$) because of their different characteristics. Namely, the $V-A$ channel does not have perturbative contributions and its dimension-4 condensate vanishes. 
In the following subsections, we choose specific weights and $s_0$ values that translate the generic master formula of~\eref{masterformula} into specific constraints on BSM couplings. The choice of weights introduced in Ref.~\cite{Cirigliano:2018dyk} is simple,  allows the separation of non-perturbative and BSM effects, and produce four BSM constraints sensitive to different theory uncertainties. As a result, correlations can be taken into account properly. We decide not to introduce additional moments, which would spoil these features and thus complicate the analysis.

\subsubsection{V+A}
The non-strange $V+A$ inclusive channel has been thoroughly studied in the literature as a QCD laboratory~\cite{Narison:1988ni,Braaten:1991qm,Davier:2005xq,Davier:2008sk,Beneke:2008ad,Beneke:2012vb,Caprini:2011ya,Abbas:2012fi,Abbas:2012py,Groote:2012jq,Baikov:2008jh,Maltman:2008nf,Boito:2012cr,Menke:2009vg,Narison:2009vy,Cvetic:2010ut,Pich:2011bb,Pich:2013sqa,Davier:2013sfa,Boito:2014sta,Pich:2016bdg,Boito:2020hvu,Caprini:2020lff,Ayala:2021mwc}. Those studies assume the absence of BSM contributions and typically use several moments of the spectral function to extract the value of the strong coupling constant $\alpha_s$ and the lowest dimensional condensates ${\cal O}^{V+A}_{2d}$. new physics terms have a weight dependence that is different to such QCD parameters, and thus we cannot just re-interpret past SM analysis as BSM constraints. Instead, we have to do the analysis again including this time BSM coefficients as free independent parameters. For that purpose, we choose the following two weights:
\bea
 \label{eqomegatau}
 \omega_{\tau}(s) &=& 
 \left(1-\frac{s}{m_{\tau}^{2}}\right)^{2}\left(1+2\frac{s}{m_{\tau}^{2}}\right) \, ,\\
 \omega_{0}(s)&=&1 \, ,
\eea
which give the total hadronic branching ratio 
and the integral of the spectral function. As we will see, the latter weight gives a relation where experimental and DV errors dominate, whereas the uncertainties of perturbative and non-perturbative OPE contributions dominate in the former case.

\paragraph{{\bf $\omega_\tau$ weight}.-}

The $I_{V+A}$ integral built with this weight and $s_0=m_\tau^2$ 
is, up to some trivial factors, nothing but the widely-studied total hadronic non-strange branching ratio ($B_{V+A}$) minus the one-pion one ($B_\pi$), {\it i.e.},
\bea
I_{V+A}^{\rm exp} 
&=&
\frac{1}{12\pi^2 |\hat{V}_{ud}|^2 S_{EW}}
\frac{B_{V+A} - B_\pi}{\hat{B}_e} = 25.049(74) \times 10^{-3}~,
\eea
where we used the HFLAV averages $B_{V+A}=0.6183(10)$ for the inclusive BR and $B_{\pi}=0.10804(52)$ for the single pion BR~\cite{HFLAV:2019otj},\footnote{The HFLAV fit is carried out summing over hadronic channels~\cite{HFLAV:2019otj}. Leptonic decays, which would potentially contaminate the results with new physics effects, are not used to reduce uncertainties in that fit. In $I_{V+A}^{\rm exp}$ we have neglected the correlation of $B_{\pi}$ and $B_{V+A}$, which, given the large theory errors, has no impact in our analysis.} which has to be removed because the lower integration limit in the $I_{V+A}$ definition is $s_{th}=4m_\pi^2$. A somewhat less precise value for $B_{V+A}$ could be obtained by integrating the ALEPH spectral function. This would not have any impact in the analysis, since theory errors are much larger than the experimental one, as we discuss below.

On the theory side, the SM prediction is
\bea
I_{V+A}^{\rm SM}
&=&-\frac{f_{\pi^\pm}^{2}}{m_\tau^2}\; \omega_\tau(m_\pi^2)
+2 A_{P}^{(\omega_\tau)}(m_\tau^2)+\frac{1}{4\pi^2}\delta_{\rm NP}^{\omega_{\tau}} \, ,
\eea
where $A^{\omega_\tau}_P=15.10(13)_{\text{pert}}(8)_{\alpha_{s}}\times 10^{-3}$ is the perturbative contribution. We have omitted the DV term, which is expected to be negligible for this weight. 
On the other hand, $\delta_{NP}^{\omega_{\tau}}$ encodes the small non-perturbative correction to the hadronic tau decay width, which is suppressed by six powers of $\Lambda_{QCD}/m_\tau$, namely
\begin{equation}
\delta_{\rm NP}^{\omega_{\tau}}= \delta_{\rm NP}^{\omega_{\tau},\, \mathcal{O}_{6}}+\delta_{\rm NP}^{\omega_{\tau}, \,\mathcal{O}_{8}} \equiv-4\pi^2\left(3\frac{\mathcal{O}_6^{V+A}}{m_\tau^6} + 2\frac{\mathcal{O}_{8,V+A}}{m_\tau^8}\right) = 0.000 \pm 0.015\,,
\end{equation}
which we have estimated using (i) $|\delta_{NP}^{\omega_{\tau}, \,\mathcal{O}_{8}}|<|\delta_{NP}^{\omega_{\tau},\, \mathcal{O}_{6}}|$; (ii)  $\mathcal{O}_6^{V+A}<|\mathcal{O}_6^{V-A}|$, which holds in the vacuum-saturation approximation~\cite{Shifman:1978bx}; and (iii) the recent determination of the V-A dimension-6 condensate, $\mathcal{O}_6^{V-A}=-0.0029(5)$ GeV$^6$~\cite{Pich:2021yll}, which we discuss in ~\sref{VminusA} in more detail.\footnote{In Ref.~\cite{Cirigliano:2018dyk} a more naive dimensional estimate was used, namely $|{\cal O}^{V+A}_{2d}|\lesssim (0.4\,\rm{GeV})^{2d}(d-1)!$, which lead to a 2x larger uncertainty in $\delta_{NP}^{\omega_{\tau}}$. This estimate is in agreement (although less precise) with the values obtained in SM analyses, which extract these non-perturbative contributions from tau data using several moments and assuming the absence of BSM effects, see {\it e.g.} Refs.~\cite{Davier:2013sfa,Boito:2014sta,Pich:2016bdg}.}

All in all, the resulting SM prediction is
\begin{equation}
I_{V+A}^{\rm SM}=24.83(39)_{\text{OPE}}(26)_{\text{pert}}(16)_{\alpha_{s}}(7)_{f_{\pi^\pm}}=24.83(50) \times 10^{-3} \, .
\end{equation}
which leads to the following new physics bound
\begin{align}\nonumber
0.0501\,\left(\epsilon_{L+R}^{d\tau}-\epsilon_{L+R}^{de}\right) - 0.0380\,\epsilon_{R}^{d\tau} +0.025(8) \hat \epsilon_{T}^{d\tau}
 &=
 0.22(39)_{\text{OPE}}(26)_{\text{pert}}(16)_{\alpha_{s}} (7)_{f_{\pi^\pm}}(7)_{\text{exp}}\times 10^{-3}\\
 &=
 0.22(50) \times 10^{-3} \; ,
\end{align}
where we see that the error is dominated by the perturbative and non-perturbative OPE uncertainties. 
As discussed above, the numerical coefficients multiplying $\epsilon_{L+R}^{d\ell}$ and $\epsilon_R^{d\tau}$ are calculated using ALEPH data~\cite{Davier:2013sfa}, whereas in the $\hat \epsilon_T^{d\tau}$ case we use $I_{VT}=0.0041(13)$ from~\tref{IVT} in~\aref{VTintegral}.

\paragraph{Integral of the $V+A$ spectral function.-}  
This observable corresponds to the $V+A$ case with $n=0$ in~\eref{masterformula}. Its SM prediction is particularly simple because ${\cal O}_2^{V+A}$ vanishes (up to negligible quark mass corrections):
\bea
I_{V+A}^{\rm SM}  = 
-\frac{f_{\pi^\pm}^{2}}{s_{0}}
+2 A_{P}^{(n=0)}(s_{0})
+ \delta^{\rm DV}_{V+A}(s_{0};n\!=\!0)
~.
\eea
We plot in~\fref{VplusA2} the difference between the experimental integral, $I^{\rm exp}_{V+A}$, and its SM value, $I^{\rm SM}_{V+A}$, for various $s_0$ values. Note that only experimental uncertainties are shown in the plot, but theory uncertainties are included as well in our analysis. 
Working with $s_{0}=2.8\, \mathrm{GeV}^{2}$, which is the last point with not-too-large experimental uncertainty, we find
\bea
\label{eq:ExpInclusive1}
I_{V+A}^{\rm exp} &=& 53.42(62)\times 10^{-3}~,\\
I_{V+A}^{\rm SM} &=& 52.45(61)_{\rm DV}(28)_{\text{pert}}(22)_{\alpha_{s}}(7)_{f_{\pi^\pm}}\times 10^{-3}=52.45(71)\times 10^{-3}~,
\eea
where we used $A_P=29.25(14)_{\text{pert}}(11)_{\alpha_{s}}\times 10^{-3}=29.25(18)\times 10^{-3}$. The weight chosen, $\omega(s)=1$, does not generate contributions from QCD vacuum condensates, which are usually not accurately known. 
On the other hand, this weight enhances experimental errors and the DV contribution because it does not suppress the $s\sim m_\tau^2$ region. Experimental errors in~\fref{VplusA2} are too large to make definite claims about the DV. 
One could assume they are negligible compared with experimental errors at $s_0=2.8$ GeV$^2$, but we have estimated conservatively the DV uncertainty from the difference between extrema in the $s_{0}\in[2.0, 2.8]\,\mathrm{GeV}^{2}$ interval of~\fref{VplusA2}. This is partly motivated by the fact that one might have DV effects that accidentally cancel $s_0$-dependent BSM contributions (even if they don't have the typical oscillatory behaviour of DVs).

\begin{figure}[t]\centering
\includegraphics[width=0.8\textwidth]{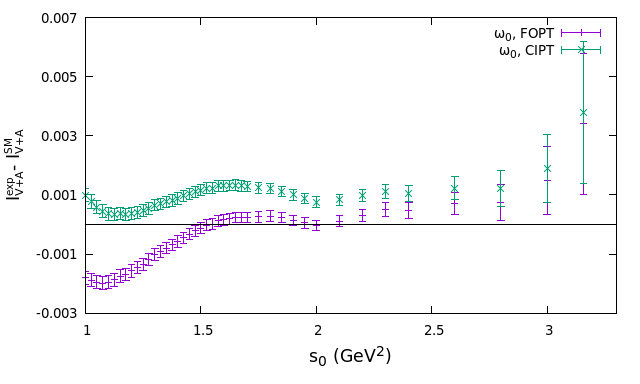}
\caption{\label{fig:VplusA2}
Difference between the experimental and SM values of the $I_{V+A}$ integrals for $\omega_0$, {\it cf.} LHS of Eq.~\eqref{eq:masterformula}. The error bars in the plot only include the experimental uncertainties.}
\end{figure}

All in all we find the following BSM bound
\begin{align}\nonumber
 0.107\,\left(\epsilon_{L+R}^{d\tau}-\epsilon_{L+R}^{de}\right) - 0.094\, \epsilon_{R}^{d\tau}+0.029(10) \hat \epsilon_{T}^{d\tau}
 &=
 1.00(62)_{\text{exp}}(61)_{\text{DV}}(28)_{\text{pert}} (22)_{\alpha_{s}}(7)_{f_{\pi^\pm}} \times 10^{-3}\\
 &=1.00(95) \times 10^{-3} \; , 
\label{eq:constraint1}
\end{align}
where the different sources of errors are shown. We see that experimental and DV errors dominate this bound. 
Once again, the numerical coefficients multiplying $\epsilon_{L+R}^{d\ell}$ and $\epsilon_R^{d\tau}$ are calculated using ALEPH data~\cite{Davier:2013sfa}, whereas in the $\hat \epsilon_T^{d\tau}$ case we use $I_{VT}=0.0048(16)$ from~\tref{IVT} in~\aref{VTintegral}.

\subsubsection{V-A}
\label{sec:VminusA}
The $V-A$ correlator would vanish if chiral symmetry were preserved beyond massless perturbative QCD. This makes the inclusive $V-A$ spectral function an excellent probe of Spontaneous Chiral Symmetry Breaking~\cite{Weinberg:1967kj,Knecht:1997ts}, which has been used to accurately determine $f_{\pi^\pm}$ and other low-energy constants of Chiral Perturbation Theory, QCD vacuum condensates ${\cal O}_D^{V-A}$ and quark-hadron DV~\cite{Boyle:2014pja,Boito:2015fra,Gonzalez-Alonso:2016ndl,Pich:2021yll}. These analyses were carried out in the absence of new physics contributions, which are the central objects of this work.  
We will be able to extract useful information about the BSM effects if we can have a good control of such non-perturbative SM contributions, which should be kept in mind when choosing the weights. Analytic weights ensure that the only low-energy parameter contributing is the pion decay constant, which is accurately known from lattice QCD. Dimension-2 and dimension-4 vacuum condensates are negligible~\cite{Braaten:1991qm} and the dimension-6 condensate can be extracted with $\lesssim 20\%$ precision from  $K\rightarrow \pi\pi$ matrix elements computed in the lattice~\cite{Pich:2021yll}. To avoid contributions from higher-dimensional condensates, which are not known from first principles, we will use polynomial weights with order smaller than three. Finally, to reduce quark-hadron DV it is convenient to work with weights that vanish for $s\approx s_0$ (sometimes known as pinched weights). These considerations lead us to using the following two weights in our analysis
\bea
\omega_1(s)&\equiv& 1-\frac{s}{s_{0}} \, ,
\label{eqomega2}\\
\omega_2(s)&\equiv&\left(1-\frac{s}{s_{0}} \right)^{2} \, ,
\eea

It is worth noting that the $\epsilon_R^{d\tau}$ and $\epsilon_T^{d\tau}$ contributions in~\eref{masterformula} are not suppressed, contrary to the SM prediction, which is suppressed because chirality is preserved at the perturbative level in the $V-A$ channel. This translates into an enhanced sensitivity to those Wilson coefficients.

\paragraph{$\boldsymbol{\omega_1(s)}$ weight.-}

In the absence of BSM effects, this weight gives nothing but a linear combination of the first and the second Weinberg Sum Rules~\cite{Weinberg:1967kj}, where the SM prediction is just the pion-pole contribution (up to small DVs):
\bea
I_{V-A}^{\rm SM} 
\,=\,
\frac{f_{\pi^\pm}^{2}}{s_{0}} \left(1-\frac{m_\pi^2}{s_{0}}\right)
+\delta^{\rm DV}_{V-A}(s_{0};\omega_1)
~,
\eea

We plot in~\fref{VminusA1} the difference between the experimental integral, $I^{\rm exp}_{V-A}$, and its SM value, $I^{\rm SM}_{V-A}$, for various $s_0$ values. Note that only experimental uncertainties are shown in the plot. 
For $s_{0}=2.8\, \mathrm{GeV}^{2}$ we have
\bea
I_{V-A}^{\rm exp} &=& 6.08(13) \times 10^{-3}~,\\
I_{V-A}^{\rm SM} &=& 6.01(60)_{\rm DV}(7)_{f_{\pi^\pm}} \times 10^{-3} ~.
\label{eq:SMpredictionInclusive3}
\eea
Since the weight suppresses the $s\sim 2.8\, \mathrm{GeV}^2$ region, one expects a small DV contribution, which is supported by the observed plateau in~\fref{VminusA1}. Thus one could just neglect the DV error in comparison with the experimental uncertainty. However, as in the integral of the $V+A$ spectral function, we opted in~\eref{SMpredictionInclusive3} to estimate conservatively the DV uncertainty from the difference between the $s_{0}=2.0$ and $2.8\, \mathrm{GeV}^{2}$ points in~\fref{VminusA1}. This gives
\bea
0.0122\,\left(\epsilon_{L+R}^{d\tau}-\epsilon_{L+R}^{de}\right) + 0.0371\, \epsilon_{R}^{d\tau}+0.023(14)\hat \epsilon_{T}^{d\tau}
&=&
0.07(60)_{\text{DV}}(13)_{\text{exp}}(7)_{f_{\pi^\pm}} \times 10^{-3}\nnl
&=&
0.07(62) \times 10^{-3} \, ,
\eea
which is clearly dominated by DV uncertainties. 
The numerical coefficients multiplying $\epsilon_{L+R}^{d\ell}$ and $\epsilon_R^{d\tau}$ are calculated using ALEPH data whereas in the $\hat \epsilon_T^{d\tau}$ case we use $I_{VT}=0.0038(24)$ from~\tref{IVT} in~\aref{VTintegral}.

\begin{figure}[t]\centering
\includegraphics[width=0.8\textwidth]{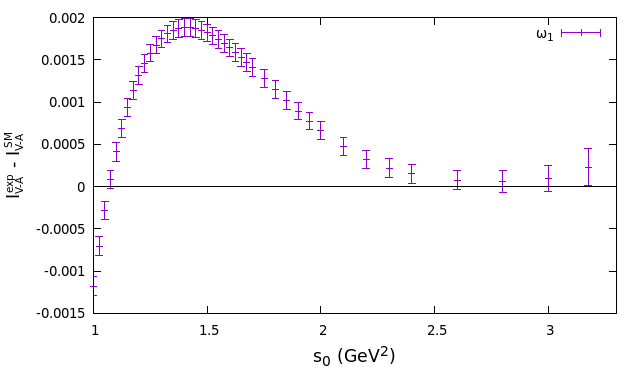}
\caption{\label{fig:VminusA1}
Difference between the experimental and SM values of the $I_{V-A}$ integrals for $\omega_1(s)$, {\it cf.} LHS of Eq.~\eqref{eq:masterformula}. The error bars in the plot only include the experimental uncertainties.}
\end{figure}

\begin{figure}[t]\centering
\includegraphics[width=0.8\textwidth]{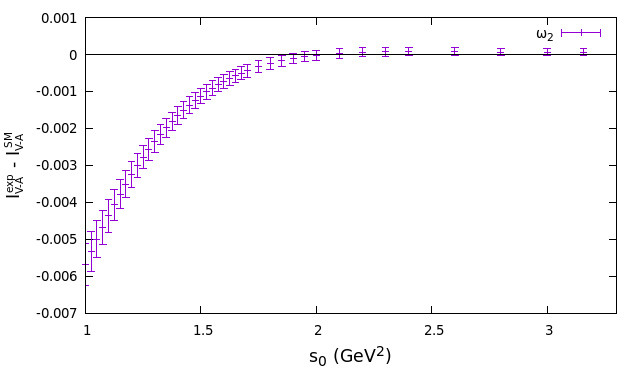}
\caption{\label{fig:VminusA2}
Difference between the experimental and SM values of the $I_{V-A}$ integrals for $\omega_2(s)$ in Eq.~\eqref{eqomega2}, {\it cf.} LHS of Eq.~\eqref{eq:masterformula}.}
\end{figure}

\paragraph{$\boldsymbol{\omega_2}(s)$ weight.-}
The SM prediction for this weight in the $V-A$ channel and using $s_0=m_\tau^2$ as upper integration limit is given by
\bea
I_{V-A}^{\rm SM}
=
\frac{f_{\pi^\pm}^{2}}{m_\tau^2} \left(1-\frac{m_\pi^2}{m_\tau^2}\right)^2
+\frac{\mathcal{O}_{6, V-A}}{m_{\tau}^{6}}
+\delta^{\rm DV}_{V-A}(m_{\tau}^{2},\omega_2)~.
\eea
Given the negligible DV expected for this weight, the only piece left to achieve a precise SM prediction is $\mathcal{O}_{6, V-A}$. Fortunately this vacuum condensate is connected with $K\to\pi\pi$ matrix elements~\cite{Donoghue:1999ku,Cirigliano:2001qw,Cirigliano:2002jy,Pich:2021yll}. Taking into account those relations, incorporating perturbative and chiral corrections and using recent lattice data~\cite{Abbott:2020hxn}, Ref.~\cite{Pich:2021yll} found\footnote{This number updates the preliminary value used in Ref.~\cite{Cirigliano:2018dyk}, $\mathcal{O}_6^{V-A}
=(-4.2 \pm 1.3) \times 10^{-3} \, \mathrm{GeV}^{6}$. The new result includes chiral corrections and new lattice results~\cite{Abbott:2020hxn}, see Ref.~\cite{Pich:2021yll} for details. The impact of this improvement on the subsequent new physics bound is very small.}
\begin{equation}
\mathcal{O}_6^{V-A}
=(-2.9 \pm 0.5) \times 10^{-3} \, \mathrm{GeV}^{6} ~,
\end{equation}
at $s_0=m_\tau^2$ (a small $s_0$-dependence appears due to the inclusion of perturbative corrections). This value leads to the following SM prediction
\bea
I_{V-A}^{\rm SM} &=& 5.212(65)_{f_{\pi^\pm}}(16)_{\mathcal{O}_{6}}\times 10^{-3}=5.212(67) \times 10^{-3}~,
\label{eq:SMpredictionInclusive4imp}
\eea
in excellent agreement with the experimental result
\begin{equation}
I_{V-A}^{\rm exp}=(5.285 \pm 0.074)\, \times 10^{-3} \, .
\end{equation}
Fig.~\ref{fig:VminusA2} shows the difference between the experimental and SM values for $s_0\leq m_\tau^2$. As expected for this weight and despite the small experimental errors there is no sign of the typical oscillatory behaviour associated to DVs. Let us note that the small $s_0$-dependence of the dimension-6 condensate was taken into account in this figure.

All in all the following BSM bound is obtained
\bea
\nonumber
0.0106\,\left(\epsilon_{L+R}^{d\tau}-\epsilon_{L+R}^{de}\right) +0.0204\, \epsilon_{R}^{d\tau}+0.017(16)\hat \epsilon_{T}^{d\tau}
&=&
0.074 (74)_{\mathrm{exp}} (65)_{f_{\pi^\pm}}(16)_{\mathcal{O}_{6}}\times 10^{-3}\\
&=&~0.07(10)\times 10^{-3}\, ,
\eea
which is dominated by the $f_{\pi^\pm}$ uncertainty. Like in the previous cases, the numerical coefficients multiplying $\epsilon_{L+R}^{d\ell}$ and $\epsilon_R^{d\tau}$ are calculated using ALEPH data whereas in the $\hat \epsilon_T^{d\tau}$ case we use $I_{VT}=0.0028(26)$ from~\tref{IVT} in~\aref{VTintegral}.

\subsubsection{Recap and SM limit}
Putting the four nonstrange inclusive constraints together and re-scaling them one finds:
\begin{align}
\epsilon_{L+R}^{d\tau}-\epsilon_{L+R}^{de}-0.76\epsilon_{R}^{d\tau}+0.49(16)\hat \epsilon_{T}^{d\tau}&\!=\!(4 \pm 10)\times 10^{-3}\!\label{eq:inclusive1},\\
\epsilon_{L+R}^{d\tau}-\epsilon_{L+R}^{de}-0.88\epsilon_{R}^{d\tau}+0.27(9)\,\hat \epsilon_{T}^{d\tau}&\!=\!(9.1\pm 8.8)\times 10^{-3}\!\label{eq:inclusive2},\\
\epsilon_{L+R}^{d\tau}-\epsilon_{L+R}^{de}+3.05\epsilon_{R}^{d\tau}+1.9(1.2)\hat \epsilon_{T}^{d\tau}&\!=\!(5\pm 51)\times 10^{-3}\!\label{eq:inclusive3},\\
\epsilon_{L+R}^{d\tau}-\epsilon_{L+R}^{de}+1.93\epsilon_{R}^{d\tau}+1.6(1.5)\hat \epsilon_{T}^{d\tau}&\!=\!(7.0\pm 9.5)\times 10^{-3}\!\label{eq:inclusive4}.
\end{align}
with the following correlation matrix
\begin{equation}
    \rho=\left(\begin{array}{cccc}
    1 & 0.12 & -0.016 & -0.09  \\
     & 1 & -0.027 & -0.11 \\
     &  & 1 & 0.23 \\
     & & & 1 \\
    \end{array}\right)~,
    \label{eq:rhoinclusive}
\end{equation}
which takes into account the main correlations between these constraints, which are of experimental origin and from the use of a common $f_{\pi^\pm}$ value. Theory uncertainties are dominated by different sources in each constraint, and thus their correlation is neglected, except for the systematic uncertainty coming from the choice of perturbative prescription, FOPT or CIPT, for which a $100 \%$ of correlation is estimated. The correlation between the $V+A$ constraints from a common $\alpha_s$ is neglected, because the associated error is subleading in both cases.

The main change of these results with respect to Ref.~\cite{Cirigliano:2018dyk} is two-fold. On one hand the uncertainty of the first constraint is $\sim 40\%$ smaller thanks to the new estimate of the non-perturbative contribution. On the other hand, we made nontrivial improvements concerning the calculation of the numerical coefficients multiplying $\hat \epsilon_T^{d\tau}$, {\it i.e.}, the $I_{VT}$ integrals defined in~\eref{IVTdef}. In Ref.~\cite{Cirigliano:2018dyk} these quantities were calculated at tree-level and leading OPE-order (quark condensate). A conservative $50\%$ uncertainty was assigned and the lower values were used in the analysis. In this work we work instead at Next-To-Leading-Log in the perturbative expansion for the quark condensante, and we include as well an estimate from the higher-dimensional condensates. The details are presented in~\aref{VTintegral} and summarized in~\tref{IVT}, where we see that the shift with respect to the tree-level LO result is significant (around $50\%$), in part because the various corrections happen to go in the same direction. 
The final $I_{VT}$ uncertainties, which are rather large and highly correlated between bounds in Eqs.~(\ref{eq:inclusive1})-(\ref{eq:inclusive4}), will be taken into account in the subsequent fits carried out in this work.

In the SM limit ($\epsilon=0$), our four dispersive relations can be used to determine the QCD parameters $\alpha_s$ and $f_{\pi^\pm}$, which enter the SM prediction: $I_{V\pm A}^{\rm SM}=f(\alpha_s,f_{\pi^\pm})$. The $V+A$ constraints, {\it i.e.}, Eqs.~(\ref{eq:inclusive1})-(\ref{eq:inclusive2}), can be translated into $\alpha_{s}$ values, which gives
\begin{equation}
\alpha_{s}(m_{\tau}^{2})=0.330 \pm 0.017 \, .
\end{equation}
in agreement with 
SM analyses~\cite{Baikov:2008jh,Beneke:2008ad,Davier:2013sfa,Boito:2014sta,Pich:2016bdg,Caprini:2020lff,Pich:2020gzz,Boito:2020xli,Ayala:2021mwc}. Our determination is less precise because we used only two moments and a rather conservative estimate of the non-perturbative contributions (instead of extracting them from tau data). It is worth noting however that our extraction is in excellent agreement with the recent review of Ref.~\cite{dEnterria:2018cye}, which has, running to the $\tau$ mass, $\alpha_{s}(m_{\tau}^{2})= 0.324(15)$ as a conservative average of hadronic tau decay analyses, which scatter around that number but with lower quoted uncertainties.

On the other hand, the second $V-A$ relation, which was built using the $\omega_2$ weight, is by far the most sensitive to the pion decay constant. In the absence of new physics contributions it gives
\begin{equation}\label{eq:fpiinc}
f_{\pi^\pm}=(131.10 \pm 0.92) \, \mathrm{MeV} \, .
\end{equation}
This is in perfect agreement with the value obtained in Ref.~\cite{Pich:2021yll}, $f_{\pi^\pm}=130.9(8) \, \mathrm{MeV}$, where the data is analyzed within the SM, using $s_{0}=2.8\, \mathrm{GeV}^2$ and a slightly different input for $V_{ud}$.

\subsection{Strange decays}

The formalism for studying the strange sector is the same as in the non-strange one, except for the change $d\rightarrow s$, the inclusion of $SU(3)$-breaking effects and the fact that $G$-parity cannot be used to separate states into $V$ and $A$ ones. The normalised invariant mass-squared distribution is then given by
\begin{align}
\frac{dN^D_\tau}{N_\tau ds}
=
\kappa_D\, \left(1-\frac{s}{m_{\tau}^{2}}\right)^2
&\Bigg[\left(1+\frac{2s}{m_{\tau}^{2}}\right)(1+2\epsilon_{L+R}^{D\tau})\operatorname{Im}\Pi_{VV,D}^{(1+0)}(s) 
\nnl
&+
\left(1+\frac{2s}{m_{\tau}^{2}}\right)(1+2\epsilon_{L+R}^{D\tau}-4\epsilon_{R}^{D\tau})\operatorname{Im}\Pi_{AA,D}^{(1+0)}(s) 
\nnl  &-\frac{2s}{m_{\tau}^{2}}\left(1+2\epsilon_{L+R}^{D\tau}+\epsilon^{D\tau}_{S}\frac{m_{\tau}}{S_{EW}^{\mathrm{had}}(m_{u}-m_{D})}\right)\operatorname{Im}\Pi_{VV,D}^{(0)}(s)
\nnl
&-\frac{2s}{m_{\tau}^{2}}\left(1+2\epsilon_{L+R}^{D\tau}-4\epsilon_{R}^{D\tau}+\epsilon^{D\tau}_{P}\frac{m_{\tau}}{S_{EW}^{\mathrm{had}}(m_{u}+m_{D})}\right)\operatorname{Im}\Pi_{AA,D}^{(0)}(s)\Bigg] 
\nnl
&+ 6\, \hat \epsilon^{D\tau}_T \, \frac{\operatorname{Im}\Pi_{VT,D}(s)}{m_{\tau}S_{EW}^{\mathrm{had}}}
\, , 
\label{eq:distVplusA}
\end{align}
for the non-strange ($D=d$) and strange ($D=s$) cases, respectively. We have also defined ${\kappa_D\equiv 12 \pi |V_{uD}|^2 \hat{B}_{e}} S_{EW} / m_{\tau}^2$ and we have added a $D$ subindex to the correlators. 
We have taken into account that $VA$, $VP$ and $AS$ correlators vanish because of parity considerations 
and we have used conservation of vector and axial currents to relate the $VS$ and $AP$ contributions with the longitudinal components of the $VV$ and $AA$ ones, respectively~\cite{Bijnens:2003rc}. The associated non-strange contributions can once again be safely neglected owing to the small value of $m_{u}$ and $m_d$, but this is not true anymore for the strange pieces. Finally, the tensor BSM term is calculated in the $SU(3)_{V}$ limit, where the $AT$ contribution vanishes.\footnote{This requires using also charge conjugation, which changes the sign of the $AT$ correlator and flips the ordering of the quark fields inside the current. In the non-strange sector this change of ordering is compensated with an extra isospin rotation so, if both are good symmetries, the $AT$ correlator changes sign after applying both transformation and hence it has to vanish. This is nothing but a $G$-parity transformation. In the strange sector the extra rotation needed is only valid when the three light masses are the same, {\it i.e.}, in the $SU(3)_V$ limit.}

Experimental resolution is worse in the strange case, mainly because of the Cabibbo suppression, 
and strange spectral functions are not publicly available. This will hopefully change soon with the arrival of Belle-II data but, in the meantime, we only work with the total strange decay width. 
We normalize it as
\beq
\hat{R}_{\tau}^D \equiv \tau_{\tau}\,\frac{\sum_{n_D}\Gamma[\tau^-\to \nu_\tau ~n_D(\gamma)]}{\hat{B}_e}=\frac{B_{D}}{\hat{B}_e} \, ,
\eeq
where $n_D$ is a hadronic system with the appropriate strangeness ({\it i.e.},  $S=0/1$ for $D=d/s$) and $B_D$ denotes the inclusive (non)strange branching ratio. In the SM limit it reduces to the usual $R_{\tau}^{D}$ definition, where $\hat{B}_e$ corresponds to SM prediction of the branching ratio associated to the $\tau^{-}\rightarrow e^{-}\nu_{\tau}\bar{\nu}_{e}$ decay mode. The hat over $R_{\tau}^D$ reminds that, in a general BSM set up, $\hat{B}_{e}\neq B_e$ due to different new physics contributions in $\tau\rightarrow e^{-}\nu_{\tau}\bar{\nu}_{e}$ with respect to $\mu\rightarrow e^{-}\nu_{\mu}\bar{\nu}_{e}$.

In the limit of $SU(3)_{V}$ conservation, the integrals of the imaginary part of the nonstrange and strange correlators are equal. Thus, in the SM one has
\beq
\frac{R_{\tau}^d}{|V_{ud}|^2} = \frac{R_{\tau}^s}{|V_{us}|^2} \,+\, \delta R_{\rm th}^{\rm SM}~,
\label{eq:strangeSM}
\eeq
where the last term denotes calculable $SU(3)$-breaking corrections. 
This relation has been used to extract $|V_{us}|$ from inclusive tau decays~\cite{Gamiz:2007qs,Hudspith:2017vew}:
\begin{equation}
|V_ {us}|^2 = \frac{R_{\tau}^s}{\frac{R_{\tau}^d}{|V_{ud}|^2}-\delta R_{\rm th}^{\rm SM}} \, .
\end{equation}
If BSM effects are present they would pollute this extraction. Comparing it with the $V_{us}$ value extracted from a different process, such as $K\to \pi\ell\nu_\ell$, we will set bounds on BSM effects that affect those two extractions differently. 
In order to do such lepton-flavor-universality test we calculate the experimentally extracted $R_\tau^D/|V_{uD}|^2$ ratio in the presence of generic nonstandard contributions
\begin{align}
\label{eq:RoverV}
\frac{\hat{R}_{\tau}^D}{|\hat{V}_{uD}|^{2}}
=
\left[\frac{R^D_{\tau}}{|V_{uD}|}^{2} \right]^{\rm SM}_{\rm th}
\left( 1+ 2 \,\delta_{\mathrm{BSM},\, D}^{\rm inc} \right)\, .
\end{align}
In analogy with the nonstrange case, we include in $\delta_{\mathrm{BSM},D}^{\rm inc}$ the potential new physics effects affecting the ratio 
$\frac{|V_{uD}|^{2}}{|\hat{V}_{uD}|^{2}}=1-2\epsilon_{L}^{De}-2\epsilon_{R}^{De}$. Integrating the inclusive invariant mass distribution of~\eref{distVplusA} we find:
\begin{align}
\delta_{\mathrm{BSM},\, D}^{\rm inc}
=\,\epsilon_{L+R}^{D\tau} - \,\epsilon_{L+R}^{De}
+c^{R}_{D}\,\epsilon_R^{D\tau}
+c^{S}_{D}\,\epsilon^{D\tau}_{S}
+c^{P}_{D}\,\epsilon_{P}^{D\tau}
+c_D^{T}\,\hat \epsilon^{D\tau}_{T} \, ,
\end{align}
where
\begin{align}\nonumber
c_D^R&=-24\pi \frac{|\hat{V}_{uD}|^{2}}{\hat{R}_{\tau}^D} \int^{m_{\tau}^{2}}_{s_{th}^{D}}\frac{ds}{m_{\tau}^{2}}\left(1+\frac{2s}{m_{\tau}^{2}}\right)\left(1-\frac{s}{m_{\tau}^{2}}\right)^{2}\operatorname{Im}\Pi_{AA,\, D}^{(1+0)}(s)
\\ &\quad\,-24\pi \frac{|\hat{V}_{uD}|^{2}}{\hat{R}_{\tau}^D} \int^{m_{\tau}^{2}}_{s_{th}^{D}}\frac{ds}{m_{\tau}^{2}}\left(\frac{-2s}{m_{\tau}^{2}}\right)\left(1-\frac{s}{m_{\tau}^{2}}\right)^{2}\operatorname{Im}\Pi_{AA,\, D}^{(0)}(s) \, ,\label{eq:alphaR}\\
c^{S}_{D}&=\frac{6\pi m_{\tau}}{m_{u}-m_{D}}\frac{|\hat{V}_{uD}|^{2}}{\hat{R}_{\tau}^D} \int^{m_{\tau}^{2}}_{s_{th}^{D}}\frac{ds}{m_{\tau}^{2}}\left(\frac{-2s}{m_{\tau}^{2}}\right)\left(1-\frac{s}{m_{\tau}^{2}}\right)^{2}\operatorname{Im}\Pi_{VV,\, D}^{(0)}(s) ,\\
c^{P}_{D}&=\frac{6\pi m_{\tau}}{m_{u}+m_{D}}\frac{|\hat{V}_{uD}|^{2}}{\hat{R}_{\tau}^D} \int^{m_{\tau}^{2}}_{s_{th}^{D}}\frac{ds}{m_{\tau}^{2}}\left(\frac{-2s}{m_{\tau}^{2}}\right)\left(1-\frac{s}{m_{\tau}^{2}}\right)^{2}\operatorname{Im}\Pi_{AA,\, D}^{(0)}(s) ,\\
c^{T}_{D}&=36\pi\frac{|\hat{V}_{uD}|^{2}}{\hat{R}_{\tau}^D} \int^{m_{\tau}^{2}}_{s_{th}^{D}}\frac{ds}{m_{\tau}^{2}}\left(1-\frac{s}{m_{\tau}^{2}}\right)^{2}\operatorname{Im}\Pi_{VT,\, D}(s) \, .\label{eq:alphaT}
\end{align}
In the expressions for the $c_D^\Gamma$ coefficients we have replaced the SM prediction of the $R_\tau^D/|V_{uD}|^2$ ratio by its experimental value, an identification that is valid up to quadratic BSM terms. In contrast with the previous subsection, we are defining $s_{th}^{D}$ in such a way that the integrals include the single pole contributions, {\it i.e.}, $s_{th}^{d}\equiv m_{\pi}^2-\epsilon$ and $s_{th}^{s}\equiv m_{K}^2-\epsilon$, mainly because it makes the connection with the SM works more straightforward. Finally, the SM prediction for $R_{\tau}^{D} / |V_{uD}|^2$ in~\eref{RoverV} can be calculated using the QCD dispersion relations that were described in the previous section. All we need to know is that the result is the same for the nonstrange and strange cases, up to calculable $SU(3)$-breaking corrections, as shown in~\eref{strangeSM}. Finally, we stress that the expression for the tensor coefficient $c_D^T$ in~\eref{alphaT} is only valid in the $SU(3)_V$ limit, as explained above.

We can now recycle the SM works of Refs.~\cite{Gamiz:2004ar,Gamiz:2007qs,Gamiz:2013wn}, which make use of strange tau data to obtain a value for $V_{us}$. In the presence of non-standard interactions the polluted $\hat{V}_{us}^{\rm inc}$ value extracted from tau decays is related to the polluted $\hat{V}_{us}$ value extracted from $K\to\pi e\nu_e$ by the following relation 
\begin{equation}
\label{eq:VusComparison}
|\hat{V}_{us}^{\rm inc}|
= \left( \frac{\hat{R}^s_{\tau}}{\frac{\hat{R}_{\tau}^d}{|\hat{V}_{ud}|^2}-\delta R_{\rm th}^{\rm SM}} \right)^{1/2}
= |\hat{V}_{us}| \left(1+ \delta_{\mathrm{BSM},\, s}^{\rm inc} - (1+\eta)\delta_{\mathrm{BSM},\, d}^{\rm inc} \right)\, ,
\end{equation}
up to quadratic BSM terms, where $\eta= \delta R_{\rm th}|\hat{V}_{us}|^2 / \hat{R}_{\tau}^s \approx 0.07$ 
is an $SU(3)$-breaking factor. 
Using $B_d=0.6183(10)$ and $B_s=0.02931(41)$ as experimental inputs~\cite{HFLAV:2019otj}, as well as $\delta R_{\rm th}^{\rm SM}= 0.237(29)$~\cite{Gamiz:2013wn} 
\begin{equation}
\label{eq:VusTau}
|\hat{V}_{us}^{\rm inc}| = 0.2192 \pm 0.0015_{\rm exp} \pm 0.0009_{\rm th} \, ,
\end{equation}
in good agreement with Ref. \cite{HFLAV:2019otj}.

Now we move to discuss the calculation of the $c_D^\Gamma$ coefficients that appear in the BSM contributions, for which we can take expressions in the SM limit, since we are neglecting quadratic new physics terms. For the $c_{R}^{d}$ we can simply use the SM limit of Eq.~(\ref{eq:distVplusA}) to rewrite $c_{R}^d=-2\,\hat{R}_{\tau,A}^d / \hat{R}_{\tau}^d$, 
which can be taken from experimental data. Once again we cannot simply use the same relation for the strange counterpart, since we cannot use $G$-parity to separate the $V$ and $A$ channels. Instead we use that the needed integral over $\mathrm{Im}\Pi_{AA,s}^{(1+0)}$ in Eq.~(\ref{eq:alphaR}) is very close to the corresponding $\mathrm{Im}\Pi_{VV,s}^{(1+0)}$ one. Deviations from the exact equality due to quark masses and spontaneous chiral symmetry breaking effects are described by OPE corrections, and their typical size is below $5\%$ of the total one \cite{Pich:1999hc}. Then we simply take $\mathrm{Im}\Pi_{AA,s}^{(1+0)}= \frac{1}{2}(\mathrm{Im}\Pi_{VV,s}^{(1+0)}+\mathrm{Im}\Pi_{AA,s}^{(1+0)})$, adding a conservative $5\%$ of estimated uncertainty, and use the SM limit of Eq.~(\ref{eq:distVplusA}) to rewrite
\begin{equation}
c_{s}^{R}=-1+12\pi\frac{|\hat{V}_{us}|^{2}}{\hat{R}_{\tau}^s} \int^{m_{\tau}^{2}}_{s_{th}^{s}}\frac{ds}{m_{\tau}^{2}}\left(\frac{-2s}{m_{\tau}^{2}}\right)\left(1-\frac{s}{m_{\tau}^{2}}\right)^{2}\operatorname{Im}(\Pi_{VV,\, s}^{(0)}-\Pi_{AA,\, s}^{(0)})(s)~.
\end{equation}

For the integrals in $c_i^\Gamma$ involving the longitudinal correlators, $\Pi^{(0)}$, we use the values obtained in Ref.~\cite{Gamiz:2002nu} for $R_{uD,V/A}^{kl,L}$, defined as
\begin{equation}
R_{uD,V/A}^{kl,L}\equiv-24\pi\int^{m_{\tau}^2}_{0}\frac{ds}{m_{\tau}^2}\left( 1-\frac{s}{m_{\tau}^2}\right)^{2+k}\left(\frac{s}{m_{\tau}^2}\right)^{1+l}\mathrm{Im}\Pi_{VV/AA, \, D}^{(0)}(s) \,.
\end{equation}
All in all we have
\begin{align}
c_{d}^{R}&=-2\frac{\hat{R}_{\tau,A}^{d}}{\hat{R}_{\tau}^{d}}=  -0.97(1) \, ,\\
c_{s}^{R}&=-1.00(5)-\frac{|\hat{V}_{us}|^2}{\hat{R}_{\tau}^{s}}(R_{us,V}^{00,L}-R_{us,A}^{00,L})=-1.03(5)    \, ,\\
c_{s}^{S}&=\frac{m_{\tau}}{m_u-m_s}\frac{|\hat{V}_{us}|^2}{2\hat{R}_{\tau}^{s}}R_{us,V}^{00,L}= \,0.08(1)   \, ,\\
c_{d}^{P}&=\frac{m_{\tau}}{m_u+m_d}\frac{|\hat{V}_{ud}|^2}{2\hat{R}_{\tau}^{d}}R_{ud,A}^{00,L}=-0.278(4)     \, ,\\
c_{s}^{P}&=\frac{m_{\tau}}{m_u+m_s}\frac{|\hat{V}_{us}|^2}{2\hat{R}_{\tau}^{s}}R_{us,A}^{00,L}=-0.38(1)   \, ,\\
c_{D}^{T}&=36\pi^2\frac{|\hat{V}_{ui}|^{2}}{\hat{R}_{\tau}^i}I_{VT}^{R_{\tau}}=0.40(13)  \, ,
\end{align}
where we have used $R_{ud,A}^{00,L}=-0.00777(8)$, $R_{us,A}^{00,L}=-0.135(3)$ and $R_{us,V}^{00,L}=-0.028(4)$ from Table 2 in Ref.~\cite{Gamiz:2002nu}. 
Note also that $c_{d}^{V}$ vanishes in the isospin limit.

Finally, we have computed the tensor coefficient $c_D^T$ in the $SU(3)_{V}$ limit using exactly the same approach as in the non-strange sector, {\it i.e.}, we use $I_{VT}^{R_\tau}=0.0041(13)$ from~\tref{IVT} in~\aref{VTintegral}. 
We expect the $SU(3)_V$ breaking piece to be negligible compared to the large $I_{VT}^{R_\tau}$ uncertainty. 
It is worth noting that our result for the tensor contribution disagrees by a factor of 2 and a minus sign with Ref.~\cite{Dighe:2019odu}, where the effect of a non-standard tensor contribution in strange tau decays was studied.\footnote{
To ease the comparison, let us write the tree-level contribution linear in $\hat\epsilon_T^{s\tau}$ (called $C_T$ in Ref.~\cite{Dighe:2019odu}) as follows:
\beq
\delta R_{\rm NP}^{\epsilon_T^{s\tau},\,\mathrm{tree}} \equiv 
\left( \frac{\hat{R}_{\tau}^d}{|\hat{V}_{ud}|^2} - \frac{\hat{R}_{\tau}^s}{|\hat{V}_{us}|^2} - \delta R_{\rm th}^{\rm SM} \right)_{\hat \epsilon_T^{s\tau},\,\mathrm{tree}} 
= - \frac{\hat{R}_{\tau}^s}{2|\hat{V}_{us}|^{2}}c_{s, \mathrm{tree}}^T \hat \epsilon_T^{s\tau} 
=144\pi^2\frac{\langle \bar{q}q\rangle_{\mu_0}}{m_\tau^3}\hat \epsilon_T^{s\tau~.}
\eeq}

Once we have calculated the  $c_D^\Gamma$ coefficients, we can use~\eref{VusComparison} to obtain a BSM constraint from the comparison of the $V_{us}$ value extracted from inclusive tau data in~\eref{VusTau}, and the $K_{\ell 3}$ value, $\hat{V}_{us} = 0.22306(56)$ (see~\sref{LEFFE}):
\begin{align}
\epsilon_{\tau\rightarrow \nu_\tau s\bar{u} }
&\equiv \delta_{\mathrm{BSM},s}^{\rm inc}-(1+\eta)\delta_{\mathrm{BSM},d}^{\rm inc}\nnl
&=1.00\,(\epsilon_{L+R}^{s\tau}-\epsilon_{L+R}^{se})-1.03\, \epsilon_{R}^{s\tau}-0.38
\,\epsilon_{P}^{s\tau}+0.40(13)\,\hat \epsilon_{T}^{s\tau}+0.08(1)\,\epsilon_{S}^{s\tau}
\nnl
&-1.07\,(\epsilon_{L+R}^{d\tau}-\epsilon_{L+R}^{de})+\,1.04\, \epsilon_{R}^{d\tau}\,+\,0.30\,\,\epsilon_{P}^{d\tau}\,-\,0.43(14)\, \hat \epsilon_{T}^{d\tau} 
\nonumber
\\&= -(0.0171 \pm 0.0085)\,,
\label{eq:vusincbound}
\end{align}
which is the main result of this section. The contributions in the second (third) line are those affecting the inclusive strange (non-strange) decay. The small difference between the numbers in those two lines is due to $SU(3)$-breaking effects. In the above result we have kept uncertainties in the new physics prefactors only when they are large ($> 10 \%$).

The observable that we have used to probe this combination of Wilson coefficients can be decomposed in four pieces: the one-pion and one-kaon channels, and the remaining inclusive non-strange and strange BRs. 
Since we have already studied the first three contributions in~\sref{TauToPnu} and~\sref{nonstrangedecays}, we can use the associated bounds in Eqs.~(\ref{eq:epsilonpi}),~(\ref{eq:epsilonK}), and (\ref{eq:inclusive1}) to disentangle the novel combination that we are now probing, which is given by
\bea
\label{eq:tildeepsilon_strange_inclusive}
\epsilon_{L+R}^{s\tau}-\epsilon_{L+R}^{se}-0.73\, \epsilon_{R}^{s\tau}-0.05(1)\,\epsilon_{P}^{s\tau}+0.5(2)\,\hat \epsilon_{T}^{s\tau}+0.10(1)\,\epsilon_{S}^{s\tau}
= -0.017(16)  
\, .
\eea
This is (half) the BSM contribution to the inclusive strange BR minus the kaon pole, i.e., the s-quark analogue of~\eref{inclusive1}. Let us stress that~\eref{tildeepsilon_strange_inclusive} does not represent a new constraint, since it is derived from~\eref{vusincbound} and the above-mentioned non-strange constraints.

\subsection{Possible future improvements}
Finally, let us briefly review some possible future improvements that would improve the BSM bounds that we have obtained from inclusive observables. On the experimental side, future spectral functions, hopefully coming from Belle II~\cite{Belle-II:2018jsg}, would translate into an improvement of the different bounds, by reducing experimental uncertainties with respect to current LEP data, which could also translate into a better knowledge of DVs. 

There is much more room for improvement in the strange sector, since publicly available spectral functions would allow us to study several integrated moments, each one sensitive to a different combinations of BSM coefficients. This would allow us to disentangle them, like we have done in the nonstrange sector. Furthermore, it was shown in Ref. \cite{RBC:2018uyk} that one can achieve a good predictive power for weight functions with poles in the Euclidean axis, once the corresponding residues are computed in the lattice.\footnote{%
Precise measurements of the relevant spectral functions would definitively help in clarifying the situation \cite{Pich:2020gzz}.
Let us note that Refs.~\cite{Hudspith:2017vew,RBC:2018uyk} quote $V_{us}$ values more compatible with $\hat{V}_{us}$ from $K_{e3}$. However, they do not directly work with the total inclusive strange BR, but with other spectral moments (that typically give more importance to the already included $\tau\to\nu_\tau K$ channel) and sometimes involve $K_{\ell 2}$ and $K_{\ell 3}$ data as well. 
\label{footnote:inclusiveVus}} 
Notice how similar weights, in combination with corresponding lattice data, could also be used to get complementary new physics bounds for the non-strange $V-A$ channel.

On the theoretical side, one of the main limitations that may be overcome in the future are the uncertainties coming from higher-order and non-perturbative corrections~\cite{Boito:2016pwf,Boito:2018rwt,Hoang:2020mkw,Hoang:2021nlz,Ayala:2021mwc}. This would decrease some of the dominating SM uncertainties in our bounds, and it would allow us to use additional moments. It would also allow for a more precise determination of the $I_{VT}$ integral, with the associated improvement on the bound over the nonstrange tensor contribution. 
Finally, long-distance radiative corrections (which currently can be neglected) should be assessed in order to achieve a per-mil level precision.

\section{Recap and combination}
\setcounter{equation}{0}
\label{sec:RecapAndFlavor}

In this section we present a likelihood function for the Wilson coefficients of the EFT Lagrangian in \eref{leff1}, combining various low-energy probes of $d(s)\to u\ell\nu_\ell$ transitions. 
We first recapitulate the bounds from the $\tau$ observables discussed in this paper. 
Next, we review and update bounds from a variety of nuclear beta, pion, and kaon decays, which probe the electron and muon charged-current interactions with light quarks. 
Finally, all these probes are combined into one global likelihood, which can be used to constrain a broad range of new physics models affecting the charged-current interactions of light quarks and leptons in \eref{leff1}. 
We discuss the SM limit of this likelihood and the phenomenological determinations of the meson decay constants and the Cabibbo angle. 
As is well known, various determinations of the latter are in tension with each other~\cite{Seng:2018yzq,Grossman:2019bzp,Crivellin:2020lzu,Crivellin:2021njn}, the fact often referred to as the {\em Cabibbo anomaly}. 
As an illustration of sensitivity to new physics, we also display constraints on the Wilson coefficients in \eref{leff1} when only one of them is present at a time.
This shows in particular  simple directions in the EFT parameter space that are favored by the Cabibbo anomaly.

\subsection{Recap of bounds from Hadronic Tau Decays}
\begin{table}[tb]\renewcommand{\arraystretch}{1.2}
\centering\begin{tabular}{|c||c|c|c|c|c|c|} \hline
& $\epsilon_L^{d\tau}\; \times\;  10^{3}$ & $\epsilon_L^{de} \; \times\;  10^{3}$
& $\epsilon_R^{d}\; \times\;  10^{3}$
& $\epsilon_P^{d\tau}\; \times\;  10^{3}$ & $\epsilon_T^{d\tau} \; \times\;  10^{3}$ & $\epsilon_S^{d\tau} \; \times\;  10^{3}$
\\ \hline \hline
$\tau \rightarrow \pi \nu$ & -0.9(7.3) & 0.9(7.3)
& 0.9(7.3)
& 0.6(5.0) & x & x
\\ \hline 

$\tau \rightarrow \pi \pi \nu$ &10(4.9) & -10(4.9) 
& x
& x & 23(12) & x
\\ \hline
$\tau \rightarrow \pi \eta \nu$ 
&x & x
&x
& x &  x & $(-21,10)$
\\ \hline
$V+A$ & 6.9(7.0)&  -6.9(7.0)
&-8.6(8.4)  
& x & 15(19) & x \\ \hline
$V-A$ & 7.0(9.5)&  -7.0(9.5) 
& 3.6(4.9)
& x & 15(17) & x \\ \hline
\multicolumn{7}{c}{} \vspace{-0.40cm}\\ \hline 
& $\epsilon_L^{s\tau}\; \times\;  10^{3}$ & $\epsilon_L^{se} \; \times\;  10^{3}$
& $\epsilon_R^{s}\; \times\;  10^{3}$
& $\epsilon_P^{s\tau}\; \times\;  10^{3}$ & $\epsilon_T^{s\tau} \; \times\;  10^{3}$ & $\epsilon_S^{s\tau} \; \times\;  10^{3}$ \\ \hline\hline
$\tau \rightarrow K \nu$ 
& -2(10) &  2(10)
&  2(10)  
&   1.2(6.1) &   x & x \\ \hline
S. Inclusive  & -17(16)&  17(16) 
&23(22)
& 340(327) & -34(35) & -170(161) \\ \hline
\end{tabular}
\caption{One-at-a-time bounds on the Wilson coefficients for each channel. For the non-strange inclusive decays, we have separated them in $V+A$ and $V-A$, cf. Eqs.~(\ref{eq:inclusive1})-(\ref{eq:inclusive2}) and~(\ref{eq:inclusive3})-(\ref{eq:inclusive4}), respectively. For the strange inclusive decays, these one-at-a-time bounds correspond to Eq.~(\ref{eq:tildeepsilon_strange_inclusive}). The cross means that the corresponding channel is not sensitive to that particular Wilson coefficient.}
\label{tab:IndividualBounds}
\end{table}
Let us recapitulate the BSM bounds that we have obtained in this work so far. 
On one hand, exclusive channels $\tau\to\pi\nu_\tau,K\nu_\tau,\pi\pi\nu_\tau$
gave us three constraints, {\it cf.} Eqs.~(\ref{eq:epsilonpi}),~(\ref{eq:epsilonK}) and~(\ref{eq:pipi}). 
On the other hand, we obtained five BSM bounds from inclusive observables, {\it cf.} Eqs.~(\ref{eq:inclusive1})-(\ref{eq:inclusive4}) and~(\ref{eq:vusincbound}).  The one-at-a-time bounds on each Wilson coefficient for each channel are displayed in Table~\ref{tab:IndividualBounds}.

Combining all these channels 
we find the following $68\%$~CL marginalized intervals for the (combinations of) Wilson coefficients:
\begin{equation}
\label{eq:tauRecap}
\left(
\begin{array}{c}
\epsilon_L^{d\tau/e}+\epsilon_R^{d\tau}-\epsilon_R^{de} \\
\epsilon_R^{d\tau} \\
\epsilon_P^{d\tau} \\
\hat \epsilon_T^{d\tau} \\
\epsilon_L^{s\tau/e} - \epsilon_R^{s\tau} - \epsilon_R^{se} - \frac{m_{K^\pm}^2}{m_\tau(m_u+m_s)} \epsilon_P^{s\tau}  \\
\epsilon_{L}^{s\tau/e} -0.03\epsilon_R^{s\tau}-\epsilon_R^{se}  + 0.08(1)  \epsilon_{S}^{s \tau} - 0.38\epsilon_{P}^{s\tau}  + 0.40(13)\hat \epsilon_{T}^{s\tau} 
\\
\end{array}
\right)
=
\left(
\begin{array}{c}
2.4 \pm 2.6  \\
0.7 \pm 1.4 \\
0.4 \pm 1.0 \\
-3.3 \pm 6.0 \\
-0.2 \pm 1.0 \\
-1.3 \pm 1.2  \\
\end{array}
\right)\times 10^{-2}~,
\end{equation}
where $\epsilon_L^{D\tau/e} \equiv \epsilon_L^{D\tau}- \epsilon_L^{D e}$, 
and the Wilson coefficients are defined in the $\overline{\rm MS}$ scheme at scale $\mu=2$ GeV.
This is the main result of this paper. 
Note that we do not have enough experimental information to disentangle the different Lorentz structures of the strange Wilson coefficients  $\epsilon_X^{s\tau}$. 
The two combinations appearing above are simply the one affecting $\tau \to K \nu_\tau$ (cf.~\eref{deltaBSMP}), and the one affecting the inclusive $\tau\to \bar{u}s\nu_\tau$ (c.f.~\eref{vusincbound}). 
The small difference between the result in~\eref{vusincbound} and the corresponding one in the global fit is due to correlation with the non-strange bounds. 
The bounds in \eref{tauRecap} take into account the correlations between inclusive non-strange constraints, {\it cf.}~\eref{rhoinclusive}, as well as between exclusive and inclusive channels due to $f_{\pi^\pm}$ and the experimental BR of $\tau\to\nu_\tau K$ (which is part of the inclusive strange BR).
The moderate loss in sensitivity for $\eL^{d \tau /e}\!+\!\eR^{d\tau}\!-\!\eR^{de}$ and $\hat \epsilon_{T}^{d\tau}$ as compared to the results in Ref.~\cite{Cirigliano:2018dyk} is a consequence of the change in the inclusive $\epsilon_{T}^{d\tau}$ prefactors, whose larger than expected corrections have opened a nearly flat direction in the subspace spanned by $\eL^{d\tau /e}\!+\!\eR^{d\tau}\!-\!\eR^{de}$ and $\hat \epsilon_{T}^{d\tau}$.

\begin{figure}
\centering
\includegraphics[width=0.5\textwidth]{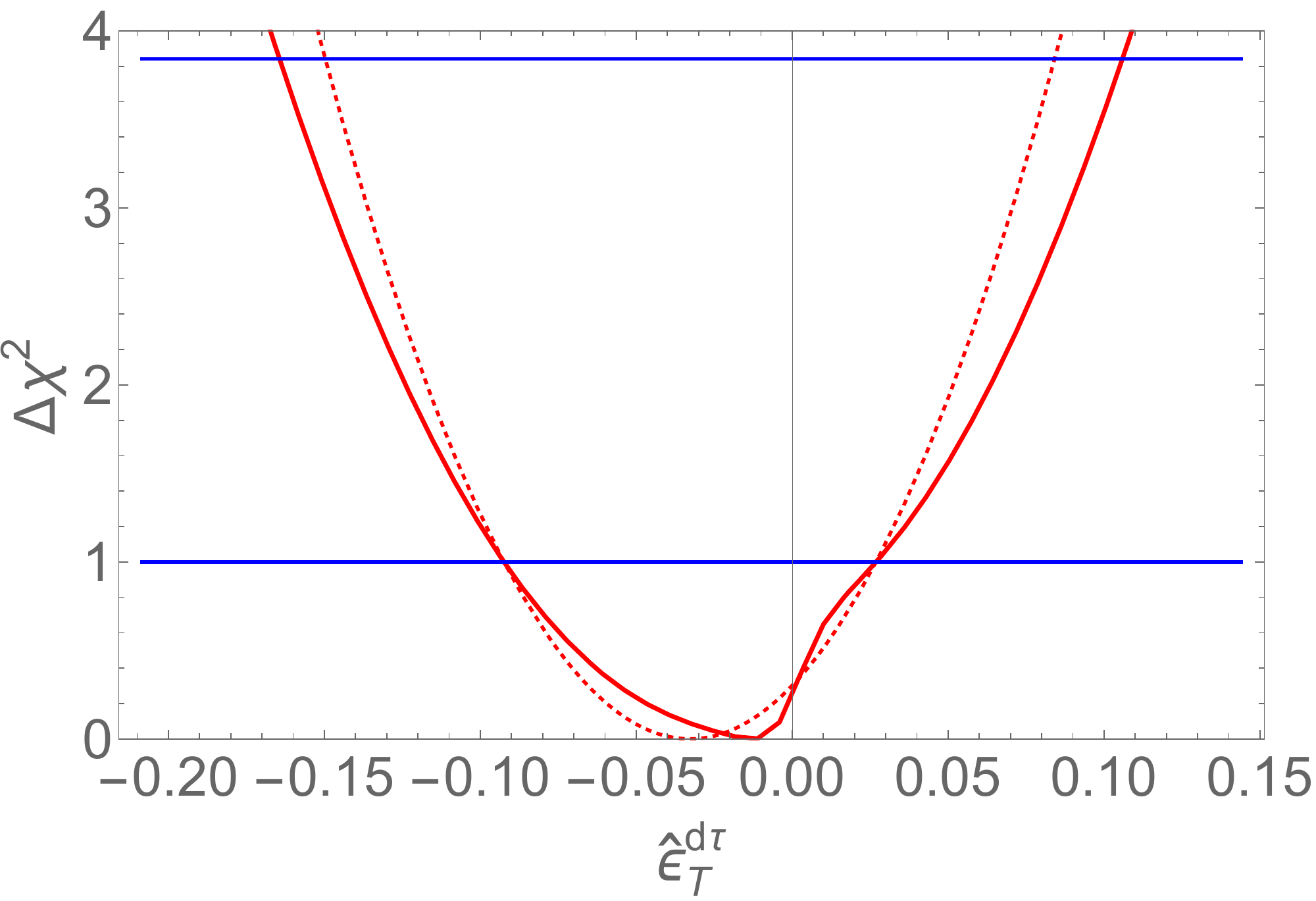}
\caption{The tau likelihood in function of the Wilson coefficient $\hat \epsilon_{T}^{d\tau}$ after marginalizing over the remaining Wilson coefficients. We show the completed non-Gaussian likelihood (solid red), and the Gaussian likelihood  (dotted red)  based on the confidence interval displayed in \eref{tauRecap}.}
\label{fig:EpsilonTdtau}
\end{figure}

In deriving \eref{tauRecap} we  have employed a nuisance parameter to take into account the correlated uncertainty of the numerical factors that multiply $\hat \epsilon_T^{d\tau}$ in the inclusive non-strange constraints, Eqs.~(\ref{eq:inclusive1})-(\ref{eq:inclusive4}).
This introduces some amount of non-Gaussianity into the likelihood.
In particular,  the confidence intervals for $\hat \epsilon_T^{d \tau}$ and $\epsilon_L^{d\tau/e}$ are not symmetric with respect to the minimum.  
For $\hat \epsilon_T^{d \tau}$ we illustrate this issue in \fref{EpsilonTdtau}.
Nevertheless, the likelihood near the minimum is quite well approximated by the Gaussian likelihood obtained from \eref{tauRecap} with the following correlation matrix:
\beq 
\label{eq:correlation}
\rho  = 
\left(
\begin{array}{cccccc}
 1 & 0.87 & -0.18 & -0.98 & -0.03 & -0.45 \\
 \text{} & 1 & -0.59 & -0.86 & 0.06 & -0.59 \\
 \text{} & \text{} & 1 & 0.18 & -0.36 & 0.38 \\
 \text{} & \text{} & \text{} & 1 & 0.04 & 0.49 \\
 \text{} & \text{} & \text{} & \text{} & 1 & 0.16 \\
 \text{} & \text{} & \text{} & \text{} & \text{} & 1 \\
\end{array}
\right).
\eeq

\subsection{Combination with $d(s)\to u\ell\nu_\ell$ transitions}
\label{sec:LEFFE}
Now we combine the hadronic tau bounds in \eref{tauRecap} with those obtained from   $d(s)\to u\ell\nu_\ell$ transitions, $\ell = e,\mu$,  which include nuclear, baryon and meson (semi)leptonic decays. 
The latter were analyzed in a global SMEFT fit in Ref.~\cite{Gonzalez-Alonso:2016etj}, which we update and enlarge in this work. 
The two datasets depend on common quantities, namely the meson decay constants $f_{\pi,K}$ and the CKM matrix elements $V_{ud,us}$. As a result, the combined fit includes by construction ``theoretically clean" ratios where meson decay constants and/or CKM elements cancel, such as e.g.  $\Gamma(\tau\to\pi\nu_\tau)/\Gamma(\pi\to\mu\nu_\mu)$.
Further interest in combining these datasets stems from the fact that in our EFT, assuming it is UV-completed by the SMEFT at $\mu \simeq m_W$, the right-handed currents are independent of lepton flavor: 
\begin{equation}
\epsilon_R^{De}=\epsilon_R^{D\mu}= \epsilon_R^{D\tau} \equiv \epsilon_R^{D}, 
\end{equation}
up to small corrections from dimension-8 operators~\cite{Bernard:2006gy,Cirigliano:2009wk}. 
We will assume this SMEFT relation in our analysis from now on. 
The consequence is that $\tau$ decays and $d(s)\to u\ell\nu_\ell$ transitions probe the same EFT parameters $\epsilon_R^{d}$ and $\epsilon_R^{s}$, 
which leads to an important synergy. 

We now describe the input observables used in the combined analysis, in addition to hadronic tau decays.
First, 
we include in the $d(s)\to u\ell\nu_\ell$ analysis the results of Ref.~\cite{Falkowski:2020pma}, where a long list of nuclear and neutron beta decay observables were studied.
In the present analysis, an older measurement of the $\beta$-$\nu$ correlation of the neutron~\cite{Darius:2017arh} by the aCORN collaboration is superseded by the new result $\tilde a_n = -0.1078(18)$~\cite{Hassan:2020hrj}. Moreover,  the latest UCN$\tau$ measurement of the neutron lifetime~\cite{UCNt:2021pcg} leads to the improved combined result $\tau_n = 878.64(59)$, where we include both bottle and beam measurements and average the errors {\it \`a la} PDG with the scale factor $S=2.2$.
Finally, we update the axial coupling of the nucleon with the latest $N_f = 2+1+1$ FLAG value $g_A = 1.246(28)$~\cite{Aoki:2021kgd,Gupta:2018qil,Chang:2018uxx,Walker-Loud:2019cif} and use Ref.~\cite{Gorchtein:2021fce} for the associated radiative corrections.  
The nuclear beta decay  data provide stringent  constraints on 
$\hat V_{ud}$, $\epsilon_R^{d}$, $\epsilon_S^{de}$, and $\hat \epsilon_T^{de}$.

We combine this beta-decay likelihood with leptonic and semileptonic pion decay data, which allows us to also constrain the pseudoscalar Wilson coefficient $\epsilon_P^{de}$ and one linear combination of the muonic Wilson coefficients  $\epsilon_X^{d\mu}$. 
The pion input is described in  Ref.~\cite{Gonzalez-Alonso:2016etj}. 
Here we  update the constraint on the tensor Wilson coefficient obtained from radiative pion decay $\pi^{-}\rightarrow e^{-}\bar{\nu}_{e}\gamma$, finding 
$\hat{\epsilon}_T^{de} =(0.5 \pm 2.4)\times 10^{-3}$. 
This result is obtained using a more precise and solid determination of the associated form factor, namely $f_T=0.232(14)$, which is based on the recent lattice determination of the magnetic susceptibility of the vacuum~\cite{Bali:2020bcn} (see~\aref{fT} for details). 
Furthermore, we also include in our analysis the pion beta decay $\pi^\pm \to \pi^0 e^\pm \nu_e$, 
although at present it has negligible impact on the fit.\footnote{We note that including in the fit the pion beta decay rate normalised by any of the $K_{\ell3}$ rates (as advocated in Ref.~\cite{Czarnecki:2019iwz}) is equivalent to simply including the pion beta decay rate, as we do in this work.
}  
Finally, in this analysis we use the $f_{\pi^\pm}$ lattice input discussed in~\sref{TauToPnu}.

The nuclear and pion data together lead to the constraints
\begin{equation}
\begin{pmatrix}
  \hat V_{ud} \\ \epsilon_R^{d} \\  \epsilon_S^{de} \\  \hat \epsilon_T^{de} \\  \epsilon_P^{de} \\  \epsilon_{LP}^{d\mu} 
\end{pmatrix}
= \begin{pmatrix}
0.97386(40)  \\ -0.012(12) \\ 0.00032(99) \\ -0.0004(11) \\ 3.9(4.3) \times 10^{-6} \\ -0.021(24)
 \end{pmatrix}, 
\qquad \rho = 
\left(
\begin{array}{cccccc}
 1 & 0.01 & 0.75 & 0.64 & 0.01 & -0.01 \\
 \text{} & 1 & 0.01 & 0. & -0.96 & 0.96 \\
 \text{} & \text{} & 1 & 0.6 & 0.01 & -0.01 \\
 \text{} & \text{} & \text{} & 1 & 0.01 & -0.01 \\
 \text{} & \text{} & \text{} & \text{} & 1 & -0.999 \\
 \text{} & \text{} & \text{} & \text{} & \text{} & 1 \\
\end{array}
\right), 
\label{eq:fitNuclearPion}
\end{equation} 
where $\epsilon_{LP}^{d\mu} \equiv   \epsilon_{L}^{d\mu/e} -    \epsilon^{d \mu}_P    { m_{\pi^\pm}^2  \over   m_\mu  (m_u + m_d) }$. 
Let us note that the above bound on $\epsilon^{d}_R$ has similar uncertainty as the corresponding tau bound in \eref{tauRecap}.

Finally, we discuss the constraints from leptonic and semileptonic kaon decays and hyperon beta decays. 
Compared to Ref.~\cite{Gonzalez-Alonso:2016etj}, we update the experimental input on semileptonic kaon decays following the recent re-analysis of Ref.~\cite{Seng:2021nar}. 
More precisely, we use the constraints on $V_{us} f_+(0)$ listed in Table~1 of that reference, however we re-interpret them as constraints on the EFT parameters in \eref{leff1} (see Ref.~\cite{Gonzalez-Alonso:2016etj} for details). 
Concerning the theory input, we use $f_+(0)=0.9698(17)$~\cite{Aoki:2021kgd,FermilabLattice:2018zqv,Carrasco:2016kpy}, while the kaon decay constant is determined from $f_{\pi^\pm}$ and $f_{K^\pm}/f_{\pi^\pm}$, as discussed in~\sref{TauToPnu}.
We obtain the following constraints from $s\to\bar{u}\ell\nu_\ell$ transitions 
\begin{equation}
\label{eq:fitKaon}
\bvec 
\hat V_{us} \\  \epsilon_L^{s \mu/e} \\   \epsilon_R^{s} \\ \epsilon_S^{s \mu} \\ 
   \epsilon_P^{s e} \\   \epsilon_P^{s \mu} \\   \hat \epsilon_T^{s \mu} \evec
    =  
    \bvec
    0.22306(56)  \\ 0.0008(22) \\  0.001(50) \\  -0.00026(44) \\ 
    -0.3(2.0) \times 10^{-5} \\  -0.0006(41) \\ 0.002(22) 
    \evec ,
    \quad 
    \rho = \left(
\begin{array}{ccccccc}
 1 & -0.11 & 0. & -0.12 & 0.03 & 0.02 & 0. \\
 \text{} & 1 & 0. & 0. & 0. & 0.02 & 0.55 \\
 \text{} & \text{} & 1 & 0. & -0.997 & -0.997 & 0. \\
 \text{} & \text{} & \text{} & 1 & -0.01 & -0.01 & 0. \\
 \text{} & \text{} & \text{} & \text{} & 1 & 0.9996 & 0. \\
 \text{} & \text{} & \text{} & \text{} & \text{} & 1 & 0.01 \\
 \text{} & \text{} & \text{} & \text{} & \text{} & \text{} & 1 \\
\end{array}
\right). 
\end{equation}

We are ready to combine the constraints from hadronic tau decays (\eref{tauRecap}), nuclear beta and pion decays (\eref{fitNuclearPion}), and kaon and hyperon decays (\eref{fitKaon}).\footnote{In fact, one of the inputs in \eref{fitNuclearPion}, namely $\Gamma(\pi \to \mu \nu_\mu)$, is replaced in the global combination by the ratio ${\Gamma(K \to \mu\nu_\mu) \over \Gamma(\pi \to \mu \nu_\mu)}$. 
The motivation is that the theoretical error on the radiative correction to the ratio~\cite{Cirigliano:2011tm} is a tad smaller than the analogous error on the individual widths. }
Our constraints are marginalized over theoretical uncertainties related to the lattice determination of the meson decay constants and calculation of the radiative corrections. 
Let us stress that we keep full track of the cross-correlations between tau and $d(s)\to u\ell\nu_\ell$ bounds due to the common CKM and meson decay constant inputs. While the polluted CKM elements $\hat V_{ud}$ and $\hat V_{us}$ are independent variables in the EFT framework,
for the sake of the presentation it is convenient to introduce a different parametrization of this subspace. 
Indeed, these two objects are related as 
\beq
\hat  V_{ud} = 
\sqrt{1-\hat V_{us}^2} \bigg [ 1  + \epsilon_L^{d s e}  + \epsilon_R^{d} 
+  {\hat V_{us}^2 \over 1- \hat V_{us}^2}   \epsilon_R^{s}  \bigg ] ,  
\eeq 
where $\epsilon_L^{d s e} \equiv \epsilon_L^{d e} + {\hat V_{us}^2 \over 1- \hat V_{us}^2}   \epsilon_L^{s e}$. 
We will use  $\epsilon_L^{d s e}$ instead of  $\hat  V_{ud}$ as a variable in our combined fit. 
In the $\overline{MS}$ scheme at $\mu=2$ GeV we obtain the following 68\% CL intervals:
\begin{equation}
\label{eq:fitGlobal}
\left(
\begin{array}{c}
\hat V_{us} \equiv  V_{us} \big (1 + \epsilon_L^{s e}  + \epsilon_R^{s}  \big ) \\
\epsilon_L^{dse}  \equiv   \epsilon_L^{d e} + {\hat V_{us}^2 \over 1- \hat V_{us}^2}   \epsilon_L^{s e} \\
\epsilon_R^d\\
\epsilon_S^{de} \\ 
\epsilon_P^{de} \\ 
\hat \epsilon_T^{de} \\ 
\epsilon_L^{s\mu/e} \\
\epsilon_R^s \\
\epsilon_P^{se} \\
\epsilon_{L}^{d\mu/e} -    \epsilon^{d \mu}_P    { m_{\pi^\pm}^2  \over   m_\mu  (m_u + m_d) }  \\
\epsilon_S^{s\mu} \\
\epsilon_P^{s\mu} \\
\hat \epsilon_T^{s\mu} \\
\epsilon_L^{d\tau/e} \\
\epsilon_P^{d\tau} \\
\hat \epsilon_T^{d\tau} \\
 \epsilon_{L}^{s\tau/e} -    \epsilon^{s \tau}_P    { m_{K^\pm}^2  \over   m_\tau  (m_u + m_s) }  \\
\epsilon_{L}^{s\tau/e}  + 0.08(1)  \epsilon_{S}^{s \tau} - 0.38\epsilon_{P}^{s\tau}  +0.40(13)\hat \epsilon_{T}^{s\tau}  \\
\end{array}
\right)
= 
\left(
\begin{array}{c}
\vphantom{\hat V_{us} \equiv  V_{us} \big (1 + \epsilon_L^{s e}  + \epsilon_R^{s}  \big )}	 
0.22306(56) \\
\vphantom{\epsilon_L^{dse}  \equiv   \epsilon_L^{d e} + {\hat V_{us}^2 \over 1- \hat V_{us}^2} }	2.2(8.6) \\
\vphantom{\epsilon_R^{d} }  -3.3(8.2) \\
\vphantom{\epsilon_S^{de}}	 3.0(9.9) \\
\vphantom{\epsilon_P^{de}}	1.3(3.4)  \\
\vphantom{\hat \epsilon_T^{de}}	 -0.4(1.1) \\
\vphantom{\epsilon_L^{s\mu/e}}	 0.8(2.2) \\
\vphantom{\epsilon_R^{s}}	 0.2(5.0) \\
\vphantom{\epsilon_P^{se}}	 -0.3(2.0) \\
\vphantom{\epsilon_{L}^{d\mu/e} -    \epsilon^{d \mu}_P    { m_{\pi^\pm}^2  \over   m_\mu  (m_u + m_d) }  }	    -0.5(1.8) \\
\vphantom{\epsilon_S^{s\mu}}	 -2.6(4.4) \\
\vphantom{\epsilon_P^{s\mu}}	 -0.6(4.1) \\
\vphantom{\hat \epsilon_T^{s\mu}}	 0.2(2.2) \\
\vphantom{\epsilon_L^{d \tau/e}}	 0.1(1.9) \\
\vphantom{\epsilon_P^{d\tau}}	 9.2(8.6) \\
\vphantom{\hat \epsilon_T^{d\tau}}	 1.9(4.5) \\
\vphantom{\epsilon^{s \tau}_P    { m_{K^\pm}^2  \over   m_\tau  (m_u + m_s) }  }	 0.0(1.0) \\
\vphantom{\epsilon_{LSPT}^{s\tau}}	 -0.7(5.2) \\
\end{array} 
\right)
\times 10^{\wedge}\left(
\begin{array}{c}
\vphantom{\hat V_{us} \equiv  V_{us} \big (1 + \epsilon_L^{s e}  + \epsilon_R^{s}  \big )}	 0 \\  
\vphantom{\epsilon_L^{dse}  \equiv   \epsilon_L^{d e} + {\hat V_{us}^2 \over 1- \hat V_{us}^2} } -3 \\
\vphantom{\epsilon_X^q} -3 \\  
\vphantom{\epsilon_X^q} -4 \\
\vphantom{\epsilon_P^{de}} -6 \\
\vphantom{\hat \epsilon_T^{de}} -3 \\
\vphantom{\epsilon_X^q} -3 \\
\vphantom{\epsilon_X^q} -2 \\
\vphantom{\epsilon_P^{se} }	  -5 \\
\vphantom{\epsilon_{L}^{d\mu/e} -\epsilon^{d \mu}_P    { m_{\pi^\pm}^2  \over   m_\mu  (m_u + m_d) } } -2 \\
\vphantom{\epsilon_X^q} -4 \\
 \vphantom{\epsilon_P^{s\mu} }     -3 \\
\vphantom{\hat \epsilon_T^{s\mu} } -2 \\
\vphantom{\epsilon_X^q} -2 \\
\vphantom{\epsilon_X^q} -3 \\
\vphantom{\epsilon_X^q} -2 \\
 \vphantom{\epsilon^{s \tau}_P    { m_{K^\pm}^2  \over   m_\tau  (m_u + m_s) }  }-1 \\
 -2 \\
\end{array}
\right),
\end{equation} 
where we recall the definition
$\epsilon_{L}^{D\ell/e}   \equiv    \epsilon_{L}^{D\ell} -  \epsilon_{L}^{D e}$. 
The correlation matrix (in the Gaussian approximation) associated with these constraints in  \eref{fitGlobal} is presented in \eref{fitCorrelationMatrix}. 
Inclusion of new physics parameters $\epsilon_X^{q \ell}$ greatly improves the quality of the fit.
We find $\chi^2_{\rm SM} - \chi^2_{\rm min} = 37.4$, 
where $\chi^2_{\rm min}$ is the value of the likelihood at the global minimum, and $\chi^2_{\rm SM}$ is the minimum on the hypersurface where all  $\epsilon_X^{q \ell}$ set to zero. 
This corresponds to $3\sigma$ preference for new physics, or $0.3\%$ p-value for the SM hypothesis.
However, a preference for particular $\epsilon_X^{q \ell}$  is not visible in \eref{fitGlobal} due to strong correlations. 
We will discuss preferred directions below, in the context of more constrained scenarios. 

\eref{fitGlobal} contains the most complete information to date about the charged-current interactions between the light quarks and leptons. 
In many cases, the real power of the constraints  is obscured by large correlations. 
As an example, the marginalized constraints on $\epsilon_L^{dse}$ and $\epsilon_R^{d}$ are both at a percent level, however the combination 
\beq
\delta_{\rm CKM} \equiv   \epsilon_L^{d s e}  + \epsilon_R^d 
+  {\hat V_{us}^2 \over 1- \hat V_{us}^2}   \epsilon_R^s 
\eeq
is much more stringently bound: 
$\delta_{\rm CKM} = -9.8(4.3) \times 10^{-4}$. 
We stress that \eref{fitGlobal} together with  \eref{fitCorrelationMatrix} contain the information allowing one to disentangle these correlations in the Gaussian approximation.\footnote{The full non-Gaussian likelihood function is available on request.} 
The preference for a non-zero value of the combination $\delta_{\rm CKM}$ is one of the facets of the Cabibbo anomaly, often called the (first row) CKM unitarity problem in the literature.

In addition to \eref{fitGlobal}, there are a few bounds on Wilson coefficients that can be obtained from their quadratic effect to certain observables.  
These are inherently non-Gaussian and uncorrelated with \eref{fitGlobal}. 
In~\sref{pieta}, we obtained the following bound using the $\tau\to\nu_\tau\pi\eta$ channel:
\begin{equation}
\epsilon^{d \tau}_S = -0.06(16) .      
\end{equation}
Furthermore, based on the differential distributions measured in $K^- \to \pi^0 e^- \nu_e$ decays~\cite{Yushchenko:2004zs}, 
Ref.~\cite{Gonzalez-Alonso:2016etj} obtained\footnote{We do not include the recent bounds on scalar and tensor interactions obtained by the OKA Collaboration from the $K_{e3}$ differential distributions~\cite{Yushchenko:2017fzv} since they are presented as preliminary.} 
\beq
\epsilon_S^{se} = -1.6(3.2) \times 10^{-3}, 
\qquad 
\hat \epsilon_T^{se} = 0.035(70). 
\eeq

\subsection{SM limit}

As a first application of the combined likelihood of \eref{fitGlobal}, we consider the SM limit, where all Wilson coefficients $\epsilon_X^{D \ell}$ are set to zero. 
There is only one independent free parameter remaining in \eref{leff1}, which we choose to be $V_{us}$. 
The other CKM element in \eref{leff1} is tied to $V_{us}$ by the unitarity relation 
$V_{ud}= \sqrt{1- V_{us}^2 - |V_{ub}|^2}$, where we use the PDG average $|V_{ub}| = 3.82(24) \times 10^{-3}$ (the precise value of $V_{ub}$ has a tiny effect on the fit).  At face value we find the constraint on the (sine of the) Cabibbo angle $V_{us}$ reads
\begin{equation}
\label{eq:VusSMunscaled} 
V_{us} = 0.22450(34). 
\end{equation}
However, as can be seen in \fref{VusSM}, this result is obtained by combining several measurements that are in strong tension with each other.
This tension is referred to as the Cabibbo anomaly. 
Note that tau decays, especially the inclusive one of Eq.~(\ref{eq:vusincbound}), further aggravate the tension (see however footnote~\ref{footnote:inclusiveVus}).

The Cabibbo anomaly can be interpreted as a hint of new physics.  
In this subsection, however, we work within the SM paradigm, and from this point of view the anomaly is simply an inconsistency between different datasets.   
Therefore, the error \eref{VusSMunscaled} does not reflect the real uncertainty on the true value of the SM Cabibbo angle, given the confusing experimental situation.  
In such a case, it is more practical to follow the PDG procedure of (artificially) inflating the errors, so as to make the different measurements compatible. 
To this end, we construct a simplified likelihood which takes into account only the most sensitive probes of the Cabibbo angle.
It includes the observables displayed in \fref{VusSM} treated as functions of $V_{us}$, $f_{\pi^\pm}$, $f_{K^\pm}/f_{\pi^\pm}$, $f_+(0)$,  and the relevant radiative corrections.
Moreover, it includes the lattice and theory constraints on the decay constants, form factor, and radiative corrections. 
We democratically inflate all the errors by the factor $S$ 
until $\chi^{2}_{\rm min}$/d.o.f is equal to one. 
Following this procedure we obtain 
\begin{equation}
\label{eq:VusSMscaled} 
V_{us} = 0.22450(67), \qquad S=2.0 \, , 
\end{equation}
from which $V_{ud} = 0.97447(15)$ follows. 
It is \eref{VusSMscaled} rather than \eref{VusSMunscaled} that better reflects the current knowledge concerning the value of Cabibbo angle, assuming the SM provides an adequate approximation for the fundamental interactions at the weak scale.   
 
\begin{figure}
    \centering
    \includegraphics[width=0.8\textwidth]{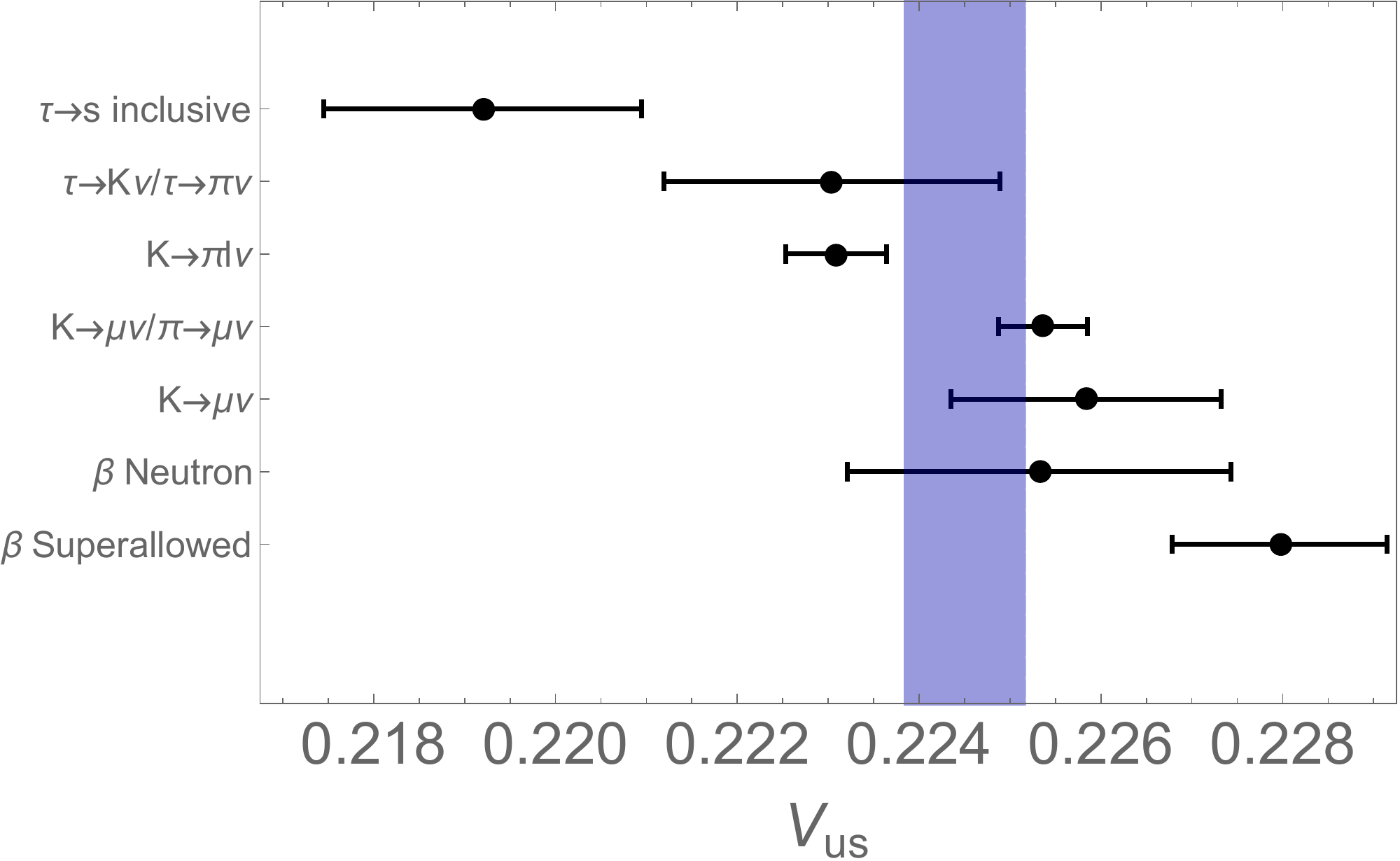}
    \caption{68\%~CL constraints on the Cabibbo angle assuming the SM is the UV completion of our EFT, which implies $\epsilon_X^{D \ell} =0$. 
    We show the separate constraints from the most precise measurement probing this parameter:  
    inclusive $\tau \to s$ decays, 2-body tau decays,  semileptonic kaon decay, the ratio $\Gamma(\pi \to \mu \nu_\mu)/\Gamma(K \to \mu \nu_\mu)$, $\Gamma(K \to \mu \nu_\mu)$, neutron beta decay,  and the superallowed $0^+ \to 0^+$ nuclear beta decays. 
    The purple band corresponds to a combination of these measurements with errors inflated \`{a} la PDG with the scale factor of $S=2.0$, so as to account for the large tension between the individual inputs.}
    \label{fig:VusSM}
\end{figure}
The results of the global fit in \eref{fitGlobal} as well as the Cabibbo angle fit in \eref{VusSMscaled} are marginalized over the uncertainties of the meson decay constants. The same likelihoods set also confidence intervals for the latter. 
In the global case these confidence intervals are not particularly revealing, because they are set by the lattice central values and errors. 
The situation changes in  the SM limit. 
Due to the limited number of free parameters, 
the meson decay constants are themselves constrained by the experimental data. We find
\begin{equation}
\label{eq:FFSMscaled}
\bvec
f_{\pi^\pm}[{\rm MeV}] \\  f_{K^\pm}/f_{\pi^\pm} \\ f_+(0)  
\evec    
= \bvec 130.54(34) \\ 1.1958(35) \\ 0.9668(28) \evec ~,\qquad S=2.0   \, . 
\end{equation}
As indicated,  both Eqs.~\eqref{eq:VusSMscaled}-\eqref{eq:FFSMscaled} come from the same fit, where we have applied the same scale factor $S=2.0$.
In spite of inflating the errors, the uncertainty on $f_{\pi^\pm}$ is reduced by more than a factor of two compared to the (face value) lattice result.  
Hadronic tau decays have a significant impact on reducing the error. 
We stress  that this more stringent constraint can only be used in the SM context, and is not valid in the presence of new physics. 
For $f_{K^\pm}/f_{\pi^\pm}$ and $f_+(0)$ the errors are actually larger compared to the (face value) lattice results, reflecting the inflated uncertainty due to the tensions in the global fit. 

\subsection{Simple new physics scenarios and perspective on Cabibbo anomaly}
 
 We move to studying the likelihood of \eref{fitGlobal} in  simplified new physics scenarios. First, we will assume that only a single Wilson coefficient $\epsilon_X^{D\ell}$ in \eref{leff1} is present at a time. 
 This exercise will allow us to identify simple directions in the parameter space where the goodness of the fit can be significantly improved compared to the SM limit. 

\begin{table}[tb]\begin{center}\renewcommand{\arraystretch}{1.2}
\begin{tabular}{|c||c|c|c|c|c|c|} \hline
& $\epsilon_X^{de}\; \times\;  10^{3}$ & $\epsilon_X^{se} \; \times\;  10^{3}$
& $\epsilon_X^{d\mu}\; \times\;  10^{3}$ & $\epsilon_X^{s\mu} \; \times\;  10^{3}$
& $\epsilon_X^{d\tau}\; \times\;  10^{3}$ & $\epsilon_X^{s\tau} \; \times\;  10^{3}$
\\ \hline \hline
$L$ &{\color{red}-0.79(25)}&-0.6(1.2)&0.40(87)&0.5(1.2)&5.0(2.5)&-18.2(6.2)
\\ \hline 
$R$ 
&-0.62(25)&{\color{red}-5.2(1.7)}&-0.62(25)&{\color{red}-5.2(1.7)}&-0.62(25)&{\color{red} -5.2(1.7)}
\\ \hline 
$S$ &1.40(65) & -1.6(3.2)  & x &  -0.51(43) & -6(16) & -270(100)
\\ \hline
$P$  &0.00018(17) & -0.00044(36) &-0.015(32) & -0.032(64) & 1.7(2.5) &  10.4(5.5)
\\ \hline
$\hat T$& 0.29(82)&  0.035(70)  &x  &  2(18) & 28(10) & -55(27) \\ \hline
\end{tabular}\end{center}
\caption{
\label{tab:oneeps} 
Constraints on the Wilson coefficients $\epsilon^{D\ell}_X$ in units of $10^{-3}$, fitting one parameter at a time. 
We highlighted in red color the entries where $3\sigma$ or larger preference for new physics is displayed. 
The cross signifies that this particular Wilson coefficient is not constrained by our analysis. 
Let us note again that we assume $\epsilon_R^{De}=\epsilon_R^{D\mu}=\epsilon_R^{D\tau}$, as predicted by the SMEFT at dimension six. 
}
\end{table}

The results are shown in \tref{oneeps}. 
First thing to see is that our likelihood constrains almost the complete set of $\epsilon_X^{d \ell}$ and $\epsilon_X^{s \ell}$ Wilson coefficients. 
The typical accuracy is percent to per mille level. 
The notable exception are $\epsilon_P^{D e}$ and $\epsilon_P^{D \mu}$ where much larger accuracy is due to the chiral enhancement of pseudoscalar interaction. 
Note that the constraints in \tref{oneeps} are often an order of magnitude better than in \eref{fitGlobal}, as in the latter case the true power of the constraints is obscured by large correlations. 
 
Furthermore, the fit shows a preference for non-zero values of several Wilson coefficients. 
This is a flip side of the Cabibbo anomaly discussed in the previous subsection.
The preference is strongest for $\epsilon_L^{de}$, $\epsilon_R^{s}$, and  $\epsilon_L^{s\tau}$, in which case a single new physics Wilson coefficients allows one to improve the fit by $\sim 9$ units of $\chi^2$.  
The reason is that these parameters alter the relation between the magnitude of the Cabibbo angle and various observables, allowing one to partially reconcile the seemingly inconsistent measurements in \fref{VusSM}. 
For example, a negative $\epsilon_L^{de}$ leads to a decrease in the pion, neutron and nuclear $\beta$ decay widths.
Consequently, $V_{ud}$ extracted from these measurements (under SM assumptions) appears smaller than the prediction of the unitarity relation (based on $V_{us}$ extracted from kaon, which is not affected by $\epsilon_L^{de}$). 

This simple analysis points to the range of possibilities for model building addressing the Cabibbo anomaly. 
$\epsilon_L^{de}$ can be generated e.g. by a vanilla charged gauge boson (W') with SM-like couplings to fermions. 
A more exotic flavor structure is needed to generate a required  $\epsilon_L^{s\tau}$, as one needs a $W'$ that is coupled much stronger to $ \bar u_L s_L$ than to $\bar u_L d_L$, and more strongly coupled to tau leptons than to electrons and muons.
On the other hand, $\epsilon_R^s$ can be generated with a $W'$ coupled to  $ \bar u_R s_R$ (and mixing with $W$ after electroweak symmetry breaking), but again the coupling to  $\bar u_R d_R$ has to be much smaller. 
We also note that there exists some preference for scalar, pseudoscalar, and tensor $\epsilon_X^{s\tau}$, opening an opportunity for leptoquark models coupled to 3rd generation leptons. 

Some of the new physics preferences displayed in \tref{oneeps} are highly correlated. 
For example, only one linear combination of  $\epsilon_L^{de}$ and $\epsilon_R^{d}$ is favored to be non-zero, while allowing these two parameters to vary independently does not improve the fit dramatically (by $2.6$ units of $\chi^2$ compared to the case with only $\epsilon_L^{de}$). 
On the other hand, some of the displayed tensions are largely independent. 
As pointed out in \cite{Grossman:2019bzp}, a scenario with new physics coupled to right-handed quarks and generating both $\epsilon_R^d$ and  $\epsilon_R^s$ is strongly favored by the data. 
We find the best fit at 
$\epsilon_R^d = -7.1(2.6) \times 10^{-4}$ and $\epsilon_R^s = -5.7(1.7) \times 10^{-3}$ with $\chi^2_{\rm SM} - \chi^2_{\rm min} = 17.5$ - formally a $3.8\sigma$ preference for this scenario with respect to the SM hypothesis. 
Another 2-parameter scenario with almost identical level of preference is the one with SM-like new physics characterized by the $\epsilon_L^{de}$, $\epsilon_L^{s\tau}$ pair. 
In this case we find the best fit at 
$\epsilon_L^{de} = -7.5(2.5) \times 10^{-4}$ and 
$\epsilon_L^{s\tau} = -1.72(62) \times 10^{-2}$, with  
$\chi^2_{\rm SM} - \chi^2_{\rm min} = 17.5$.  
The left panel of \fref{twobyone} clarifies where this preference comes from.
The presence of $\epsilon_L^{de}$ puts the Cabibbo angle measured in nuclear beta decays in good agreement with the one measured in kaon decays, while $\epsilon_L^{s\tau}$ achieves a similar feat with the Cabibbo angle measured in hadronic tau decays.
Some tension remains in the two-parameter scenarios,  notably between the inclusive tau and other determinations in the first case, and between the semileptonic and leptonic kaon decays in the second case. 

The tension can be completely eradicated in multi-parameter scenarios.
The right panel of \fref{twobyone} shows the Cabibbo angle in an example with 3 parameters: 
$\epsilon_R^d$, $\epsilon_R^s$, and $\epsilon_L^{s\tau}$.
In this case the best fit is
$\epsilon_R^d = -6.8(2.6) \times 10^{-4}$, 
$\epsilon_R^s = -5.9(1.7) \times 10^{-3}$, and 
$\epsilon_L^{s\tau} = -1.81(62) \times 10^{-2}$.
At the minimum of the likelihood the different datasets now point to perfectly compatible values of the Cabibbo angle, 
for which the best fit is $V_{us}= 0.22432(36)$. 
Thanks to removing the tension, the 3-parameter scenario improves the goodness of fit by $\chi^2_{\rm SM} - \chi^2_{\rm min} = 26.1$ -  
a whopping $4.4 \sigma$ preference compared to the SM hypothesis.

Finally, another interesting scenario is the one in which all Wilson Coefficients are zero except a universal left-handed one, as this is the situation generated in the SMEFT with $U(3)^5$ flavor symmetry. In this case, which was studied in Ref.~\cite{Cirigliano:2009wk}, all channels receive the same universal global correction that is hidden in the ``BSM polluted" $\hat{V}_{ud}$ and $\hat{V}_{us}$ elements. As a result the only observable consequence is an apparent violation of unitarity. In our notation, this means that the only non-zero coefficient in~\eref{fitGlobal} is $\epsilon_L^{dse}=-(8.0\pm2.7)\times 10^{-4}$.

\begin{figure}
\centering
\includegraphics[width=0.48\textwidth]{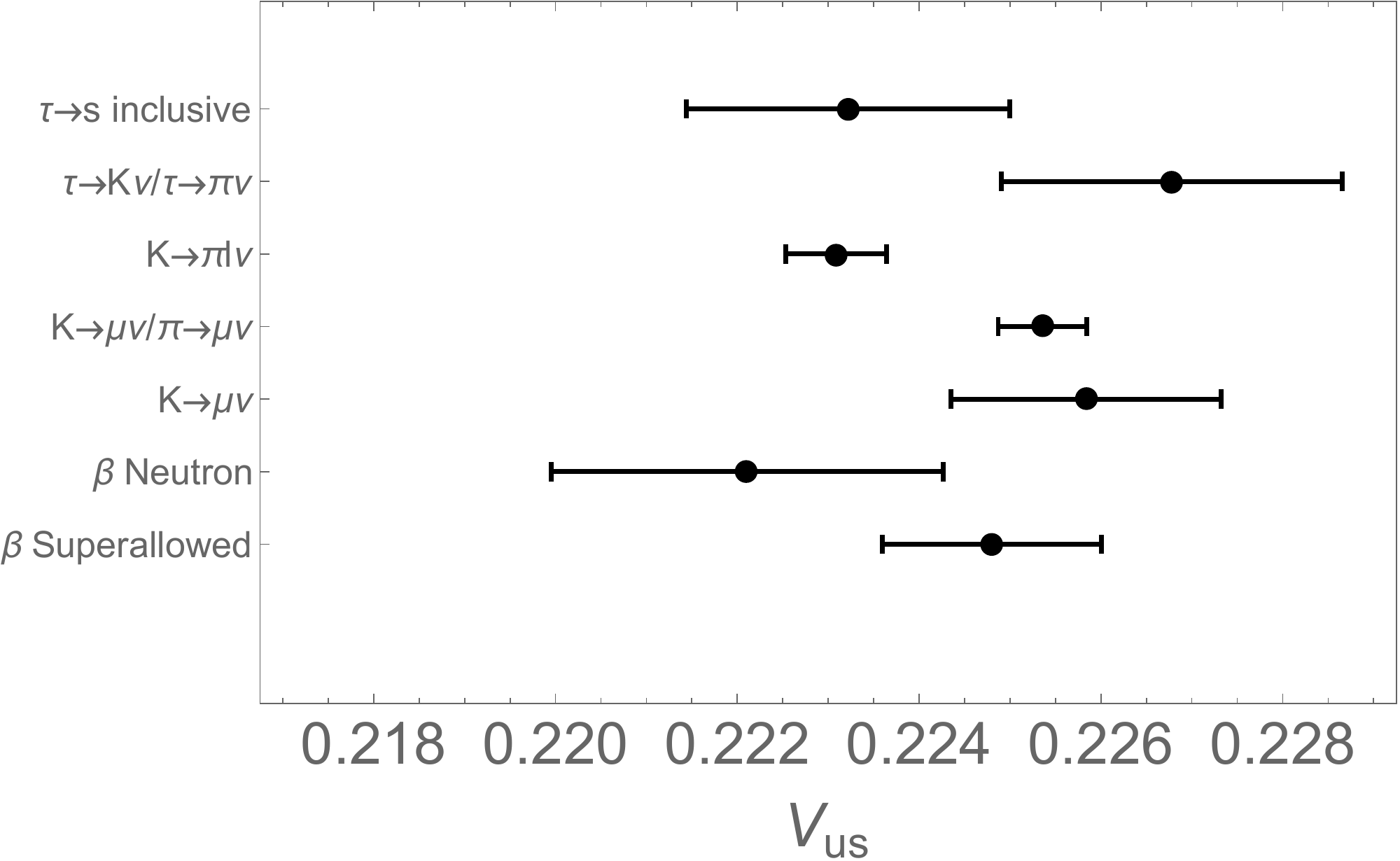}
\quad 
\includegraphics[width=0.48\textwidth]{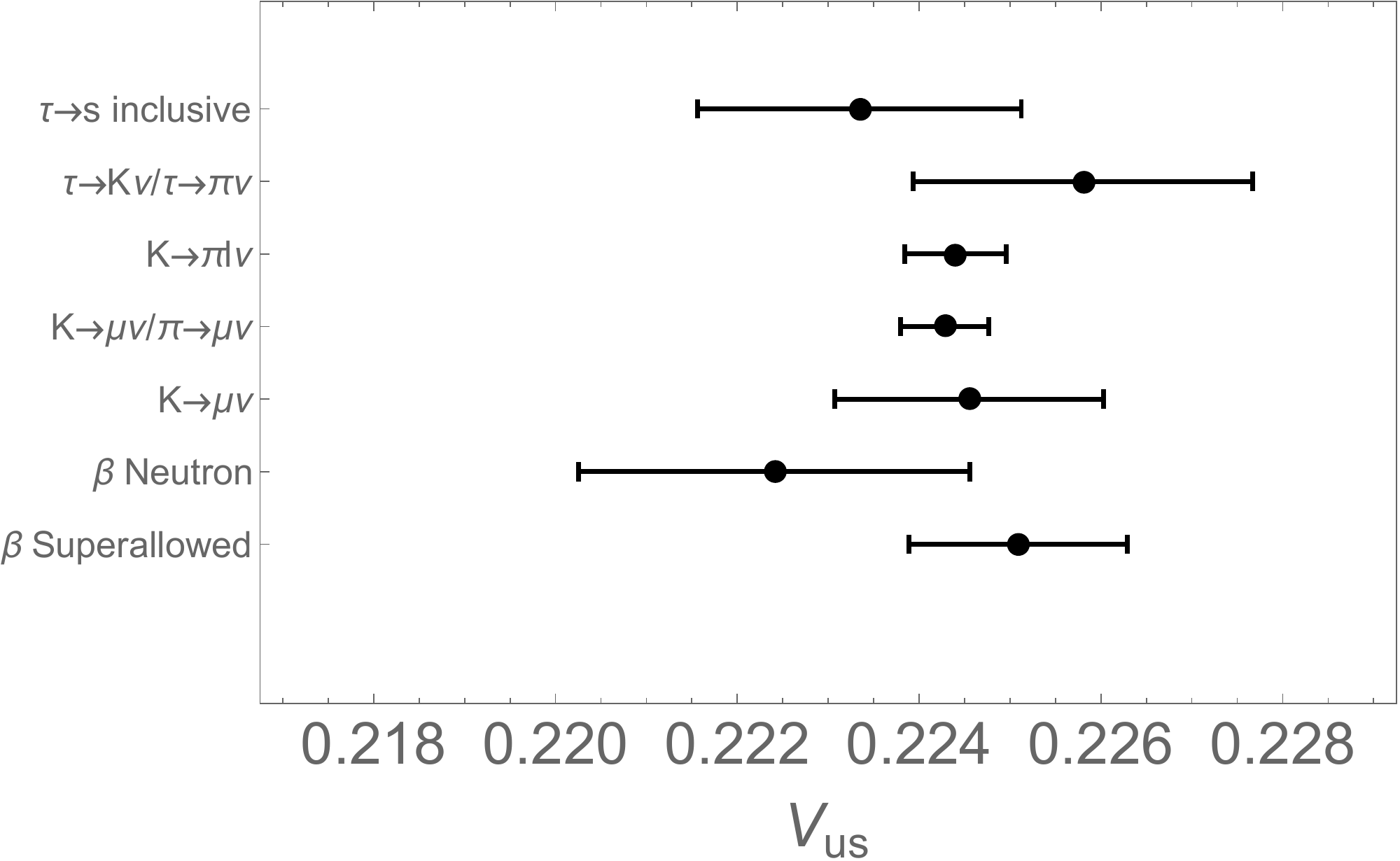}
\caption{
The Cabibbo angle beyond the SM. 
Black error bars show the determination of $V_{us}$ using different subsets of experimental data, see the caption of \fref{VusSM} for details. 
{\em Left:} Determination of $V_{us}$ in the presence of new physics characterized by the Wilson coefficients $\epsilon_L^{de} = -7.5 \times 10^{-4}$ and $\epsilon_L^{s\tau} = -1.7\times 10^{-2}$, 
with the remaining  $\epsilon_X^{D\ell}$ set to zero. 
Such a configuration partly improves compatibility between different datasets, 
removing the largest tensions present in the SM fit. 
However, some tensions remain, notably between semileptonic and leptonic kaon decays.  
{\em Right:} 
The same in the presence of three new physics Wilson coefficients: 
$\epsilon_R^{d} = -6.8 \times 10^{-4}$, $\epsilon_R^{s} = -5.9 \times 10^{-3}$,  and $\epsilon_L^{s\tau} = -1.8\times 10^{-2}$.
In a relatively simple scenario where these 3 parameters are generated by new physics,  all the datasets point to perfectly compatible values of the Cabibbo angle, with the combined value $V_{us} = 0.22432(36)$.
}
    \label{fig:twobyone}
\end{figure}

\section{Conclusions and outlook}
\label{sec:conclusions}
\setcounter{equation}{0}

In this paper we studied hadronic tau decays in the framework of an EFT for light SM degrees of freedom.
This EFT describes the low-energy dynamics of the SM, as well as subleading effects of hypothetical non-SM particles with masses larger than $2$~GeV. 
Focusing on the charged-current interactions between light quarks and leptons, the leading non-standard effects are parametrized by a set of Wilson coefficients $\epsilon_X^{q \ell}$, cf. \eref{leff1}. 
The main new result of this paper is \eref{tauRecap} summarizing the constraints on $\epsilon_X^{q \ell}$ from a large set of hadronic tau observables, 
which include the 2-body $\tau \to \pi(K)\nu_\tau$, 3-body $\tau \to \pi \pi \nu_\tau$, and inclusive $\tau \to \nu_\tau \bar{u}d(s)$ decays. 
There we quote percent level marginalized constraints on six linear combinations of $\epsilon_X^{D \tau}$, $D = d,s$, and we provide the correlation matrix in \eref{correlation}. These bounds reach the per mille level when only one operator is present.

The 2-body channels are theoretically simple, involving only the non-perturbative meson decay constants $f_{\pi^\pm}$ and $f_{K^\pm}$ and calculable radiative corrections. 
For this reason they have been commonly used in the literature for constraining new physics or the CKM elements. 
On the other hand, the multi-body and inclusive channels are theoretically more challenging,  and the present paper provides the most comprehensive  discussion to date of the resulting constraints on new physics.  
Compared to Ref.~\cite{Cirigliano:2018dyk}, we extend the analysis to include strange decays ($\tau \to K \nu_\tau$, $\tau \to K \pi \nu_\tau$, $\tau \to \bar{u} s \nu_\tau$). 
We also update and improve the analysis of the remaining channels with the most recent theoretical and experimental input, and we provide the details of theoretical calculations that allow us to determine the new physics dependence of hadronic tau observables.  

We expect the constraints from hadronic tau decays to be further improved in the near future.  
On the experimental front, the old LEP measurements of the spectral functions should be improved by Belle II.  
For our type of analysis, publicly available high-quality (inclusive) data in the strange sector would be especially welcome,  as they would allow us to define additional integrated observables and disentangle various $\epsilon_X^{s \tau}$ Wilson coefficients. 
On the theoretical side, we expect further progress in estimating higher-order and non-perturbative corrections, so as to reduce the dominant  uncertainties in the SM predictions. 
Concerning the exclusive decay channels, we expect significant experimental and theoretical progress in several of them. In the 2-pion channel there is an ongoing effort due to the connection with the $g-2$ anomaly. Expected progress in the $\tau \to \eta \pi \nu_\tau$ channel~\cite{Belle-II:2018jsg,Petar:1369,Moussallam:2021flg} 
would allow us to achieve sensitivity to linear (rather than quadratic) effects in $\epsilon_S^{d \tau}$, allowing us to incorporate this parameter into the global Gaussian likelihood.

Tau data can be used to extract $V_{us}$ and new physics contributions simultaneously. Such analysis does not show any significant preference for new physics, except for the $\sim 2\sigma$ tension in the $\tau\to\nu_\tau \pi\pi$ channel. The situation changes abruptly when the information from nuclear $\beta$, baryon, pion, and kaon  decays is included in the picture. 
Within the SM paradigm, various observables in this larger dataset exhibit the Cabibbo anomaly, that is they point to mutually inconsistent values of the Cabibbo angle. 
Beyond the SM, this tension may be interpreted as a hint for new particles coupled to the SM quarks and leptons.  
This paper provides a complete and unbiased characterization of the nuclear, baryon, pion, kaon and tau data  within a general EFT framework.  
Compared to earlier EFT analyses~\cite{Grossman:2019bzp,Coutinho:2019aiy}, we allow all leading order deformations of the SM to be simultaneously present.   
In particular, the non-standard scalar, pseudoscalar, and tensor interactions (induced e.g. in leptoquark models) are taken into account in our analysis. 
The global likelihood defined by \eref{fitGlobal} and \eref{fitCorrelationMatrix} can be used to constrain parameters of any new physics models with new particles heavier than the weak scale. 
In this general likelihood, the Cabibbo anomaly is reflected as a $3.0 \sigma$ preference for new physics (non-zero values of $\epsilon_X^{q \ell}$) with respect to the SM hypothesis ($\epsilon_X^{q \ell} =0$).  
The preference is strengthened in certain constrained scenarios, for example in some of the single-$\epsilon$ 
scenarios displayed in \tref{oneeps}. 

All in all, our study of hadronic tau decays as new physics probes has allowed us to provide for the first time a model-independent and global analysis of semileptonic charged-current decays of light quarks involving {\em all three} lepton families (that is, $\bar{u}d\bar{e}\nu_e,\bar{u}d\bar{\mu}\nu_\mu,\bar{u}d\bar{\tau}\nu_\tau,\bar{u}s\bar{e}\nu_e,\bar{u}s\bar{\mu}\nu_\mu,\bar{u}s\bar{\tau}\nu_\tau$). Our results provide on one hand guidance for model building and on the other hand an unbiased tool to test the implications of new physics models in this wide set of transitions.

\section*{Acknowledgements}

We thank Elvira G\'amiz for valuable discussions. 
AF and ARS are partially supported by the Agence Nationale de la Recherche (ANR) under grant ANR-19-CE31-0012 (project MORA).
AF is supported by the European Union’s Horizon 2020 research and innovation programme
under the Marie Sklodowska-Curie grant agreement No. 860881 (HIDDe$\nu$ network). 
MGA and DDC are supported by the {\it Generalitat Valenciana} (Spain) through the {\it plan GenT} program (CIDEGENT/2018/014), and MCIN/AEI/10.13039/501100011033 Grant No. PID2020-114473GB-I00.
VC is supported  by the US Department of Energy through  
the Office of Nuclear Physics and  the  
LDRD program at Los Alamos National Laboratory. Los Alamos National Laboratory is operated by Triad National Security, LLC, for the National Nuclear Security Administration of U.S.\ Department of Energy (Contract No. 89233218CNA000001).

\appendix
\renewcommand{\theequation}{\Alph{section}.\arabic{equation}}

\section{QCD computation of inclusive integrals}\label{app:inc}
 \setcounter{equation}{0}
In this appendix we give technical details on the QCD calculation of the different integrals entering in our analysis.

\subsection{Standard Model contribution}
\label{app:inclSMprediction}

\begin{figure}[t]\centering
\includegraphics[width=0.3\textwidth]{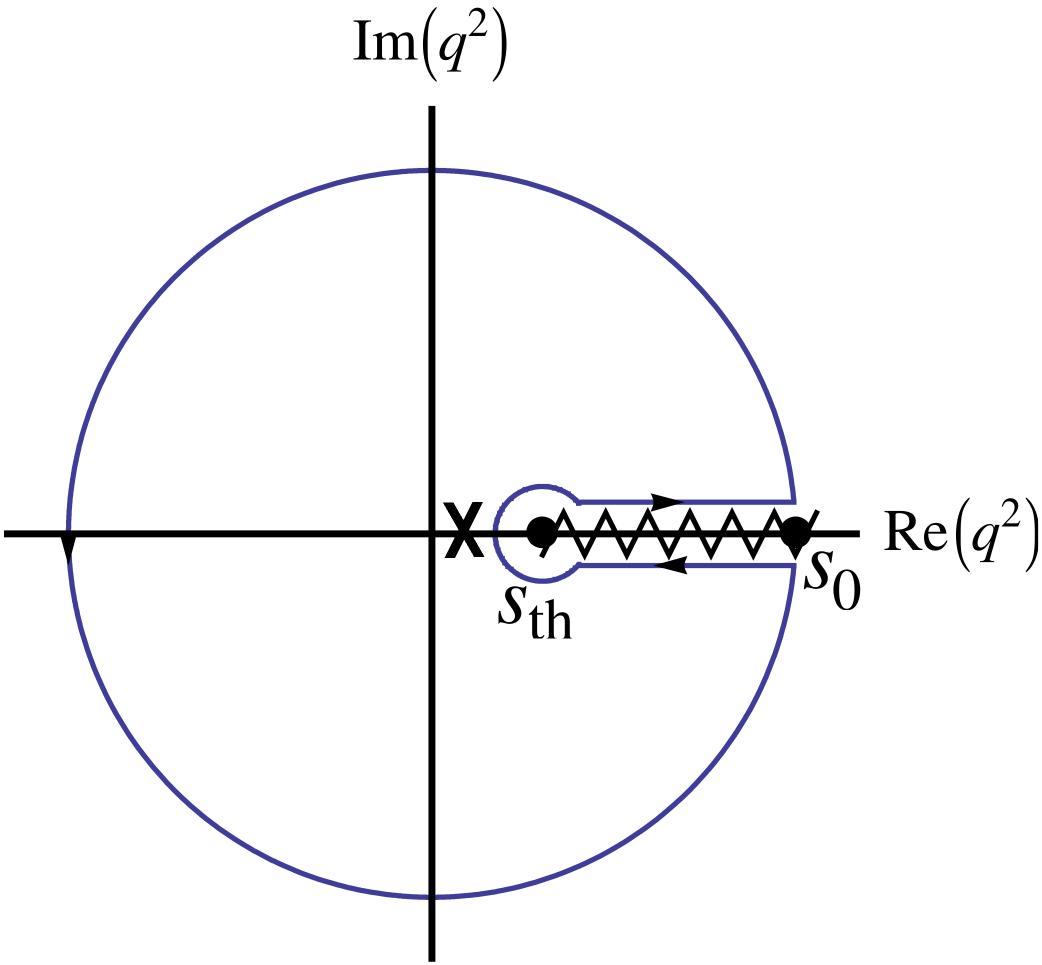}
\caption{\label{fig:circuit} Contour
of integration for Eq. (\ref{eq:disprel})}
\end{figure}

In the SM limit, the main QCD objects entering  our inclusive analysis are the two-point correlation functions of quark currents,
\begin{equation}\label{eq:eqcurrcorrcopy}
i \int d^{4}x\; e^{iqx} \;
\langle 0 |T[J^{\mu}(x)J^{\nu \dagger}(0)]|0\rangle
\; =(-g^{\mu\nu}q^{2}+q^{\mu}q^{\nu})\; \Pi^{(1)}_{JJ}(q^{2})
+ q^{\mu}q^{\nu}\;\Pi^{(0)}_{JJ}(q^{2}) \, ,
\end{equation}
where $J=\{V,A\}$, $V^{\mu}=\bar{d}\gamma^{\mu}u$, $A^{\mu}=\bar{d}\gamma^{\mu}\gamma^{5}u$.
Eqs.~(\ref{eq:distV}) and~(\ref{eq:distA}) connect the experimental tau distributions to $\Pi^{(1+0)}_{JJ}(q^2)\equiv \Pi^{(1)}_{JJ}(q^2)+\Pi^{(0)}_{JJ}(q^2)$, which is an
analytic function in all the complex plane except for the physical cut, which lies on the Minkowskian axis. The continuum threshold for the $(1+0)$ correlator is $s_{th}=4m_{\pi}^{2}$. As a consequence, if we integrate that correlator times any monomial function 
$(s/s_0)^n$ along the contour of Fig.~\ref{fig:circuit}, the only contribution comes from the residue at the pion pole. 
Equating the pion pole contribution to the integral along the different parts of the circuit leads to
\begin{align}
I_{V\pm A}^{\rm SM} (s_0;n)
= 
\mp \frac{f_{\pi^\pm}^{2}}{s_{0}}\left(\frac{m_{\pi}^{2}}{s_{0}}\right)^{n}
+\frac{i}{2\pi}\oint_{|s|=s_{0}}\frac{ds}{s_{0}}\left(\frac{s}{s_{0}}\right)^{n}\Pi_{V\pm A}^{(1+0),\,\text{OPE}}\label{eq:disprel}
+\delta^{\rm DV}_{V\pm A}(s_{0};n)~,
\end{align}
where we have approximated 
$\Pi^{\mathrm{OPE}}_{V\pm A}(s)$
along the complex 
circle $|s|=s_0$
by the analytic continuation of its OPE  expression~\cite{Shifman:1978bx}
\be\label{eq:ope}
\Pi^{\mathrm{OPE}}_{V/A}(s=-Q^{2})\; =\; \sum_{D}\frac{1}{(Q^2)^{D/2}}\sum_{\mathrm{dim} \, \mathcal{O}=D} C_{D, V/A}(Q^{2},\mu)\;\langle\mathcal{O}(\mu)\rangle
\;\equiv\; \sum_{D}\;\dfrac{\mathcal{O}_{D,\, V/A}}{(Q^2)^{D/2}}\, .
\ee
This approximation works very precisely if the upper limit of the integral, $s_{0}$, is large enough, except maybe near the positive real axis. In the previous expression, $\delta^{\rm DV}_{V\pm A}(s_{0};n)$ accounts for the small deviations from it, known as quark-hadron Duality Violations (DV) ~\cite{Braaten:1991qm,Chibisov:1996wf,Shifman:2000jv,Cata:2005zj,GonzalezAlonso:2010xf,Boito:2014sta,Pich:2016bdg,Boito:2017cnp}.

The $D=0$ part of the OPE corresponds to the massless perturbative-QCD prediction, which will be denoted with an index $P$. Since chirality is preserved in the massless QCD Lagrangian, this contribution, which only depends on $\alpha_{s}$, is identical for the $VV$ and the $AA$ correlators. In fact, we can recycle previous studies within the SM to obtain this contribution. Following the same notation as in Ref.~\cite{Pich:2016bdg}, this purely perturbative part can be computed using the Adler function~\cite{Adler:1974gd},\footnote{The $(1+0)$ superscript will be omitted from now on.}
\begin{equation}\label{adler}
D(s)\;\equiv\; -s\,\frac{d\,\Pi^{P}(s)}{ds}\; =\;\frac{1}{4\pi^{2}}\;\sum_{n=0}\tilde K_{n}(\xi)\;
a_{s}^{n}(-\xi^2 s) \, ,
\end{equation}
where
$\xi$ parameterizes the QCD renormalization scale and
$a_{s}(s)\equiv \alpha_{s}(s)/\pi$  satisfies the renormalization-group equation,\footnote{Different normalizations for the $\beta_n$ coefficients can be found in the literature. This form of the RGE corresponds to $\beta_{1}=-9/2$. }
\beqa\label{running}
2\,\frac{s}{a_s}\,\frac{d\, a_{s}(s)}{d s}\; =\; \sum_{n=1} \beta_{n}\, a_{s}^{n}(s) \, . \label{eq:asrunning}
\eeqa
The perturbative coefficients $K_n\equiv \tilde K_{n}(\xi=1)$ are  known up to $n\le 4$~\cite{Baikov:2008jh}. The homogeneous 
renormalization-group equation satisfied by the Adler function determines the corresponding 
scale-dependent parameters $\tilde K_{n}(\xi)$.
Although the dependence on the renormalization scale cancels order by order, 
the truncation to a finite perturbative order leads to a scale dependence from the missing
higher-order terms, which must be taken into account when estimating perturbative uncertainties.

Let us define\footnote{Notice how $A_{P}^{(n)}$ has been re-scaled by $\pi$ with respect to Ref.~\cite{Pich:2016bdg}.}
\beqa\label{pertu}
A^{(n)}_{P}(s_{0})\;\equiv\; \; \frac{i}{2\pi}\;\oint_{|s|=s_{0}}
\frac{ds}{s_{0}}\;\left(\frac{s}{s_{0}}\right)^{n}\, \Pi^{P}(s)\, .
\eeqa
Using integration
by parts,
\begin{align}\nonumber
A^{(n)}_{P}(s_{0})&=\frac{i}{2\pi(n+1)}\left((\Pi(s_{0}-i\epsilon)-\Pi(s_{0}+i\epsilon))+\oint_{|s|=s_{0}}\frac{ds}{s_{0}}\left(\frac{s}{s_{0}}\right)^{n}D(s)  \right)
\\&=\frac{i}{2\pi(n+1)}\oint_{|s|=s_{0}}\frac{ds}{s_{0}}
\left[\left(\frac{s}{s_{0}} \right)^{n}-\frac{s_{0}}{s}  \right]D(s) \, , 
\end{align}
inserting the perturbative Adler function and parameterizing 
the 
circle $|s|=s_0$
as $s=-s_{0}\, e^{i\varphi}$, one finds:
\begin{equation}\label{aomegapert}
A^{(n)}_{P}(s_{0})=-\frac{1}{8\pi^{3}(n+1)}\sum_{m=0}\tilde K_{m}(\xi)
\int^{\pi}_{-\pi}d\varphi \left((-1)^{n+1}e^{i\varphi(n+1)}-1\right)a_{s}^{m}(\xi^{2}s_{0}e^{i\varphi}) \, .
\end{equation}
We take $\alpha_{s}(M_{Z}^{2})=0.1184(8)$ from the lattice~~\cite{Aoki:2021kgd,Maltman:2008bx,PACS-CS:2009zxm,McNeile:2010ji,Chakraborty:2014aca,Bruno:2017gxd,Bazavov:2019qoo,Cali:2020hrj,Ayala:2020odx},\footnote{Even when potentially contaminated by new physics $f_{\pi^\pm}$ values have been used in some lattice determinations to set the scale, $r_{0}$, $r_{1}$ or $\sqrt{t_{0}}$, we have checked that it has no significant impact on our bounds, since alternative inputs for them ($r_{0}$, $r_{1}$ or $\sqrt{t_{0}}$) based on resonance masses would have not modified the extracted strong coupling value significantly.} 
then we run it up to $a_{s}(\xi^2 s_{0})$. and perform the integral, both truncating the integrand to a fixed 
perturbative order 
in $\alpha_s(\xi^2 s_{0})$ (fixed-order perturbation theory, FOPT), and solving
exactly the differential $\beta$-function equation in the $\beta_{n>n_{\mathrm{max}}}=0$ approximation
(contour-improved perturbation theory, CIPT).

Again, as in Ref.~\cite{Pich:2016bdg}, within a given perturbative approach, either CIPT or FOPT, we will estimate the perturbative uncertainty varying the renormalization scale in the interval $\xi^2 \in [0.5\, ,\, 2]$. Additionally, we will truncate the perturbative series at $n=5$, taking $K_{5}=275 \pm 400$ as an educated guess of the maximal range of variation of the unknown fifth-order contribution. These two sources of theoretical uncertainty will be combined quadratically. 

In order to give a combined determination for the observables, we will finally average the CIPT 
and FOPT results. Since the previously estimated perturbative uncertainties do not fully account 
for the difference between these two prescriptions, we will conservatively assess the final error 
adding in quadrature half the difference between the CIPT and FOPT values to
the smallest of the CIPT and FOPT errors.\footnote{Some recent works aimed to understand higher-order corrections, which eventually may lead to a significant reduction on the perturbative uncertainty, can be found in the literature~\cite{Boito:2016pwf,Boito:2018rwt,Hoang:2020mkw,Hoang:2021nlz,Ayala:2021mwc}.} 

For the tiny non-perturbative part of the OPE, we work at leading order in $\alpha_{s}$, with the exception of $\mathcal{O}_6^{V-A}$, where $\alpha_s$ corrections are incorporated. Thus, $\mathcal{O}_{D}$ is independent on $Q^{2}$. Its contribution 
to $I_{V\pm A}^{\rm SM} (s_0;n)$
is
\begin{equation}\label{eq:cond}
\frac{i}{2\pi}\;\oint_{|s|=s_{0}}
\frac{ds}{s_{0}}\left(\frac{s}{s_{0}}\right)^{n}\;\,\Pi^{\mathrm{OPE}}_{V\pm A}(s)
=-\frac{\mathcal{O}_{2(n+1)}}{(-s_{0})^{n+1}} \, .
\end{equation}
Some of the condensates entering into the bounds are unknown. When this is the case, we take a conservative dimensional guess based on the ones that are known. More details on it are given in the main text.

Finally, the DV term can be written as
\bea
 \delta^{\rm DV}_{V\pm A}(s_{0};n)
 &\equiv&
 \frac{i}{2\pi}\oint_{|s|=s_{0}}\frac{ds}{s_{0}}\left(\frac{s}{s_{0}}\right)^{n}
 \left(\Pi_{V\pm A}-\Pi_{V\pm A}^{\text{OPE}})(s)\right)\nn
 &=&
 \label{dvs}
-\int^{\infty}_{s_{0}}\frac{ds}{s_{0}}\left(\frac{s}{s_{0}}\right)^{n}\,\frac{1}{\pi} \left(\text{Im}\,\Pi_{V\pm A}(s) - \text{Im}\,\Pi_{V\pm A}^{\text{OPE}}(s) \right) \, .
\eea
One 
common
tool to reduce these effects is
pinching, {\it i.e.}, using weight functions that reduce the contributions of the integrals near 
the Minkowskian axis.
We know DV effects go to zero very fast with the opening of the hadronic multiplicity, typically
in an oscillatory way. Then, DV uncertainties should be strongly dominated by the contributions
near the upper limit of the integral. 
Thus, we take $\delta^{\rm DV}_{V\pm A}=0$ and estimate the associated uncertainty from the size of the small fluctuations in the predictions for the different dispersion relations when changing $s_0$ in moderate intervals.

\subsection{Computation of the $\Pi_{VT}$ integral}
\label{app:VTintegral}
In the presence of non-standard interactions, the vector-tensor correlation function enters  our analysis, connected to the invariant mass distribution of Eq.~(\ref{eq:distV}),
\begin{equation}\label{eq:eqcurrcorrcopy2}
i \int d^{4}x\; e^{iqx} \; 
\langle 0 |T[V^{\mu}(x)T^{\alpha\beta \dagger}(0)]|0\rangle
\; =\;
i(g^{\mu\alpha}q^{\beta}-g^{\mu\beta}q^{\alpha})\; \Pi_{VT}(q^{2})\, ,
\end{equation}
where $V^{\mu}=\bar{d}\gamma^{\mu}u$ and $T^{\mu\nu}=\bar{d}\sigma^{\mu\nu}u$.

A dispersion relation analogous to Eq. (\ref{eq:disprel}) follows from the analytic properties of the $\Pi_{VT}(s)$ correlator,
\begin{equation}\label{eq:VTdisp}
I_{VT} (s_0,n)  =I_{VT}^{(1)}(s_0,n)  +I_{VT}^{(2)} (s_0,n)   \; ,
\end{equation} 
where $I_{VT}^{(1)}$ is the contribution from the complex 
circle
(from now on for ease of notation we drop the arguments $(s_0,n)$ in $I_{VT}^{(1),(2)}$),
\begin{equation}\label{eq:IVT1}
I_{VT}^{(1)}\equiv -\frac{m_{\tau}}{4\pi i}\oint_{|s|=s_{0}}\frac{ds}{s_{0}}\left(\frac{s}{s_{0}}\right)^{n}\frac{\Pi_{VT}(s)}{s+\frac{m_{\tau}^{2}}{2}}\, ,
\end{equation}and $I_{VT}^{(2)}$ the contribution from the kinematic pole,
\begin{equation}\label{eq:IVT2}
I_{VT}^{(2)}\equiv-\left(-\frac{m_{\tau}^{2}}{2 s_{0}}\right)^{n+1}\, \frac{\Pi_{VT}(Q^2=\frac{m_{\tau}^{2}}{2}) }{m_{\tau}} \, .
\end{equation}where $Q^{2}$ is the Euclidean momenta ($s=Q^{2}e^{i\pi}$). The OPE of $\Pi_{VT}(s)$ should provide a good approximation in both 
$I_{VT}^{(1)}$ and $I_{VT}^{(2)}$
for $s_{0} \sim m_{\tau}^{2}$.

In the following we describe the OPE contributions at several degrees of approximation. We start with the tree-level, lower-dimension, estimates made in Ref.~\cite{Cirigliano:2018dyk} and add several improvements. Our final value is based on a full Next-to-Leading Log (NLL) evaluation in $\alpha_{s}$ for the lowest dimensional term, including the running of $\alpha_{s}$ along the complex circle. Uncertainties coming from higher-dimensional contributions are also discussed.

\paragraph{Tree-level.-}
$\Pi_{VT}(Q^{2})$ vanishes at all orders in massless perturbative-QCD. This is a direct consequence of the chirality flipping nature of the tensor current, $\bar{d}\sigma^{\mu\nu}u=\bar{d}_L\sigma^{\mu\nu}u_R+\bar{d}_R\sigma^{\mu\nu}u_L$, since chiral symmetry is preserved by the massless QCD Lagrangian.

Chiral symmetry is, however, spontaneously broken, and the two-point correlation function $\Pi_{VT}(Q^{2})$ does not vanish in QCD. Taking into account the tiny size of the light quark masses, the leading OPE contribution comes from the quark condensate, $\langle \bar{q} q \rangle$. Disregarding small $SU(3)_{V}$-breaking corrections, the tree-level result is~\cite{Craigie:1981jx,Jamin:2008rm},\footnote{There is a misprint in the global sign in Eq. (2.8) of Ref.~\cite{Craigie:1981jx}}$^,$\footnote{In the absence of explicit sources of $SU(3)_{V}$ breaking, such as the light quark masses and electromagnetism, vacuum is invariant under those transformations and then only $SU(3)_{V}$ singlet operators can acquire a nonzero vev. As a consequence, $\langle \bar{u}u \rangle\approx\langle \bar{d}d \rangle\approx\langle \bar{s}s \rangle$.}
\begin{equation}\label{eq:opevttree}
\Pi_{VT, \, \mathrm{Tree}}^{\mathrm{OPE}}(Q^{2})\approx-\frac{2}{Q^{2}}\langle0| \bar{q}q|0\rangle_{\mu} \, .
\end{equation}This expression is only expected to give a reliable first estimate when the quark condensate is evaluated at a scale $\mu$ that is close both to the matching point $\mu_0$ and to $\sqrt{s_0}$. Taking $\mu=\mu_{0}\equiv 2 \, \mathrm{GeV}$, this requirement is approximately satisfied. Using the ansatz of Eq. (\ref{eq:opevttree}) one finds,
\begin{align}
I_{VT, \mathrm{Tree}}^{(1)}&=(1-\delta_{n,0}) \frac{2 \langle \bar{q} q \rangle_{\mu_{0}}}{s_{0} \, m_{\tau}}\left(-\frac{m_{\tau}^{2}}{2 s_{0}}  \right)^{n}\, ,\\
I_{VT, \mathrm{Tree}}^{(2)}&=-\frac{2\langle \bar{q} q \rangle_{\mu_{0}}}{s_{0}m_{\tau}}\left(-\frac{m_{\tau}^{2}}{2 s_{0}}  \right)^{n}\, .
\end{align}
Adding both pieces,
\begin{equation}
I_{VT, \, \mathrm{Tree}}=-\frac{2 \langle \bar{q}q \rangle_{\mu_{0}} }{m_{\tau}s_{0}}\delta_{n,0} \, ,
\end{equation}which corresponds to the expression given in Ref.~\cite{Cirigliano:2018dyk}.

\paragraph{Leading Logarithmic (LL) resummation.-} A first improvement in the calculation consists in keeping track of the scale dependence of the correlator, resumming the logs from the matching scale in Eq.~(\ref{eq:masterformula}), which we choose  to be $\mu_{0}= 2 \, \mathrm{GeV}$, to the most convenient choice $\mu$ to cancel logarithms with lower-energy scales, resumming the cascade of $\sim \alpha^n_{s}(\mu)\log^n\left( \frac{\mu_{0}^2}{\mu^2}\right)$ contributions.

When taking into account QCD corrections, $I_{VT}(s_0,n)$ becomes dependent on the renormalization scale, as a consequence of the non-zero anomalous dimension of the tensor current. In contrast to an already scale independent $I^{SM}_{V\pm A}(s)$, this scale dependence only cancels in the $\epsilon_{T}\, I_{VT}$ product. The scale dependence of $I_{VT}$ is inherited by $\Pi_{VT}^{OPE}(s)$, for which both the vacuum condensates ($\langle \bar{q} q \rangle(\mu)$ at lower dimension) and the Wilson coefficients in front ($C_{\langle\bar{q}q\rangle}(\mu) $) are scale dependent~\cite{Jamin:2008rm}.

Let us define our convention for the anomalous dimension of any operator $\mathcal{O}$,
\begin{equation}
\gamma_{\mathcal{O}}=\sum_{n} \gamma_{\mathcal{O}}^{(n)}\left(\frac{\alpha_{s}}{\pi}\right)^{n} \, ,
\end{equation}through the identity
\begin{equation}\label{eq:anomdef}
\left( \mu \frac{d}{d\mu}+\gamma_{\mathcal{O}} \right)\mathcal{O}(\mu) \equiv 0 \, .
\end{equation}
The leading order anomalous dimension  ($N_{F}=3$) of $C_{\langle \bar{q} q \rangle}$ can be obtained 
either (i) by combining the anomalous dimension of the quark condensate $\gamma^{(1)}_{\langle \bar{q} q \rangle}=-\gamma^{(1)}_{m}=-2$ with the one from $\epsilon_{T}^{d\tau}(\mu)$, $\gamma^{(1)}_{\epsilon_{T}^{d\tau}}=-\frac{2}{3}$~\cite{Gonzalez-Alonso:2017iyc} and requiring that the $\epsilon^{d\tau}_{T}(\mu)\,\Pi_{VT}(\mu)$ product must be scale-independent; (ii) or directly from the one-loop calculation of $C_{\langle \bar{q} q \rangle}(\mu)$~\cite{Jamin:2008rm}. The result is the same, $\gamma^{(1)}_{C_{\langle \bar{q}q\rangle}}=\frac{8}{3}$. Then, starting from a matching scale $\mu_{0}$ in Eq. (\ref{eq:masterformula}), we can re-express $\Pi_{VT}^{\mathrm{OPE}}(\mu_{0})$ as a function of $C_{\bar{q}q}(\mu)$ at any other scale $\mu$ by solving  the leading order version of the RGE for $C_{\bar{q}q}(\mu)$. One obtains, up to $\alpha_{s}^n\log^{n-1}\frac{\mu_{0}^{2}}{\mu^{2}}$ corrections (starting at $n=1$), 
\begin{equation}\label{eq:piopell}
\Pi^{\mathrm{OPE}}_{VT, \, \mathrm{LL}}(Q^{2}, \mu_{0})=\frac{-2}{Q^{2}}\left( \frac{\alpha_{s}(\mu_{0}^{2})}{\alpha_{s}(\mu^{2})} \right)^{-\frac{\gamma^{(1)}}{\beta_{1}}} \langle \bar{q}q\rangle_{\mu_{0}} \, ,
\end{equation}
where $\gamma^{(1)}$ refers to $C_{\langle \bar{q}q\rangle}$ and $\beta_{1}=-\frac{9}{2}$. Now we have the freedom to set the most suitable scale $\mu$ to avoid large logarithms with low-energy scales in both terms of Eq. (\ref{eq:VTdisp}). Inspecting Eqs. (\ref{eq:IVT1}) and (\ref{eq:IVT2}), natural choices for $I_{VT}^{(1)}$ and $I_{VT}^{(2)}$ are, respectively, 
$\mu_{1}^2=s_{0}\,  \xi(x)$ and $\mu_{2}^{2}=\frac{m_{\tau}^{2}}{2}$,
where $x\equiv\frac{Q^{2}}{s_{0}}$ and $\xi(x)$ is a function that depends on whether a logarithmic resummation along the 
$|s| =s_0$
circle is performed,  $\xi^{\mathrm{CIPT}}(x)=x$, or not, $\xi^{\mathrm{FOPT}}(x)=1$. One finds:
\begin{align}
&I_{VT, \, \mathrm{LL}}^{(1)\, \mathrm{FOPT}}(\mu_{0})=\left( \frac{\alpha_{s}(\mu_{0}^{2})}{\alpha_{s}(s_{0})} \right)^{-\frac{\gamma^{(1)}}{\beta_{1}}}I^{(1)}_{VT, \, \mathrm{Tree}} \, ,\\
&I_{VT,\, \mathrm{LL}}^{(1)\, \mathrm{CIPT}}(\mu_{0})=\frac{(-1)^{n}m_{\tau} \langle \bar{q} q \rangle_{\mu_{0}}}{2\pi i s_{0}^{2}}\left( \frac{\alpha_{s}(\mu_{0}^{2})}{\alpha_{s}(s_{0})} \right)^{-\frac{\gamma^{(1)}}{\beta_{1}}} \oint_{|x|=1} dx \frac{x^{n-1}}{x-\frac{m_{\tau}^{2}}{2s_{0}}}\left(1-\frac{\beta_{1}\alpha_{s}(s_{0})}{2\pi}\log x\right)^{-\frac{\gamma^{(1)}}{\beta_{1}}} \, ,\\
&I_{VT, \, \mathrm{LL}}^{(2)}(\mu_{0})=\left( \frac{\alpha_{s}(\mu_{0}^{2})}{\alpha_{s}(m_{\tau}^{2}/2)} \right)^{-\frac{\gamma^{(1)}}{\beta_{1}}}I^{(2)}_{VT, \, \mathrm{Tree}}\, .
\end{align}

\paragraph{Full $\mathcal{O}(\alpha_{s})$ corrections plus LL resummation.-} We can add the non-logarithmic contribution of order $\alpha_{s}$ computed in Ref.~\cite{Jamin:2008rm} to the cascade of $\alpha_{s}^{n}\log^{n}\frac{\mu_{0}}{\mu}$ contributions. Including this correction one has, up to (and not including) corrections of order $\alpha_{s}^{n}\log^{m}\frac{\mu_{0}}{\mu}$ with $0 \leq m<n$, starting at $n=2$,
\begin{equation}\label{eq:piopellals}
\Pi^{\rm OPE}_{VT}(Q^{2}, \mu_{0})=\frac{-2}{Q^{2}}\left( \frac{\alpha_{s}(\mu_{0}^{2})}{\alpha_{s}(\mu^{2})} \right)^{-\frac{\gamma^{(1)}}{\beta_{1}}}\left[1-  \frac{4\alpha_{s}(\mu^2)}{3\pi}\left( 1- \log\frac{Q^{2}}{\mu^{2}} \right) \right]\langle\bar{q}q\rangle_{\mu_{0}} \, .
\end{equation}which with the previous scale choices introduces a correction that goes (up to NLL corrections) as:
\begin{align}
\Delta I_{VT,\, \mathrm{no\, logs}}^{\alpha_{s}, \, \mathrm{FOPT} }&=-\frac{4\alpha_{s}(\mu_{0}^2)}{3\pi}I_{VT, \mathrm{Tree}}
+\frac{(-1)^{n}m_{\tau}\langle \bar{q}q \rangle_{\mu_{0}}}{2\pi i s_{0}^{2}}\, \frac{4\alpha_{s}(\mu_{0}^2)}{3\pi}\oint dx \frac{x^{n-1}}{x-\frac{m_{\tau}^{2}}{2s_{0}}}\log x \, ,\\
\Delta I_{VT,\, \mathrm{no\, logs}}^{\alpha_{s}, \, \mathrm{CIPT} }&=-\frac{4\alpha_{s}(\mu_{0}^2)}{3\pi}I_{VT, \mathrm{Tree}} \, .
\end{align}
\paragraph{NLL analysis.-} Since we have the full $D=3$ contribution at NLO in $\alpha_{s}$ available, the only remaining piece for a full NLL analysis is the NLO anomalous dimension of the Wilson coefficient $C_{\langle qq \rangle}(\mu)$. Once again, scale independence of Eq. (\ref{eq:masterformula}) implies $\gamma^{(2)}_{\epsilon_{T}^{d\tau}}+\gamma^{(2)}_{C_{\langle \bar{q}q\rangle}}+\gamma^{(2)}_{\langle \bar{q}q\rangle}=0$. Taking into account that $\gamma_{\langle qq \rangle}^{(2)}=-\gamma_{m}^{(2)}=-\frac{91}{12}$ and $\gamma^{(2)}_{\epsilon^{d\tau}_{T}}=-\frac{155}{36}$~\cite{Gonzalez-Alonso:2017iyc}, one finds $\gamma^{(2)}_{C_{\langle \bar{q}q\rangle}}=\frac{107}{9}$. Solving~\cite{Buchalla:1995vs} the NLO version of the RGE equation (\ref{eq:anomdef}) for $C_{\langle \bar{q}q\rangle}$ with the initial condition:
\begin{equation}
C_{\langle \bar{q}q\rangle}(\mu^{2})=1-\frac{4}{3}\frac{\alpha_{s}(\mu^{2})}{\pi}\left( 1- \log\frac{Q^{2}}{\mu^{2}} \right) \, ,
\end{equation}one obtains:
\begin{align} \nonumber
\Pi^{\mathrm{OPE}}_{VT,\, \mathrm{NLL}}(Q^{2}, \mu_{0})=\frac{-2}{Q^{2}}\left( \frac{\alpha_{s}(\mu_{0}^{2})}{\alpha_{s}(\mu^{2})} \right)^{-\frac{\gamma^{(1)}}{\beta_{1}}}&\left[1+\frac{\alpha_{s}(\mu_{0}^2)-\alpha_{s}(\mu^{2})}{\pi\beta_{1}}\left(\frac{\beta_{2}}{\beta_{1}}\gamma^{(1)}-\gamma^{(2)}\right)     \right] \\ \times &\left[1-\frac{4}{3}\frac{\alpha_{s}(\mu^{2})}{\pi}\left( 1- \log\frac{Q^{2}}{\mu^{2}} \right) \right]\langle\bar{q}q\rangle_{\mu_{0}} \, ,\label{eq:piopennll}
\end{align}where again $\gamma^{(i)}$ refers to $C_{\langle \bar{q}q \rangle}$ and $\beta_{2}=-8$. Setting again the renormalization scales as above one has

\begin{equation}
I_{VT,\, \mathrm{NLL}}^{(1)}=-\frac{(-1)^{n}m_{\tau}}{4\pi i\, s_{0}}\oint_{|x|=1}dx \, x^{n}\, \frac{\Pi^{\mathrm{OPE}}_{VT,\, \mathrm{NLL}}(Q^{2}=s_{0}x, \mu_{0}, \mu^2=\xi(x)s_{0}) }{x-\frac{m_{\tau}^{2}}{2s_{0}}}\, ,
\end{equation}

\begin{align}\nonumber
I_{VT,\, \mathrm{NLL}}^{(2)}=\left(-\frac{m_{\tau}^{2}}{2 s_{0}}\right)^{n+1} \frac{4}{m_{\tau}^{3}}\left( \frac{\alpha_{s}(\mu_{0}^{2})}{\alpha_{s}(m_{\tau}^{2}/2)} \right)^{-\frac{\gamma^{(1)}}{\beta_{1}}}&\left[1+\frac{\alpha_{s}(\mu_{0}^2)-\alpha_{s}(m_{\tau}^{2}/2)}{\pi\beta_{1}}\left(\frac{\beta_{2}}{\beta_{1}}\gamma^{(1)}-\gamma^{(2)}\right)     \right] \\ \times &\left[1-\frac{4}{3}\frac{\alpha_{s}(m_{\tau}^{2}/2)}{\pi}\right]\langle\bar{q}q\rangle_{\mu_{0}}
\end{align}
where once again $\xi(x)=1$ corresponds to FOPT and $\xi(x)=x$ to CIPT.\footnote{The two-loop analytic continuation of the running coupling must be implemented for the first term in the rhs of Eq. (\ref{eq:piopennll}).}

\paragraph{Higher dimensional corrections.-} The following dimensional contribution comes from dimension $D=5$~\cite{Cata:2008zc}. At tree-level,
\begin{equation}
\Pi_{VT}^{\mathrm{OPE},\, D=5}(Q^{2})=\frac{2}{3Q^{4}} \langle 0| g_{s}\bar{q} G^{\mu\nu}\sigma_{\mu\nu} q|0\rangle \, ,
\end{equation}which gives
\begin{equation}
I_{VT}^{\mathrm{OPE},\, D=5}=\frac{2\langle 0| g_{s}\, \bar{q}\, G^{\mu\nu}\sigma_{\mu\nu}\, q|0\rangle}{3 m_{\tau} s_{0}^{2}} (2\delta_{n,0}-\delta_{n,1}) \, .
\end{equation}

\paragraph{Numerical implementation.-}
The main input needed for the numerical evaluation of $I_{VT}$ is then the quark condensate. We take as input the latest $N_{f}= 2 + 1$ lattice estimate, $\langle \bar{q}q\rangle^{\frac{1}{3}}=-272(5) \mathrm{MeV}$~\cite{Bazavov:2010yq,Borsanyi:2012zv,Durr:2013goa,Boyle:2015exm,Cossu:2016eqs,Aoki:2017paw,Aoki:2021kgd}. 

Unfortunately, no modern precise determination of the $D=5$ condensate is available. 
We will take conservatively,
\begin{equation}
\langle 0| g_{s}\bar{q} G^{\mu\nu}\sigma_{\mu\nu} q|0\rangle=0 \pm 0.8 \, \mathrm{GeV}^{2} \langle \bar{q} q \rangle \, ,
\end{equation}
where we used the 40-year-old result of Ref.~\cite{Belyaev:1982sa} (obtained from baryon sum rules) as an estimate of the uncertainty of the quark-gluon condensate and of the neglected higher-order OPE contributions  ($D>3$). 

The $I_{VT}$ values obtained in the above-described approximations are shown in Table~\ref{tab:IVT}. The different corrections are rather large (notice how the leading contribution for most of the monomial functions only starts at NLO in $\alpha_{s}$) and go in the same direction. Our final values correspond to the NLL (CIPT and FOPT average) ones taking their difference with the NLO+LL ones as estimate of the perturbative ($D=3$) uncertainty.

\begin{table}[tbh]\begin{center}
\begin{tabular}{|c|c|c|c|c|c}
\cline{1-5}
Weight         & $\omega=1$    & $\omega_{\tau}$   &  $\omega_1$ & $\omega_2$  &  \\ \cline{1-5}
Tree           & $8.1$         & $7.2$        & $8.1$            & $7.2$            &  \\ \cline{1-5}
LL             & $6.6$         & $6.2$        & $6.0$            & $4.7$            &  \\ \cline{1-5}
NLO+LL         & $5.4$         & $4.6$        & $4.6$            & $3.6$            &  \\ \cline{1-5}
NLL            & $4.8$         & $4.1$        & $3.8$            & $2.8$            &  \\ \cline{1-5}
$\sigma_{D=5}$ & $0\pm1.5$         & $0\pm1.2$        & $0\pm2.3$            & $0\pm2.4$            &  \\ \cline{1-5}
Final          & $4.8 \pm 1.6$ & $4.1\pm 1.3$ & $3.8 \pm 2.4$    & $2.8 \pm 2.6$    &  \\ \cline{1-5}
\end{tabular}\end{center}\caption{Theoretical values of $I_{VT}$ for the four weight functions used in this work and obtained working at different levels of approximation (see main text). The results are given in $10^{-3}$ units. The $s_0$ values are those chosen in~\sref{nonstrangedecays} for each moment, {\it i.e.}, $2.8$ GeV$^2$, $m_\tau^2$, $2.8$ GeV$^2$, and $m_\tau^2$, respectively.
\label{tab:IVT}}
\end{table}

\section{Tensor form factor $f_T$ in radiative pion decays}
\label{app:fT}
\setcounter{equation}{0}

In this appendix we describe an improved evaluation of the tensor form-factor appearing in the radiative pion  decay $\pi^{-}\rightarrow e^{-}\bar{\nu}_{e}\gamma$,
\begin{equation}\label{eq:tensorrad}
\langle \gamma(k, \epsilon) | \bar{u} \sigma^{\mu\nu}\gamma_{5}d|\pi^{-}\rangle=-\frac{e}{2}f_{T}(k_{\mu}\epsilon_{\nu}-k_{\nu}\epsilon_{\mu}) \, .
\end{equation}
In Ref.~\cite{Voloshin:1992sn} a connection between the tensor form factor and the magnetic susceptibility of the vacuum $\chi$~\cite{Ioffe:1983ju} was derived  by using current algebra (see Ref.~\cite{Mateu:2007tr} for a more detailed re-derivation). Let us first
revisit this connection by using instead the chiral Lagrangian of Ref.~\cite{Cata:2007ns},

\begin{equation}\label{O4}
{\cal{L}}_4^{\chi PT}=\Lambda_1\,\langle\, t_+^{\mu\nu}\,f_{+\mu\nu}\,\rangle\,-\,i\,\Lambda_{2}\,\langle\, t_+^{\mu\nu}\,u_{\mu}u_{\nu}\,\rangle\,+\,\Lambda_{3}\,\langle\, t_+^{\mu\nu}\,t_{\mu\nu}^+\,\rangle\,+\,\Lambda_{4}\,\langle\, t_+^{\mu\nu}\,\rangle^2\, , 
\end{equation}
derived by adding  a new term involving tensor sources to the QCD Lagrangian and by building the lowest order  chiral operators   with 
the same transformation properties under $SU(N_{F})\times SU(N_{F})$  (in our case $N_{F}=2$). The effective low-energy realization of the tensor quark current at leading order is obtained by equating functional derivatives of the action with respect to the tensor sources for both Lagrangians. Taking the derivative of the generating functional with respect to $e_{i}\bar{t}_{\mu\nu}^{ii}$, where $e_{i}$ is the electric charge of the associated light quark and $\bar{t}^{\mu\nu}$ is the tensor source as defined in Ref.~\cite{Cata:2007ns}, and contracting with an initial photon state, one finds at leading order
\begin{equation}
\chi\langle \bar{q}{q} \rangle \langle0|F_{\mu\nu}|\gamma(q,\epsilon)\rangle \equiv \frac{1}{e_{i}}\langle 0|\bar{q}_{i}\sigma_{\mu\nu}q_{i}|\gamma(q,\epsilon)\rangle=-2\Lambda_{1}\langle0|F_{\mu\nu}|\gamma(q,\epsilon)\rangle \, ,
\end{equation}
where the first identity corresponds to the definition of the magnetic susceptibility, $\chi$. One then finds $\Lambda_{1}=-\frac{\chi\langle \bar{q}{q} \rangle}{2}$.\footnote{$\Lambda_{2}$ is also known, since it can be shown to be proportional to the tensor form factor of Eq. (\ref{eq:tensorshapepipi}).} Taking now the derivative of the action with respect to $\bar{t}_{\mu\nu}^{12}$ and contracting with a  photon in the final state and a pion in the  initial state one obtains,
\begin{equation}
\langle \gamma(k, \epsilon) | \bar{u} \sigma^{\mu\nu}\gamma_{5}d|\pi^{-}\rangle=-\frac{ie\sqrt{2}\Lambda_{1}}{3F}\langle \gamma(k, \epsilon) | F_{\mu\nu}\pi^{-}|\pi^{-}\rangle=\frac{e\sqrt{2}\Lambda_{1}}{3F}(k_{\mu}\epsilon_{\nu}-k_{\nu}\epsilon_{\mu}) \, ,
\end{equation}
from which, comparing with Eq. (\ref{eq:tensorrad}) one finds,
\begin{equation}
f_{T}=\frac{\sqrt{2}\chi \langle \bar{q} q \rangle}{3F} \, ,
\end{equation}
in perfect agreement with Eq. (51) of Ref.~\cite{Mateu:2007tr}. The magnetic susceptibility $\chi$ was estimated in that reference by modeling the $\Pi_{VT}$ correlator assuming dominance of one vector resonance (the $\rho$) and using that $\Pi_{VT}(0)$ is proportional to $\chi$ (see also Refs.~\cite{Craigie:1981jx,Balitsky:1985aq,Knecht:2001xc,Mateu:2007tr,Bijnens:2020xnl}). Fortunately, the quantity of interest has been precisely computed in the lattice~\cite{Bali:2020bcn}
\begin{equation}
\chi\langle \bar{q}q\rangle=(45.4 \pm 1.5)\, \mathrm{MeV}
\end{equation}
Using that result and taking $F=F_{\pi}=(130.50\pm 0.13)/\sqrt{2}$~MeV~\cite{ParticleDataGroup:2020ssz}, which it is valid at the working chiral order, one obtains 
\begin{equation}
f_{T}=0.232 \pm 0.012 \pm 0.008 \, ,
\end{equation}
at $\mu=2$ GeV in the $\overline{MS}$ scheme. The first uncertainty corresponds to an estimate of higher-order chiral corrections ($\sim 5 \%$), and the second one corresponds to the one coming from the lattice input. This updates the bound of Ref.~\cite{Gonzalez-Alonso:2016etj} to:
\begin{equation}
\hat{\epsilon}_{T}^{de}=(0.5\pm 2.4) \times 10^{-3} \; .
\end{equation}

\section{Correlation matrix}
\setcounter{equation}{0}

In this appendix we present the correlation matrix (in the Gaussian approximation) associated with the global constraints in  \eref{fitGlobal}:
{\tiny
\begin{equation}
\label{eq:fitCorrelationMatrix}
\left(
\begin{array}{cccccccccccccccccc}
 1 & 0.01 & 0. & 0. & 0. & 0. & -0.11 & 0. & 0.03 & 0. & -0.12 & 0.02 & 0. & 0. & 0. & 0. & -0.03 & -0.05 \\
 \text{} & 1 & -0.97 & 0. & 0.91 & 0. & 0. & -0.26 & 0.25 & -0.92 & 0. & 0.25 & 0. & -0.83 & 0.48 & 0.85 & -0.25 & -0.16 \\
 \text{} & \text{} & 1 & 0.04 & -0.95 & 0.03 & 0. & 0. & 0. & 0.95 & 0. & 0. & 0. & 0.86 & -0.5 & -0.88 & 0. & -0.1 \\
 \text{} & \text{} & \text{} & 1 & -0.02 & 0.6 & 0. & 0. & 0. & 0.02 & 0. & 0. & 0. & 0.03 & 0. & -0.04 & 0. & 0. \\
 \text{} & \text{} & \text{} & \text{} & 1 & -0.01 & 0. & 0. & 0.02 & -0.999 & 0. & 0.02 & 0. & -0.75 & 0.68 & 0.76 & -0.02 & 0.09 \\
 \text{} & \text{} & \text{} & \text{} & \text{} & 1 & 0. & 0. & 0. & 0.01 & 0. & 0. & 0. & 0.02 & 0. & -0.03 & 0. & 0. \\
 \text{} & \text{} & \text{} & \text{} & \text{} & \text{} & 1 & 0. & 0. & 0. & 0. & 0.02 & 0.55 & 0. & 0. & 0. & 0. & 0.01 \\
 \text{} & \text{} & \text{} & \text{} & \text{} & \text{} & \text{} & 1 & -0.997 & 0. & 0. & -0.997 & 0. & 0. & 0. & 0. & 0.99 & 0.98 \\
 \text{} & \text{} & \text{} & \text{} & \text{} & \text{} & \text{} & \text{} & 1 & -0.02 & -0.01 & 0.9996 & 0. & 0.02 & 0.04 & -0.02 & -0.997 & -0.98 \\
 \text{} & \text{} & \text{} & \text{} & \text{} & \text{} & \text{} & \text{} & \text{} & 1 & 0. & -0.02 & 0. & 0.75 & -0.68 & -0.76 & 0.02 & -0.09 \\
 \text{} & \text{} & \text{} & \text{} & \text{} & \text{} & \text{} & \text{} & \text{} & \text{} & 1 & -0.01 & 0. & 0. & 0. & 0. & 0.01 & 0.01 \\
 \text{} & \text{} & \text{} & \text{} & \text{} & \text{} & \text{} & \text{} & \text{} & \text{} & \text{} & 1 & 0.01 & 0.02 & 0.04 & -0.02 & -0.997 & -0.98 \\
 \text{} & \text{} & \text{} & \text{} & \text{} & \text{} & \text{} & \text{} & \text{} & \text{} & \text{} & \text{} & 1 & 0. & 0. & 0. & 0. & 0. \\
 \text{} & \text{} & \text{} & \text{} & \text{} & \text{} & \text{} & \text{} & \text{} & \text{} & \text{} & \text{} & \text{} & 1 & -0.06 & -0.97 & -0.01 & -0.06 \\
 \text{} & \text{} & \text{} & \text{} & \text{} & \text{} & \text{} & \text{} & \text{} & \text{} & \text{} & \text{} & \text{} & \text{} & 1 & 0.11 & -0.04 & 0.05 \\
 \text{} & \text{} & \text{} & \text{} & \text{} & \text{} & \text{} & \text{} & \text{} & \text{} & \text{} & \text{} & \text{} & \text{} & \text{} & 1 & 0.02 & 0.08 \\
 \text{} & \text{} & \text{} & \text{} & \text{} & \text{} & \text{} & \text{} & \text{} & \text{} & \text{} & \text{} & \text{} & \text{} & \text{} & \text{} & 1 & 0.98 \\
 \text{} & \text{} & \text{} & \text{} & \text{} & \text{} & \text{} & \text{} & \text{} & \text{} & \text{} & \text{} & \text{} & \text{} & \text{} & \text{} & \text{} & 1 \\
\end{array}
\right).
\end{equation} 
}

\bibliographystyle{JHEP}
\bibliography{longtau}

\providecommand{\href}[2]{#2}\begingroup\raggedright\begin{thebibliography}{100}

\bibitem{Pich:2013lsa}
A.~Pich {\em Prog. Part. Nucl. Phys.} {\bf 75} (2014) 41--85,
  [\href{http://arxiv.org/abs/1310.7922}{{\tt arXiv:1310.7922}}].

\bibitem{Schael:2005am}
{\bf ALEPH} Collaboration, S.~Schael et~al. {\em Phys. Rept.} {\bf 421} (2005)
  191--284, [\href{http://arxiv.org/abs/hep-ex/0506072}{{\tt hep-ex/0506072}}].

\bibitem{Braaten:1991qm}
E.~Braaten, S.~Narison, and A.~Pich {\em Nucl. Phys.} {\bf B373} (1992)
  581--612.

\bibitem{Boito:2014sta}
D.~Boito, M.~Golterman, K.~Maltman, J.~Osborne, and S.~Peris {\em Phys. Rev.}
  {\bf D91} (2015), no.~3 034003, [\href{http://arxiv.org/abs/1410.3528}{{\tt
  arXiv:1410.3528}}].

\bibitem{Pich:2016bdg}
A.~Pich and A.~Rodriguez-Sanchez {\em Phys. Rev.} {\bf D94} (2016), no.~3
  034027, [\href{http://arxiv.org/abs/1605.06830}{{\tt arXiv:1605.06830}}].

\bibitem{Gamiz:2002nu}
E.~G\'amiz et~al. {\em JHEP} {\bf 01} (2003) 060,
  [\href{http://arxiv.org/abs/hep-ph/0212230}{{\tt hep-ph/0212230}}].

\bibitem{Gamiz:2004ar}
E.~G\'amiz et~al. {\em Phys. Rev. Lett.} {\bf 94} (2005) 011803,
  [\href{http://arxiv.org/abs/hep-ph/0408044}{{\tt hep-ph/0408044}}].

\bibitem{Boito:2015fra}
D.~Boito, A.~Francis, M.~Golterman, R.~Hudspith, R.~Lewis, K.~Maltman, and
  S.~Peris {\em Phys. Rev.} {\bf D92} (2015), no.~11 114501,
  [\href{http://arxiv.org/abs/1503.03450}{{\tt arXiv:1503.03450}}].

\bibitem{Gonzalez-Alonso:2016ndl}
M.~Gonz\'alez-Alonso, A.~Pich, and A.~Rodr\'\i{}guez-S\'anchez {\em Phys. Rev.
  D} {\bf 94} (2016), no.~1 014017,
  [\href{http://arxiv.org/abs/1602.06112}{{\tt arXiv:1602.06112}}].

\bibitem{Aoki:2021kgd}
Y.~Aoki et~al. \href{http://arxiv.org/abs/2111.09849}{{\tt arXiv:2111.09849}}.

\bibitem{Seng:2018yzq}
C.-Y. Seng, M.~Gorchtein, H.~H. Patel, and M.~J. Ramsey-Musolf {\em Phys. Rev.
  Lett.} {\bf 121} (2018), no.~24 241804,
  [\href{http://arxiv.org/abs/1807.10197}{{\tt arXiv:1807.10197}}].

\bibitem{Grossman:2019bzp}
Y.~Grossman, E.~Passemar, and S.~Schacht {\em JHEP} {\bf 07} (2020) 068,
  [\href{http://arxiv.org/abs/1911.07821}{{\tt arXiv:1911.07821}}].

\bibitem{Coutinho:2019aiy}
A.~M. Coutinho, A.~Crivellin, and C.~A. Manzari {\em Phys. Rev. Lett.} {\bf
  125} (2020), no.~7 071802, [\href{http://arxiv.org/abs/1912.08823}{{\tt
  arXiv:1912.08823}}].

\bibitem{HFLAV:2019otj}
{\bf HFLAV} Collaboration, Y.~S. Amhis et~al. {\em Eur. Phys. J. C} {\bf 81}
  (2021), no.~3 226, [\href{http://arxiv.org/abs/1909.12524}{{\tt
  arXiv:1909.12524}}].

\bibitem{LHCb:2021trn}
{\bf LHCb} Collaboration, R.~Aaij et~al.
  \href{http://arxiv.org/abs/2103.11769}{{\tt arXiv:2103.11769}}.

\bibitem{LHCb:2017avl}
{\bf LHCb} Collaboration, R.~Aaij et~al. {\em JHEP} {\bf 08} (2017) 055,
  [\href{http://arxiv.org/abs/1705.05802}{{\tt arXiv:1705.05802}}].

\bibitem{Bernard:2007cf}
V.~Bernard, M.~Oertel, E.~Passemar, and J.~Stern {\em JHEP} {\bf 01} (2008)
  015, [\href{http://arxiv.org/abs/0707.4194}{{\tt arXiv:0707.4194}}].

\bibitem{Garces:2017jpz}
E.~A. Garc\'es, M.~Hern\'andez~Villanueva, G.~L\'opez~Castro, and P.~Roig {\em
  JHEP} {\bf 12} (2017) 027, [\href{http://arxiv.org/abs/1708.07802}{{\tt
  arXiv:1708.07802}}].

\bibitem{Cirigliano:2017tqn}
V.~Cirigliano, A.~Crivellin, and M.~Hoferichter {\em Phys. Rev. Lett.} {\bf
  120} (2018), no.~14 141803, [\href{http://arxiv.org/abs/1712.06595}{{\tt
  arXiv:1712.06595}}].

\bibitem{Cirigliano:2018dyk}
V.~Cirigliano, A.~Falkowski, M.~Gonz\'alez-Alonso, and
  A.~Rodr\'\i{}guez-S\'anchez {\em Phys. Rev. Lett.} {\bf 122} (2019), no.~22
  221801, [\href{http://arxiv.org/abs/1809.01161}{{\tt arXiv:1809.01161}}].

\bibitem{Cirigliano:2009wk}
V.~Cirigliano, J.~Jenkins, and M.~Gonzalez-Alonso {\em Nucl. Phys.} {\bf B830}
  (2010) 95--115, [\href{http://arxiv.org/abs/0908.1754}{{\tt
  arXiv:0908.1754}}].

\bibitem{Cirigliano:2012ab}
V.~Cirigliano, M.~Gonzalez-Alonso, and M.~L. Graesser {\em JHEP} {\bf 02}
  (2013) 046, [\href{http://arxiv.org/abs/1210.4553}{{\tt arXiv:1210.4553}}].

\bibitem{Buchmuller:1985jz}
W.~Buchmuller and D.~Wyler {\em Nucl. Phys.} {\bf B268} (1986) 621--653.

\bibitem{Gonzalez-Alonso:2016etj}
M.~Gonzalez-Alonso and J.~Martin~Camalich {\em JHEP} {\bf 12} (2016) 052,
  [\href{http://arxiv.org/abs/1605.07114}{{\tt arXiv:1605.07114}}].

\bibitem{Falkowski:2019xoe}
A.~Falkowski, M.~Gonz\'alez-Alonso, and Z.~Tabrizi {\em JHEP} {\bf 05} (2019)
  173, [\href{http://arxiv.org/abs/1901.04553}{{\tt arXiv:1901.04553}}].

\bibitem{ParticleDataGroup:2020ssz}
{\bf Particle Data Group} Collaboration, P.~A. Zyla et~al. {\em PTEP} {\bf
  2020} (2020), no.~8 083C01.

\bibitem{Gonzalez-Alonso:2017iyc}
M.~Gonzalez-Alonso, J.~Martin~Camalich, and K.~Mimouni {\em Phys. Lett.} {\bf
  B772} (2017) 777--785, [\href{http://arxiv.org/abs/1706.00410}{{\tt
  arXiv:1706.00410}}].

\bibitem{Dorsner:2016wpm}
I.~Dor\v{s}ner, S.~Fajfer, A.~Greljo, J.~F. Kamenik, and N.~Ko\v{s}nik {\em
  Phys. Rept.} {\bf 641} (2016) 1--68,
  [\href{http://arxiv.org/abs/1603.04993}{{\tt arXiv:1603.04993}}].

\bibitem{Angelescu:2021lln}
A.~Angelescu, D.~Be\v{c}irevi\'c, D.~A. Faroughy, F.~Jaffredo, and O.~Sumensari
  {\em Phys. Rev. D} {\bf 104} (2021), no.~5 055017,
  [\href{http://arxiv.org/abs/2103.12504}{{\tt arXiv:2103.12504}}].

\bibitem{Descotes-Genon:2018foz}
S.~Descotes-Genon, A.~Falkowski, M.~Fedele, M.~Gonz\'alez-Alonso, and J.~Virto
  {\em JHEP} {\bf 05} (2019) 172, [\href{http://arxiv.org/abs/1812.08163}{{\tt
  arXiv:1812.08163}}].

\bibitem{Filipuzzi:2012mg}
A.~Filipuzzi, J.~Portoles, and M.~Gonzalez-Alonso {\em Phys. Rev.} {\bf D85}
  (2012) 116010, [\href{http://arxiv.org/abs/1203.2092}{{\tt
  arXiv:1203.2092}}].

\bibitem{Rodriguez-Sanchez:2018dzw}
A.~Rodriguez-Sanchez, {\em {Precision physics in Hadronic Tau Decays}}.
\newblock PhD thesis, Valencia U., 2018.

\bibitem{Dowdall:2013rya}
R.~Dowdall, C.~Davies, G.~Lepage, and C.~McNeile {\em Phys. Rev. D} {\bf 88}
  (2013) 074504, [\href{http://arxiv.org/abs/1303.1670}{{\tt
  arXiv:1303.1670}}].

\bibitem{Carrasco:2014poa}
N.~Carrasco et~al. {\em Phys. Rev. D} {\bf 91} (2015), no.~5 054507,
  [\href{http://arxiv.org/abs/1411.7908}{{\tt arXiv:1411.7908}}].

\bibitem{Bazavov:2017lyh}
A.~Bazavov et~al. {\em Phys. Rev. D} {\bf 98} (2018), no.~7 074512,
  [\href{http://arxiv.org/abs/1712.09262}{{\tt arXiv:1712.09262}}].

\bibitem{Miller:2020xhy}
N.~Miller et~al. {\em Phys. Rev. D} {\bf 102} (2020), no.~3 034507,
  [\href{http://arxiv.org/abs/2005.04795}{{\tt arXiv:2005.04795}}].

\bibitem{Follana:2007uv}
{\bf HPQCD, UKQCD} Collaboration, E.~Follana, C.~T.~H. Davies, G.~P. Lepage,
  and J.~Shigemitsu {\em Phys. Rev. Lett.} {\bf 100} (2008) 062002,
  [\href{http://arxiv.org/abs/0706.1726}{{\tt arXiv:0706.1726}}].

\bibitem{Bazavov:2010hj}
{\bf MILC} Collaboration, A.~Bazavov et~al. {\em PoS} {\bf LATTICE2010} (2010)
  074, [\href{http://arxiv.org/abs/1012.0868}{{\tt arXiv:1012.0868}}].

\bibitem{Durr:2010hr}
S.~Durr, Z.~Fodor, C.~Hoelbling, S.~Katz, S.~Krieg, T.~Kurth, L.~Lellouch,
  T.~Lippert, A.~Ramos, and K.~Szabo {\em Phys. Rev. D} {\bf 81} (2010) 054507,
  [\href{http://arxiv.org/abs/1001.4692}{{\tt arXiv:1001.4692}}].

\bibitem{Blum:2014tka}
{\bf RBC, UKQCD} Collaboration, T.~Blum et~al. {\em Phys. Rev.} {\bf D93}
  (2016), no.~7 074505, [\href{http://arxiv.org/abs/1411.7017}{{\tt
  arXiv:1411.7017}}].

\bibitem{Durr:2016ulb}
S.~D\"urr et~al. {\em Phys. Rev. D} {\bf 95} (2017), no.~5 054513,
  [\href{http://arxiv.org/abs/1601.05998}{{\tt arXiv:1601.05998}}].

\bibitem{Bornyakov:2016dzn}
{\bf QCDSF\textendash{}UKQCD} Collaboration, V.~Bornyakov, R.~Horsley,
  Y.~Nakamura, H.~Perlt, D.~Pleiter, P.~Rakow, G.~Schierholz, A.~Schiller,
  H.~St\"uben, and J.~Zanotti {\em Phys. Lett. B} {\bf 767} (2017) 366--373,
  [\href{http://arxiv.org/abs/1612.04798}{{\tt arXiv:1612.04798}}].

\bibitem{Blossier:2009bx}
{\bf ETM} Collaboration, B.~Blossier et~al. {\em JHEP} {\bf 07} (2009) 043,
  [\href{http://arxiv.org/abs/0904.0954}{{\tt arXiv:0904.0954}}].

\bibitem{Arroyo-Urena:2021nil}
M.~A. Arroyo-Ure\~na, G.~Hern\'andez-Tom\'e, G.~L\'opez-Castro, P.~Roig, and
  I.~Rosell {\em Phys. Rev. D} {\bf 104} (2021), no.~9 L091502,
  [\href{http://arxiv.org/abs/2107.04603}{{\tt arXiv:2107.04603}}].

\bibitem{Cirigliano:2007xi}
V.~Cirigliano and I.~Rosell {\em Phys.Rev.Lett.} {\bf 99} (2007) 231801,
  [\href{http://arxiv.org/abs/0707.3439}{{\tt arXiv:0707.3439}}].

\bibitem{Rosner:2015wva}
J.~L. Rosner, S.~Stone, and R.~S. Van~de Water {\em Submitted to: Particle Data
  Book} (2015) [\href{http://arxiv.org/abs/1509.02220}{{\tt
  arXiv:1509.02220}}].

\bibitem{Roig:2019rwf}
P.~Roig {\em EPJ Web Conf.} {\bf 212} (2019) 08002,
  [\href{http://arxiv.org/abs/1903.02682}{{\tt arXiv:1903.02682}}].

\bibitem{Cerri:2018ypt}
A.~Cerri et~al. {\em CERN Yellow Rep. Monogr.} {\bf 7} (2019) 867--1158,
  [\href{http://arxiv.org/abs/1812.07638}{{\tt arXiv:1812.07638}}].

\bibitem{Lueck-talk-ICHEP2018}
T.~Lueck XXXIX International Conference on High Energy Physics (ICHEP 2018),
  Seoul, South Korea, 2018.

\bibitem{Aubert:2009qj}
{\bf BaBar} Collaboration, B.~Aubert et~al. {\em Phys. Rev. Lett.} {\bf 105}
  (2010) 051602, [\href{http://arxiv.org/abs/0912.0242}{{\tt
  arXiv:0912.0242}}].

\bibitem{Miranda:2018cpf}
J.~A. Miranda and P.~Roig {\em JHEP} {\bf 11} (2018) 038,
  [\href{http://arxiv.org/abs/1806.09547}{{\tt arXiv:1806.09547}}].

\bibitem{Marciano:1988vm}
W.~J. Marciano and A.~Sirlin {\em Phys. Rev. Lett.} {\bf 61} (1988) 1815--1818.

\bibitem{Braaten:1990ef}
E.~Braaten and C.-S. Li {\em Phys. Rev.} {\bf D42} (1990) 3888--3891.

\bibitem{Erler:2002mv}
J.~Erler {\em Rev. Mex. Fis.} {\bf 50} (2004) 200--202,
  [\href{http://arxiv.org/abs/hep-ph/0211345}{{\tt hep-ph/0211345}}].

\bibitem{Cirigliano:2002pv}
V.~Cirigliano, G.~Ecker, and H.~Neufeld {\em JHEP} {\bf 08} (2002) 002,
  [\href{http://arxiv.org/abs/hep-ph/0207310}{{\tt hep-ph/0207310}}].

\bibitem{Davier:2009zi}
M.~Davier, A.~Hoecker, B.~Malaescu, C.~Z. Yuan, and Z.~Zhang {\em Eur. Phys.
  J.} {\bf C66} (2010) 1--9, [\href{http://arxiv.org/abs/0908.4300}{{\tt
  arXiv:0908.4300}}].

\bibitem{Antonelli:2013usa}
M.~Antonelli, V.~Cirigliano, A.~Lusiani, and E.~Passemar {\em JHEP} {\bf 10}
  (2013) 070, [\href{http://arxiv.org/abs/1304.8134}{{\tt arXiv:1304.8134}}].

\bibitem{Flores-Baez:2013eba}
F.~V. Flores-Ba\'ez and J.~R. Morones-Ibarra {\em Phys. Rev. D} {\bf 88}
  (2013), no.~7 073009, [\href{http://arxiv.org/abs/1307.1912}{{\tt
  arXiv:1307.1912}}].

\bibitem{Davier:2013sfa}
M.~Davier, A.~H\"ocker, B.~Malaescu, C.-Z. Yuan, and Z.~Zhang {\em Eur. Phys.
  J. C} {\bf 74} (2014), no.~3 2803,
  [\href{http://arxiv.org/abs/1312.1501}{{\tt arXiv:1312.1501}}].

\bibitem{Davier:2017zfy}
M.~Davier, A.~Hoecker, B.~Malaescu, and Z.~Zhang {\em Eur. Phys. J. C} {\bf 77}
  (2017), no.~12 827, [\href{http://arxiv.org/abs/1706.09436}{{\tt
  arXiv:1706.09436}}].

\bibitem{Keshavarzi:2018mgv}
A.~Keshavarzi, D.~Nomura, and T.~Teubner {\em Phys. Rev. D} {\bf 97} (2018),
  no.~11 114025, [\href{http://arxiv.org/abs/1802.02995}{{\tt
  arXiv:1802.02995}}].

\bibitem{Jegerlehner:2011ti}
F.~Jegerlehner and R.~Szafron {\em Eur. Phys. J.} {\bf C71} (2011) 1632,
  [\href{http://arxiv.org/abs/1101.2872}{{\tt arXiv:1101.2872}}].

\bibitem{Eidelman:1995ny}
S.~Eidelman and F.~Jegerlehner {\em Z. Phys.} {\bf C67} (1995) 585--602,
  [\href{http://arxiv.org/abs/hep-ph/9502298}{{\tt hep-ph/9502298}}].

\bibitem{Brodsky:1967sr}
S.~J. Brodsky and E.~De~Rafael {\em Phys. Rev.} {\bf 168} (1968) 1620--1622.

\bibitem{Bouchiat:1957zz}
C.~Bouchiat and L.~Michel {\em Phys. Rev.} {\bf 106} (1957) 170--172.

\bibitem{Cirigliano:2001er}
V.~Cirigliano, G.~Ecker, and H.~Neufeld {\em Phys. Lett. B} {\bf 513} (2001)
  361--370, [\href{http://arxiv.org/abs/hep-ph/0104267}{{\tt hep-ph/0104267}}].

\bibitem{Zhang:2009ag}
Z.~Zhang, M.~Davier, A.~Hoecker, G.~Lopez~Castro, B.~Malaescu, X.~H. Mo,
  G.~Toledo~Sanchez, P.~Wang, and C.~Z. Yuan {\em PoS} {\bf EPS-HEP2009} (2009)
  373.

\bibitem{Aoyama:2020ynm}
T.~Aoyama et~al. {\em Phys. Rept.} {\bf 887} (2020) 1--166,
  [\href{http://arxiv.org/abs/2006.04822}{{\tt arXiv:2006.04822}}].

\bibitem{Guerrero:1997ku}
F.~Guerrero and A.~Pich {\em Phys. Lett.} {\bf B412} (1997) 382--388,
  [\href{http://arxiv.org/abs/hep-ph/9707347}{{\tt hep-ph/9707347}}].

\bibitem{Celis:2013xja}
A.~Celis, V.~Cirigliano, and E.~Passemar {\em Phys. Rev.} {\bf D89} (2014)
  013008, [\href{http://arxiv.org/abs/1309.3564}{{\tt arXiv:1309.3564}}].

\bibitem{Ecker:1988te}
G.~Ecker, J.~Gasser, A.~Pich, and E.~de~Rafael {\em Nucl. Phys.} {\bf B321}
  (1989) 311.

\bibitem{Baum:2011rm}
I.~Baum, V.~Lubicz, G.~Martinelli, L.~Orifici, and S.~Simula {\em Phys. Rev.}
  {\bf D84} (2011) 074503, [\href{http://arxiv.org/abs/1108.1021}{{\tt
  arXiv:1108.1021}}].

\bibitem{Hoferichter:2018zwu}
M.~Hoferichter, B.~Kubis, J.~Ruiz~de Elvira, and P.~Stoffer {\em Phys. Rev.
  Lett.} {\bf 122} (2019), no.~12 122001,
  [\href{http://arxiv.org/abs/1811.11181}{{\tt arXiv:1811.11181}}]. [Erratum:
  Phys.Rev.Lett. 124, 199901 (2020)].

\bibitem{Mateu:2007tr}
V.~Mateu and J.~Portol\'es {\em Eur. Phys. J.} {\bf C52} (2007) 325--338,
  [\href{http://arxiv.org/abs/0706.1039}{{\tt arXiv:0706.1039}}].

\bibitem{Cata:2008zc}
O.~Cata and V.~Mateu {\em Phys. Rev.} {\bf D77} (2008) 116009,
  [\href{http://arxiv.org/abs/0801.4374}{{\tt arXiv:0801.4374}}].

\bibitem{Davier:2019can}
M.~Davier, A.~Hoecker, B.~Malaescu, and Z.~Zhang {\em Eur. Phys. J. C} {\bf 80}
  (2020), no.~3 241, [\href{http://arxiv.org/abs/1908.00921}{{\tt
  arXiv:1908.00921}}]. [Erratum: Eur.Phys.J.C 80, 410 (2020)].

\bibitem{Keshavarzi:2019abf}
A.~Keshavarzi, D.~Nomura, and T.~Teubner {\em Phys. Rev. D} {\bf 101} (2020),
  no.~1 014029, [\href{http://arxiv.org/abs/1911.00367}{{\tt
  arXiv:1911.00367}}].

\bibitem{Miranda:2020wdg}
J.~A. Miranda and P.~Roig {\em Phys. Rev. D} {\bf 102} (2020) 114017,
  [\href{http://arxiv.org/abs/2007.11019}{{\tt arXiv:2007.11019}}].

\bibitem{Bruno:2018ono}
M.~Bruno, T.~Izubuchi, C.~Lehner, and A.~Meyer {\em PoS} {\bf LATTICE2018}
  (2018) 135, [\href{http://arxiv.org/abs/1811.00508}{{\tt arXiv:1811.00508}}].

\bibitem{Escribano:2016ntp}
R.~Escribano, S.~Gonzalez-Solis, and P.~Roig {\em Phys. Rev.} {\bf D94} (2016),
  no.~3 034008, [\href{http://arxiv.org/abs/1601.03989}{{\tt
  arXiv:1601.03989}}].

\bibitem{Roig:private}
P.~Roig. Private communication.

\bibitem{delAmoSanchez:2010pc}
{\bf BaBar} Collaboration, P.~del Amo~Sanchez et~al. {\em Phys. Rev.} {\bf D83}
  (2011) 032002, [\href{http://arxiv.org/abs/1011.3917}{{\tt
  arXiv:1011.3917}}].

\bibitem{Belle-II:2018jsg}
{\bf Belle-II} Collaboration, W.~Altmannshofer et~al. {\em PTEP} {\bf 2019}
  (2019), no.~12 123C01, [\href{http://arxiv.org/abs/1808.10567}{{\tt
  arXiv:1808.10567}}]. [Erratum: PTEP 2020, 029201 (2020)].

\bibitem{Petar:1369}
P.~Rados {\em BELLE2-TALK-CONF-2019-026} (2019).

\bibitem{Moussallam:2021flg}
B.~Moussallam, {\it {Deriving experimental constraints on the scalar form
  factor in the second-class $\tau \to\eta \pi \nu$ mode}},  in {\em {16th
  International Workshop on Tau Lepton Physics~}}, 12, 2021.
\newblock \href{http://arxiv.org/abs/2112.04429}{{\tt arXiv:2112.04429}}.

\bibitem{Hayasaka:2009zz}
{\bf Belle} Collaboration, K.~Hayasaka {\em PoS} {\bf EPS-HEP2009} (2009) 374.

\bibitem{BABAR:2011aa}
{\bf BaBar} Collaboration, J.~P. Lees et~al. {\em Phys. Rev.} {\bf D85} (2012)
  031102, [\href{http://arxiv.org/abs/1109.1527}{{\tt arXiv:1109.1527}}].
  [Erratum: Phys. Rev.D85,099904(2012)].

\bibitem{Grossman:2011zk}
Y.~Grossman and Y.~Nir {\em JHEP} {\bf 04} (2012) 002,
  [\href{http://arxiv.org/abs/1110.3790}{{\tt arXiv:1110.3790}}].

\bibitem{Bigi:2005ts}
I.~I. Bigi and A.~I. Sanda {\em Phys. Lett.} {\bf B625} (2005) 47--52,
  [\href{http://arxiv.org/abs/hep-ph/0506037}{{\tt hep-ph/0506037}}].

\bibitem{Rendon:2019awg}
J.~Rend\'on, P.~Roig, and G.~Toledo~S\'anchez {\em Phys. Rev. D} {\bf 99}
  (2019), no.~9 093005, [\href{http://arxiv.org/abs/1902.08143}{{\tt
  arXiv:1902.08143}}].

\bibitem{Gonzalez-Solis:2020jlh}
S.~Gonz\`alez-Sol\'\i{}s, A.~Miranda, J.~Rend\'on, and P.~Roig {\em Phys. Lett.
  B} {\bf 804} (2020) 135371, [\href{http://arxiv.org/abs/1912.08725}{{\tt
  arXiv:1912.08725}}].

\bibitem{Gonzalez-Solis:2019lze}
S.~Gonz\`alez-Sol\'\i{}s, A.~Miranda, J.~Rend\'on, and P.~Roig {\em Phys. Rev.
  D} {\bf 101} (2020), no.~3 034010,
  [\href{http://arxiv.org/abs/1911.08341}{{\tt arXiv:1911.08341}}].

\bibitem{Pich:2021yll}
A.~Pich and A.~Rodr\'\i{}guez-S\'anchez {\em JHEP} {\bf 06} (2021) 005,
  [\href{http://arxiv.org/abs/2102.09308}{{\tt arXiv:2102.09308}}].

\bibitem{Gamiz:2007qs}
E.~G\'amiz et~al. {\em PoS} {\bf KAON} (2008) 008,
  [\href{http://arxiv.org/abs/0709.0282}{{\tt arXiv:0709.0282}}].

\bibitem{Hudspith:2017vew}
R.~J. Hudspith, R.~Lewis, K.~Maltman, and J.~Zanotti {\em Phys. Lett.} {\bf
  B781} (2018) 206--212, [\href{http://arxiv.org/abs/1702.01767}{{\tt
  arXiv:1702.01767}}].

\bibitem{Weinberg:1995mt}
S.~Weinberg, {\em The Quantum Theory of Fields. Vol. 1: Foundations.}
\newblock Cambridge University Press, 1995.

\bibitem{deRafael:1997ea}
E.~de~Rafael, {\it {An Introduction to sum rules in QCD: Course}},  in {\em
  {Les Houches Summer School in Theoretical Physics, Session 68: Probing the
  Standard Model of Particle Interactions}}, pp.~1171--1218, 7, 1997.
\newblock \href{http://arxiv.org/abs/hep-ph/9802448}{{\tt hep-ph/9802448}}.

\bibitem{Gonzalez-Alonso:2010vnm}
M.~Gonzalez-Alonso, {\em {Low-energy tests of the Standard Model}}.
\newblock PhD thesis, Valencia U., 2010.

\bibitem{Bijnens:2003rc}
J.~Bijnens, E.~Gamiz, E.~Lipartia, and J.~Prades {\em JHEP} {\bf 04} (2003)
  055, [\href{http://arxiv.org/abs/hep-ph/0304222}{{\tt hep-ph/0304222}}].

\bibitem{Shifman:1978bx}
M.~A. Shifman, A.~I. Vainshtein, and V.~I. Zakharov {\em Nucl. Phys.} {\bf
  B147} (1979) 385--447.

\bibitem{Maltman:2008bx}
K.~Maltman, D.~Leinweber, P.~Moran, and A.~Sternbeck {\em Phys. Rev.} {\bf D78}
  (2008) 114504, [\href{http://arxiv.org/abs/0807.2020}{{\tt
  arXiv:0807.2020}}].

\bibitem{PACS-CS:2009zxm}
{\bf PACS-CS} Collaboration, S.~Aoki et~al. {\em JHEP} {\bf 10} (2009) 053,
  [\href{http://arxiv.org/abs/0906.3906}{{\tt arXiv:0906.3906}}].

\bibitem{McNeile:2010ji}
C.~McNeile, C.~T.~H. Davies, E.~Follana, K.~Hornbostel, and G.~P. Lepage {\em
  Phys. Rev.} {\bf D82} (2010) 034512,
  [\href{http://arxiv.org/abs/1004.4285}{{\tt arXiv:1004.4285}}].

\bibitem{Chakraborty:2014aca}
B.~Chakraborty, C.~T.~H. Davies, B.~Galloway, P.~Knecht, J.~Koponen, G.~C.
  Donald, R.~J. Dowdall, G.~P. Lepage, and C.~McNeile {\em Phys. Rev.} {\bf
  D91} (2015), no.~5 054508, [\href{http://arxiv.org/abs/1408.4169}{{\tt
  arXiv:1408.4169}}].

\bibitem{Bruno:2017gxd}
{\bf ALPHA} Collaboration, M.~Bruno, M.~Dalla~Brida, P.~Fritzsch, T.~Korzec,
  A.~Ramos, S.~Schaefer, H.~Simma, S.~Sint, and R.~Sommer {\em Phys. Rev.
  Lett.} {\bf 119} (2017), no.~10 102001,
  [\href{http://arxiv.org/abs/1706.03821}{{\tt arXiv:1706.03821}}].

\bibitem{Bazavov:2019qoo}
{\bf TUMQCD} Collaboration, A.~Bazavov, N.~Brambilla, X.~Garcia~i Tormo,
  P.~Petreczky, J.~Soto, A.~Vairo, and J.~H. Weber {\em Phys. Rev. D} {\bf 100}
  (2019), no.~11 114511, [\href{http://arxiv.org/abs/1907.11747}{{\tt
  arXiv:1907.11747}}].

\bibitem{Cali:2020hrj}
S.~Cali, K.~Cichy, P.~Korcyl, and J.~Simeth {\em Phys. Rev. Lett.} {\bf 125}
  (2020) 242002, [\href{http://arxiv.org/abs/2003.05781}{{\tt
  arXiv:2003.05781}}].

\bibitem{Ayala:2020odx}
C.~Ayala, X.~Lobregat, and A.~Pineda {\em JHEP} {\bf 09} (2020) 016,
  [\href{http://arxiv.org/abs/2005.12301}{{\tt arXiv:2005.12301}}].

\bibitem{Narison:1988ni}
S.~Narison and A.~Pich {\em Phys. Lett.} {\bf B211} (1988) 183.

\bibitem{Davier:2005xq}
M.~Davier, A.~Hocker, and Z.~Zhang {\em Rev. Mod. Phys.} {\bf 78} (2006)
  1043--1109, [\href{http://arxiv.org/abs/hep-ph/0507078}{{\tt
  hep-ph/0507078}}].

\bibitem{Davier:2008sk}
M.~Davier et~al. {\em Eur. Phys. J.} {\bf C56} (2008) 305--322,
  [\href{http://arxiv.org/abs/0803.0979}{{\tt arXiv:0803.0979}}].

\bibitem{Beneke:2008ad}
M.~Beneke and M.~Jamin {\em JHEP} {\bf 09} (2008) 044,
  [\href{http://arxiv.org/abs/0806.3156}{{\tt arXiv:0806.3156}}].

\bibitem{Beneke:2012vb}
M.~Beneke, D.~Boito, and M.~Jamin {\em JHEP} {\bf 01} (2013) 125,
  [\href{http://arxiv.org/abs/1210.8038}{{\tt arXiv:1210.8038}}].

\bibitem{Caprini:2011ya}
I.~Caprini and J.~Fischer {\em Phys. Rev. D} {\bf 84} (2011) 054019,
  [\href{http://arxiv.org/abs/1106.5336}{{\tt arXiv:1106.5336}}].

\bibitem{Abbas:2012fi}
G.~Abbas, B.~Ananthanarayan, I.~Caprini, and J.~Fischer {\em Phys. Rev. D} {\bf
  87} (2013), no.~1 014008, [\href{http://arxiv.org/abs/1211.4316}{{\tt
  arXiv:1211.4316}}].

\bibitem{Abbas:2012py}
G.~Abbas, B.~Ananthanarayan, and I.~Caprini {\em Phys. Rev. D} {\bf 85} (2012)
  094018, [\href{http://arxiv.org/abs/1202.2672}{{\tt arXiv:1202.2672}}].

\bibitem{Groote:2012jq}
S.~Groote, J.~G. Korner, and A.~A. Pivovarov {\em Phys. Part. Nucl.} {\bf 44}
  (2013) 285--298, [\href{http://arxiv.org/abs/1212.5346}{{\tt
  arXiv:1212.5346}}].

\bibitem{Baikov:2008jh}
P.~A. Baikov, K.~G. Chetyrkin, and J.~H. Kuhn {\em Phys. Rev. Lett.} {\bf 101}
  (2008) 012002, [\href{http://arxiv.org/abs/0801.1821}{{\tt
  arXiv:0801.1821}}].

\bibitem{Maltman:2008nf}
K.~Maltman and T.~Yavin {\em Phys. Rev.} {\bf D78} (2008) 094020,
  [\href{http://arxiv.org/abs/0807.0650}{{\tt arXiv:0807.0650}}].

\bibitem{Boito:2012cr}
D.~Boito, M.~Golterman, M.~Jamin, A.~Mahdavi, K.~Maltman, J.~Osborne, and
  S.~Peris {\em Phys. Rev. D} {\bf 85} (2012) 093015,
  [\href{http://arxiv.org/abs/1203.3146}{{\tt arXiv:1203.3146}}].

\bibitem{Menke:2009vg}
S.~Menke \href{http://arxiv.org/abs/0904.1796}{{\tt arXiv:0904.1796}}.

\bibitem{Narison:2009vy}
S.~Narison {\em Phys. Lett. B} {\bf 673} (2009) 30--36,
  [\href{http://arxiv.org/abs/0901.3823}{{\tt arXiv:0901.3823}}].

\bibitem{Cvetic:2010ut}
G.~Cvetic, M.~Loewe, C.~Martinez, and C.~Valenzuela {\em Phys. Rev. D} {\bf 82}
  (2010) 093007, [\href{http://arxiv.org/abs/1005.4444}{{\tt
  arXiv:1005.4444}}].

\bibitem{Pich:2011bb}
A.~Pich, {\it {Tau Decay Determination of the QCD Coupling}},  in {\em
  {Workshop on Precision Measurements of $\alpha_s$}}, 7, 2011.
\newblock \href{http://arxiv.org/abs/1107.1123}{{\tt arXiv:1107.1123}}.

\bibitem{Pich:2013sqa}
A.~Pich {\em PoS} {\bf ConfinementX} (2012) 022,
  [\href{http://arxiv.org/abs/1303.2262}{{\tt arXiv:1303.2262}}].

\bibitem{Boito:2020hvu}
D.~Boito and F.~Oliani {\em Phys. Rev. D} {\bf 101} (2020), no.~7 074003,
  [\href{http://arxiv.org/abs/2002.12419}{{\tt arXiv:2002.12419}}].

\bibitem{Caprini:2020lff}
I.~Caprini {\em Phys. Rev. D} {\bf 102} (2020), no.~5 054017,
  [\href{http://arxiv.org/abs/2006.16605}{{\tt arXiv:2006.16605}}].

\bibitem{Ayala:2021mwc}
C.~Ayala, G.~Cvetic, and D.~Teca {\em Eur. Phys. J. C} {\bf 81} (2021), no.~10
  930, [\href{http://arxiv.org/abs/2105.00356}{{\tt arXiv:2105.00356}}].

\bibitem{Weinberg:1967kj}
S.~Weinberg {\em Phys. Rev. Lett.} {\bf 18} (1967) 507--509.

\bibitem{Knecht:1997ts}
M.~Knecht and E.~de~Rafael {\em Phys. Lett.} {\bf B424} (1998) 335--342,
  [\href{http://arxiv.org/abs/hep-ph/9712457}{{\tt hep-ph/9712457}}].

\bibitem{Boyle:2014pja}
P.~A. Boyle, L.~Del~Debbio, N.~Garron, R.~J. Hudspith, E.~Kerrane, K.~Maltman,
  and J.~M. Zanotti {\em Phys. Rev. D} {\bf 89} (2014), no.~9 094510,
  [\href{http://arxiv.org/abs/1403.6729}{{\tt arXiv:1403.6729}}].

\bibitem{Donoghue:1999ku}
J.~F. Donoghue and E.~Golowich {\em Phys. Lett.} {\bf B478} (2000) 172--184,
  [\href{http://arxiv.org/abs/hep-ph/9911309}{{\tt hep-ph/9911309}}].

\bibitem{Cirigliano:2001qw}
V.~Cirigliano, J.~F. Donoghue, E.~Golowich, and K.~Maltman {\em Phys. Lett.}
  {\bf B522} (2001) 245--256, [\href{http://arxiv.org/abs/hep-ph/0109113}{{\tt
  hep-ph/0109113}}].

\bibitem{Cirigliano:2002jy}
V.~Cirigliano, J.~F. Donoghue, E.~Golowich, and K.~Maltman {\em Phys. Lett.}
  {\bf B555} (2003) 71--82, [\href{http://arxiv.org/abs/hep-ph/0211420}{{\tt
  hep-ph/0211420}}].

\bibitem{Abbott:2020hxn}
{\bf RBC, UKQCD} Collaboration, R.~Abbott et~al. {\em Phys. Rev. D} {\bf 102}
  (2020), no.~5 054509, [\href{http://arxiv.org/abs/2004.09440}{{\tt
  arXiv:2004.09440}}].

\bibitem{Pich:2020gzz}
A.~Pich {\em Prog. Part. Nucl. Phys.} {\bf 117} (2021) 103846,
  [\href{http://arxiv.org/abs/2012.04716}{{\tt arXiv:2012.04716}}].

\bibitem{Boito:2020xli}
D.~Boito, M.~Golterman, K.~Maltman, S.~Peris, M.~V. Rodrigues, and W.~Schaaf
  {\em Phys. Rev. D} {\bf 103} (2021), no.~3 034028,
  [\href{http://arxiv.org/abs/2012.10440}{{\tt arXiv:2012.10440}}].

\bibitem{dEnterria:2018cye}
D.~d'Enterria {\em PoS} {\bf DIS2018} (2018) 109,
  [\href{http://arxiv.org/abs/1806.06156}{{\tt arXiv:1806.06156}}].

\bibitem{Gamiz:2013wn}
E.~Gamiz, {\it {$|V_{us}|$ from hadronic $\tau$ decays}},  in {\em {7th
  International Workshop on the CKM Unitarity Triangle}}, 1, 2013.
\newblock \href{http://arxiv.org/abs/1301.2206}{{\tt arXiv:1301.2206}}.

\bibitem{Pich:1999hc}
A.~Pich and J.~Prades {\em JHEP} {\bf 10} (1999) 004,
  [\href{http://arxiv.org/abs/hep-ph/9909244}{{\tt hep-ph/9909244}}].

\bibitem{Dighe:2019odu}
A.~Dighe, S.~Ghosh, G.~Kumar, and T.~S. Roy
  \href{http://arxiv.org/abs/1902.09561}{{\tt arXiv:1902.09561}}.

\bibitem{RBC:2018uyk}
{\bf RBC, UKQCD} Collaboration, P.~Boyle, R.~J. Hudspith, T.~Izubuchi,
  A.~J\"uttner, C.~Lehner, R.~Lewis, K.~Maltman, H.~Ohki, A.~Portelli, and
  M.~Spraggs {\em Phys. Rev. Lett.} {\bf 121} (2018), no.~20 202003,
  [\href{http://arxiv.org/abs/1803.07228}{{\tt arXiv:1803.07228}}].

\bibitem{Boito:2016pwf}
D.~Boito, M.~Jamin, and R.~Miravitllas {\em Phys. Rev. Lett.} {\bf 117} (2016),
  no.~15 152001, [\href{http://arxiv.org/abs/1606.06175}{{\tt
  arXiv:1606.06175}}].

\bibitem{Boito:2018rwt}
D.~Boito, P.~Masjuan, and F.~Oliani {\em JHEP} {\bf 08} (2018) 075,
  [\href{http://arxiv.org/abs/1807.01567}{{\tt arXiv:1807.01567}}].

\bibitem{Hoang:2020mkw}
A.~H. Hoang and C.~Regner \href{http://arxiv.org/abs/2008.00578}{{\tt
  arXiv:2008.00578}}.

\bibitem{Hoang:2021nlz}
A.~H. Hoang and C.~Regner {\em The European Physical Journal Special Topics}
  {\bf 230} (2021), no.~12 [\href{http://arxiv.org/abs/2105.11222}{{\tt
  arXiv:2105.11222}}].

\bibitem{Crivellin:2020lzu}
A.~Crivellin and M.~Hoferichter {\em Phys. Rev. Lett.} {\bf 125} (2020), no.~11
  111801, [\href{http://arxiv.org/abs/2002.07184}{{\tt arXiv:2002.07184}}].

\bibitem{Crivellin:2021njn}
A.~Crivellin, M.~Hoferichter, and C.~A. Manzari {\em Phys. Rev. Lett.} {\bf
  127} (2021), no.~7 071801, [\href{http://arxiv.org/abs/2102.02825}{{\tt
  arXiv:2102.02825}}].

\bibitem{Bernard:2006gy}
V.~Bernard, M.~Oertel, E.~Passemar, and J.~Stern {\em Phys. Lett.} {\bf B638}
  (2006) 480--486, [\href{http://arxiv.org/abs/hep-ph/0603202}{{\tt
  hep-ph/0603202}}].

\bibitem{Falkowski:2020pma}
A.~Falkowski, M.~Gonz\'alez-Alonso, and O.~Naviliat-Cuncic {\em JHEP} {\bf 04}
  (2021) 126, [\href{http://arxiv.org/abs/2010.13797}{{\tt arXiv:2010.13797}}].

\bibitem{Darius:2017arh}
G.~Darius et~al. {\em Phys. Rev. Lett.} {\bf 119} (2017), no.~4 042502.

\bibitem{Hassan:2020hrj}
M.~T. Hassan et~al. {\em Phys. Rev. C} {\bf 103} (2021), no.~4 045502,
  [\href{http://arxiv.org/abs/2012.14379}{{\tt arXiv:2012.14379}}].

\bibitem{UCNt:2021pcg}
{\bf UCN\ensuremath{\tau}} Collaboration, F.~M. Gonzalez et~al. {\em Phys. Rev.
  Lett.} {\bf 127} (2021), no.~16 162501,
  [\href{http://arxiv.org/abs/2106.10375}{{\tt arXiv:2106.10375}}].

\bibitem{Gupta:2018qil}
R.~Gupta, Y.-C. Jang, B.~Yoon, H.-W. Lin, V.~Cirigliano, and T.~Bhattacharya
  {\em Phys. Rev. D} {\bf 98} (2018) 034503,
  [\href{http://arxiv.org/abs/1806.09006}{{\tt arXiv:1806.09006}}].

\bibitem{Chang:2018uxx}
C.~C. Chang et~al. {\em Nature} {\bf 558} (2018), no.~7708 91--94,
  [\href{http://arxiv.org/abs/1805.12130}{{\tt arXiv:1805.12130}}].

\bibitem{Walker-Loud:2019cif}
A.~Walker-Loud et~al. {\em PoS} {\bf CD2018} (2020) 020,
  [\href{http://arxiv.org/abs/1912.08321}{{\tt arXiv:1912.08321}}].

\bibitem{Gorchtein:2021fce}
M.~Gorchtein and C.-Y. Seng {\em JHEP} {\bf 10} (2021) 053,
  [\href{http://arxiv.org/abs/2106.09185}{{\tt arXiv:2106.09185}}].

\bibitem{Bali:2020bcn}
G.~S. Bali, G.~Endr\H{o}di, and S.~Piemonte {\em JHEP} {\bf 07} (2020) 183,
  [\href{http://arxiv.org/abs/2004.08778}{{\tt arXiv:2004.08778}}].

\bibitem{Czarnecki:2019iwz}
A.~Czarnecki, W.~J. Marciano, and A.~Sirlin {\em Phys. Rev. D} {\bf 101}
  (2020), no.~9 091301, [\href{http://arxiv.org/abs/1911.04685}{{\tt
  arXiv:1911.04685}}].

\bibitem{Seng:2021nar}
C.-Y. Seng, D.~Galviz, W.~J. Marciano, and U.-G. Mei\ss{}ner
  \href{http://arxiv.org/abs/2107.14708}{{\tt arXiv:2107.14708}}.

\bibitem{FermilabLattice:2018zqv}
{\bf Fermilab Lattice, MILC} Collaboration, A.~Bazavov et~al. {\em Phys. Rev.
  D} {\bf 99} (2019), no.~11 114509,
  [\href{http://arxiv.org/abs/1809.02827}{{\tt arXiv:1809.02827}}].

\bibitem{Carrasco:2016kpy}
N.~Carrasco, P.~Lami, V.~Lubicz, L.~Riggio, S.~Simula, and C.~Tarantino {\em
  Phys. Rev. D} {\bf 93} (2016), no.~11 114512,
  [\href{http://arxiv.org/abs/1602.04113}{{\tt arXiv:1602.04113}}].

\bibitem{Cirigliano:2011tm}
V.~Cirigliano and H.~Neufeld {\em Phys. Lett.} {\bf B700} (2011) 7--10,
  [\href{http://arxiv.org/abs/1102.0563}{{\tt arXiv:1102.0563}}].

\bibitem{Yushchenko:2004zs}
O.~P. Yushchenko et~al. {\em Phys. Lett.} {\bf B589} (2004) 111--117,
  [\href{http://arxiv.org/abs/hep-ex/0404030}{{\tt hep-ex/0404030}}].

\bibitem{Yushchenko:2017fzv}
{\bf OKA} Collaboration, O.~P. Yushchenko et~al. {\em JETP Lett.} {\bf 107}
  (2018), no.~3 139--142, [\href{http://arxiv.org/abs/1708.09587}{{\tt
  arXiv:1708.09587}}]. [Pisma Zh. Eksp. Teor. Fiz.107,no.3,147(2018)].

\bibitem{Chibisov:1996wf}
B.~Chibisov, R.~D. Dikeman, M.~A. Shifman, and N.~Uraltsev {\em Int. J. Mod.
  Phys.} {\bf A12} (1997) 2075--2133,
  [\href{http://arxiv.org/abs/hep-ph/9605465}{{\tt hep-ph/9605465}}].

\bibitem{Shifman:2000jv}
M.~A. Shifman, {\it {Quark hadron duality}},  in {\em {8th International
  Symposium on Heavy Flavor Physics}}, vol.~3, (Singapore), pp.~1447--1494,
  World Scientific, 7, 2000.
\newblock \href{http://arxiv.org/abs/hep-ph/0009131}{{\tt hep-ph/0009131}}.

\bibitem{Cata:2005zj}
O.~Cat\`a, M.~Golterman, and S.~Peris {\em JHEP} {\bf 08} (2005) 076,
  [\href{http://arxiv.org/abs/hep-ph/0506004}{{\tt hep-ph/0506004}}].

\bibitem{GonzalezAlonso:2010xf}
M.~Gonzalez-Alonso, A.~Pich, and J.~Prades {\em Phys. Rev.} {\bf D82} (2010)
  014019, [\href{http://arxiv.org/abs/1004.4987}{{\tt arXiv:1004.4987}}].

\bibitem{Boito:2017cnp}
D.~Boito, I.~Caprini, M.~Golterman, K.~Maltman, and S.~Peris {\em Phys. Rev.}
  {\bf D97} (2018), no.~5 054007, [\href{http://arxiv.org/abs/1711.10316}{{\tt
  arXiv:1711.10316}}].

\bibitem{Adler:1974gd}
S.~L. Adler {\em Phys. Rev.} {\bf D10} (1974) 3714.

\bibitem{Craigie:1981jx}
N.~S. Craigie and J.~Stern {\em Phys. Rev.} {\bf D26} (1982) 2430.

\bibitem{Jamin:2008rm}
M.~Jamin and V.~Mateu {\em JHEP} {\bf 04} (2008) 040,
  [\href{http://arxiv.org/abs/0802.2669}{{\tt arXiv:0802.2669}}].

\bibitem{Buchalla:1995vs}
G.~Buchalla, A.~J. Buras, and M.~E. Lautenbacher {\em Rev. Mod. Phys.} {\bf 68}
  (1996) 1125--1144, [\href{http://arxiv.org/abs/hep-ph/9512380}{{\tt
  hep-ph/9512380}}].

\bibitem{Bazavov:2010yq}
A.~Bazavov et~al. {\em PoS} {\bf LATTICE2010} (2010) 083,
  [\href{http://arxiv.org/abs/1011.1792}{{\tt arXiv:1011.1792}}].

\bibitem{Borsanyi:2012zv}
S.~Borsanyi, S.~Durr, Z.~Fodor, S.~Krieg, A.~Schafer, E.~E. Scholz, and K.~K.
  Szabo {\em Phys. Rev.} {\bf D88} (2013) 014513,
  [\href{http://arxiv.org/abs/1205.0788}{{\tt arXiv:1205.0788}}].

\bibitem{Durr:2013goa}
{\bf Budapest-Marseille-Wuppertal} Collaboration, S.~Dürr et~al. {\em Phys.
  Rev.} {\bf D90} (2014), no.~11 114504,
  [\href{http://arxiv.org/abs/1310.3626}{{\tt arXiv:1310.3626}}].

\bibitem{Boyle:2015exm}
P.~A. Boyle et~al. {\em Phys. Rev.} {\bf D93} (2016), no.~5 054502,
  [\href{http://arxiv.org/abs/1511.01950}{{\tt arXiv:1511.01950}}].

\bibitem{Cossu:2016eqs}
G.~Cossu, H.~Fukaya, S.~Hashimoto, T.~Kaneko, and J.-I. Noaki {\em PTEP} {\bf
  2016} (2016), no.~9 093B06, [\href{http://arxiv.org/abs/1607.01099}{{\tt
  arXiv:1607.01099}}].

\bibitem{Aoki:2017paw}
{\bf JLQCD} Collaboration, S.~Aoki, G.~Cossu, H.~Fukaya, S.~Hashimoto, and
  T.~Kaneko {\em PTEP} {\bf 2018} (2018), no.~4 043B07,
  [\href{http://arxiv.org/abs/1705.10906}{{\tt arXiv:1705.10906}}].

\bibitem{Belyaev:1982sa}
V.~M. Belyaev and B.~L. Ioffe {\em Sov. Phys. JETP} {\bf 56} (1982) 493--501.
  [Zh. Eksp. Teor. Fiz.83,876(1982)].

\bibitem{Voloshin:1992sn}
M.~B. Voloshin {\em Phys. Lett.} {\bf B283} (1992) 120--122.

\bibitem{Ioffe:1983ju}
B.~L. Ioffe and A.~V. Smilga {\em Nucl. Phys.} {\bf B232} (1984) 109--142.

\bibitem{Cata:2007ns}
O.~Cat\`a and V.~Mateu {\em JHEP} {\bf 09} (2007) 078,
  [\href{http://arxiv.org/abs/0705.2948}{{\tt arXiv:0705.2948}}].

\bibitem{Balitsky:1985aq}
I.~I. Balitsky, A.~V. Kolesnichenko, and A.~V. Yung {\em Sov. J. Nucl. Phys.}
  {\bf 41} (1985) 178.

\bibitem{Knecht:2001xc}
M.~Knecht and A.~Nyffeler {\em Eur. Phys. J.} {\bf C21} (2001) 659--678,
  [\href{http://arxiv.org/abs/hep-ph/0106034}{{\tt hep-ph/0106034}}].

\bibitem{Bijnens:2020xnl}
J.~Bijnens, N.~Hermansson-Truedsson, L.~Laub, and A.~Rodr\'\i{}guez-S\'anchez
  {\em JHEP} {\bf 10} (2020) 203, [\href{http://arxiv.org/abs/2008.13487}{{\tt
  arXiv:2008.13487}}].

\end{thebibliography}\endgroup

\end{document}